\documentclass[11pt,usletterpaper]{article}
\usepackage[margin=1in]{geometry}
\usepackage[margin=5pt,font=small,labelfont=bf]{caption}
\usepackage{graphicx,amsmath,amssymb,fge,natbib,setspace,lineno,xspace,color}
\definecolor{darkblue}{rgb}{0.05,0.0,0.6}
\usepackage[colorlinks=true,breaklinks=true,linkcolor=darkblue,urlcolor=darkblue,anchorcolor=darkblue,citecolor=darkblue]{hyperref}
\usepackage{fancyhdr}
\newcommand{\T}{\rule{0pt}{3.2ex}} 
\newcommand{\B}{\rule[-1.2ex]{0pt}{0pt}} 

\newcommand{\trp}[1]{\left[ #1 \right]^\mathrm{T}}

\newcommand{\ddt}[1]{\dfrac{\partial #1}{\partial t}}
\newcommand{\DiDt}[1]{\dfrac{D_\phi^i #1}{Dt}}

\newcommand{\DstarDt}[1]{\dfrac{D^* #1}{Dt}}
\newcommand{\DbDt}[1]{\dfrac{\bar{D} #1}{Dt}}

\newcommand{\Div}{\Grad\!\cdot}
\newcommand{\Grad}{\mbox{\boldmath $\nabla$}}

\newcommand{\normsum}[2]{\left[#1\right]_{\Sigma_{#2}}}

\newcommand{\vel}{\mathbf{v}}

\newcommand{\vs}{\mathbf{v}^s}
\newcommand{\vl}{\mathbf{v}^{\ell}}
\newcommand{\vi}{\mathbf{v}^i}

\newcommand{\vb}{\bar{\mathbf{v}}}
\newcommand{\vdi}{\mathbf{v}_\Delta^i}
\newcommand{\vdl}{\mathbf{v}_\Delta^\ell}
\newcommand{\vds}{\mathbf{v}_\Delta^s}
\newcommand{\vdv}{\mathbf{v}_\Delta^v}

\newcommand{\phii}{\phi^i}
\newcommand{\phik}{\phi^k}
\newcommand{\phil}{\phi^\ell}
\newcommand{\phis}{\phi^s}
\newcommand{\phiv}{\phi^v}

\newcommand{\q}{\mathbf{q}}

\newcommand{\etai}{\eta^i}

\newcommand{\etas}{\eta^s}
\newcommand{\etal}{\eta^\ell}
\newcommand{\etav}{\eta^v}

\newcommand{\zetas}{\zeta^s}

\newcommand{\Ps}{P^s}
\newcommand{\Pl}{P^\ell}

\newcommand{\Pdi}{P_\Delta^i}
\newcommand{\Pdl}{P_\Delta^\ell}
\newcommand{\Pds}{P_\Delta^s}

\newcommand{\Pb}{\bar{P}}

\newcommand{\ei}{e^i}
\newcommand{\ui}{u^i}
\newcommand{\si}{s^i}
\newcommand{\Ti}{T^i}
\renewcommand{\Pi}{P^i}
\newcommand{\rhoi}{\rho^i}

\newcommand{\rhos}{\rho^s}
\newcommand{\rhol}{\rho^{\ell}}

\newcommand{\rhov}{\rho^v}
\newcommand{\rhob}{\bar{\rho}}

\newcommand{\gvec}{\mathbf{g}}
\newcommand{\xvec}{\mathbf{x}}

\newcommand{\Di}{\mathbf{\underline{D}}^i}
\newcommand{\Dstar}{\mathbf{\underline{D}}^*}
\newcommand{\Ds}{\mathbf{\underline{D}}^s}
\newcommand{\Dl}{\mathbf{\underline{D}}^\ell}

\newcommand{\Wi}{\mathbf{\underline{W}}^i}
\newcommand{\Vi}{\mathbf{\underline{V}}^i}

\newcommand{\I}{\mathbf{I}}
\newcommand{\cji}{c_j^i}

\newcommand{\qvi}{\mathbf{\underline{q}}_v^i}
\newcommand{\qei}{\mathbf{q}_e^i}
\newcommand{\qji}{\mathbf{q}_j^i}
\newcommand{\Qai}{\mathbf{Q}_a^i}

\newcommand{\Qji}{Q_j^i}
\newcommand{\Qvi}{\mathbf{Q}_v^i}
\newcommand{\Qei}{Q_e^i}

\newcommand{\Qsi}{Q_s^i}
\newcommand{\Qmi}{Q_\rho^i}
\newcommand{\Uai}{\boldsymbol{\Upsilon}_a^i}

\newcommand{\Usi}{\Upsilon_s^i}

\newcommand{\qai}{\mathbf{\underline{q}}_a^i}

\newcommand{\qsi}{\mathbf{q}_s^i}
\newcommand{\qfi}{\mathbf{q}_\phi^i}

\newcommand{\half}{\dfrac{1}{2}}

\newcommand{\third}{\dfrac{1}{3}}

\newcommand{\rhostar}{\rho^*}

\newcommand{\Pstar}{P^*}
\newcommand{\GPstar}{(\Grad P)^* }
\newcommand{\DGPstar}{\Delta (\Grad P)^{i*} }
\newcommand{\Gphistar}{(\Grad \phi)^*}
\newcommand{\DGphistar}{\Delta (\Grad \phi)^{i*}}
\newcommand{\vstar}{\mathbf{v}^*}
\newcommand{\Tstar}{T^*}

\newcommand{\mjstar}{\mu_j^*}
\newcommand{\mjstarP}{\tilde{\mu}_j^*}

\newcommand{\Gmstari}{(\Grad \mu^i)_*}
\newcommand{\DGmstari}{\Delta (\Grad \mu^i)_{j*}}
\newcommand{\Gcstari}{(\Grad c^i)_*}
\newcommand{\DGcstari}{\Delta (\Grad c^i)_{j*}}
\newcommand{\mstari}{\mu_*^i}
\newcommand{\Gji}{\Gamma_j^i}
\newcommand{\Gai}{\boldsymbol{\Gamma}_a^i}
\newcommand{\Gmi}{\Gamma_\rho^i}
\newcommand{\Gvi}{\boldsymbol{\Gamma}_v^i}
\newcommand{\Gei}{\Gamma_e^i}

\newcommand{\Gsi}{\Gamma_s^i}
\newcommand{\Gfi}{\Gamma_\phi^i}
\newcommand{\ai}{\mathbf{a}^i}

\newcommand{\oai}{\omega_{C_a}^i}

\newcommand{\ovi}{\omega_{C_v}^i}

\newcommand{\ofi}{\omega_{C_\phi}^i}

\newcommand{\Dai}{\Delta \mathbf{a}^{i*}}
\newcommand{\DTi}{\Delta T^{i*}}
\newcommand{\Dvi}{\Delta \mathbf{v}^{i*}}

\newcommand{\DPi}{\Delta P^{i*}}

\newcommand{\Dmji}{\Delta \mu_j^{i*}}
\newcommand{\DmjiP}{\Delta \tilde{\mu}_j^{i*}}

\newcommand{\mji}{\mu_j^i}
\newcommand{\Csi}{C_s^i}
\newcommand{\Cvi}{C_v^i}
\newcommand{\Cji}{C_j^i}
\newcommand{\Cfi}{C_\phi^i}
\newcommand{\Ksi}{K_s^i}
\newcommand{\Kvi}{K_v^i}
\newcommand{\Kji}{K_j^i}
\newcommand{\Kfi}{K_\phi^i}

\newcommand{\Drho}{\Delta \rho}

\title{A continuum model of multi-phase reactive transport \\ in igneous systems}

\author{Tobias Keller$^{1,*}$ \& Jenny Suckale$^1$ \\
{\small $^1$Stanford University, Geophysics Department, 397 Panama Mall, Stanford CA 94305, USA.} \\
{\small $^*$corresponding author: tokeller@stanford.edu. }}

\begin{document}
\maketitle 

\begin{abstract}
Multi-phase reactive transport processes are ubiquitous in igneous systems. A challenging aspect of modelling igneous phenomena is that they range from solid-dominated porous to liquid-dominated suspension flows and therefore entail a wide spectrum of rheological conditions, flow speeds, and length scales. Most previous models have been restricted to the two-phase limits of porous melt transport in deforming, partially molten rock and crystal settling in convecting magma bodies. The goal of this paper is to develop a framework that can capture igneous system from source to surface at all phase proportions including not only rock and melt but also an exsolved volatile phase. Here, we derive an $n$-phase reactive transport model building on the concepts of Mixture Theory, along with principles of Rational Thermodynamics and procedures of Non-equilibrium Thermodynamics. Our model operates at the macroscopic system scale and requires constitutive relations for fluxes within and transfers between phases, which are the processes that together give rise to reactive transport phenomena. We introduce a phase- and process-wise symmetrical formulation for fluxes and transfers of entropy, mass, momentum, and volume, and propose phenomenological coefficient closures that determine how fluxes and transfers respond to mechanical and thermodynamic forces. Finally, we demonstrate that the known limits of two-phase porous and suspension flow emerge as special cases of our general model and discuss some ramifications for modelling pertinent two- and three-phase flow problems in igneous systems. \\ \\
This preprint has been submitted for publication in \textit{Geophysical Journal International}.
\end{abstract}


\section{Introduction \label{sec:intro}}

The complex interplay between mechanics and thermodynamics in reactive transport processes involving multiple material phases---solids, liquids, and gases---is a common theme in many natural systems, as well as various applied science and engineering contexts. Here, we are primarily interested in volcanoes and their deep magmatic roots, collectively known as igneous systems. Other examples include methane seepage through thawing permafrost or along geological faults \citep[e.g.,][]{Christensen2004,Etiope2009}, the evolution of hydrocarbon reservoirs \citep[e.g.,][]{Roure2005}, and hydrothermal ore formation \citep[e.g.,][]{Sillitoe2003,Sillitoe2010}, as well as the related engineering problems of carbon sequestration \citep[e.g.,][]{Gaus2005}, hydraulic fracturing \citep[e.g.,][]{Jha2014}, and ore concentration \citep[e.g.,][]{Cariaga2005}. Reactive multi-phase systems are highly non-linear with reactions driving transport and transport enhancing reaction. The resulting complex feedbacks may cause localisation of natural phenomena in space and time, mediate chemical differentiation, and drive selective element concentration.

The objective of this study is to develop a framework for formulating custom-built, hypothesis-driven multi-phase reactive transport models of igneous processes. The framework will address these phenomena at the scale of volcanic conduits, shallow magma reservoirs and trans-crustal mush bodies, partially molten regions of the upper mantle, and up to entire planetary bodies. The aim is to produce testable predictions for comparison against field data and thus to advance our understanding of volcanic activity and associated hazards, of the planetary differentiation that created a habitable Earth as well as the deep volatile cycles maintaining it \citep[e.g.,]{Lenardic2016}, and of the magmatic-hydrothermal generation of economic deposits of iron, copper, gold and other important metals \citep{Sillitoe2003,Sillitoe2010}. Although we are focusing on igneous systems, we cast the model in a general form to allow application to other natural and engineering contexts, some of which are more accessible to observation or experiment than subsurface igneous systems. 

Developing macroscopic models of coupled thermodynamic and mechanical behavior of systems much larger than their microscopic constituents builds on a long line of scientific inquiry \citep[e.g.,][]{degroot84,truesdell84}. One common approach is known as Mixture Theory \citep[e.g.,][]{Bowen1976}. The theory assumes that the macroscopic behaviour resulting from microscopic phase interactions may be modeled as a set of interpenetrating and interacting continuum fields. Implicit to the approach is the concept of scale separation, namely that it is possible---indeed, appropriate and meaningful---to average over processes at the local scale to derive a model of system-scale behavior without capturing any of the scales in between \citep[e.g.,][]{anderson1967fluid, slattery1967flow}. 

\begin{figure}[htb]
  \centering
  \includegraphics[width=0.8\textwidth]{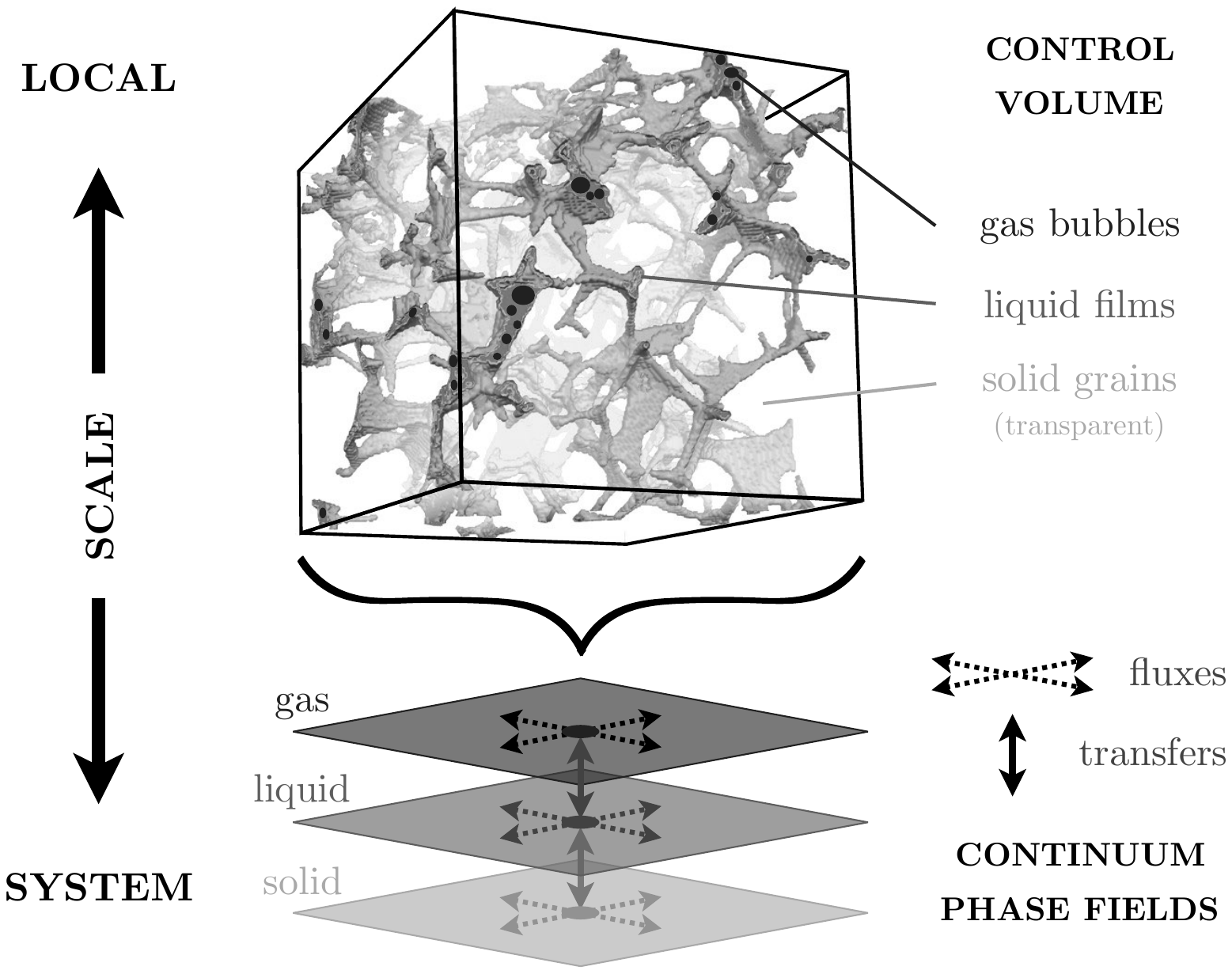}
  \caption{Schematic of a control volume (cube) with phase constituents in an igneous three-phase aggregate at the local scale, with corresponding points (circles) on a set of continuum phase fields (squares) at the system scale. Image is a microtomographic scan of a partially molten rock modified from \cite{Zhu2011}; vapour bubbles (black) added to illustrate a three-phase system. Dashed and solid arrows denote fluxes within and transfers between phases, representing system-scale effects of local-scale phase interactions.}
  \label{fig:consv}
\end{figure}

In mixture models, the distributed microscopic properties of a volume of multi-phase aggregate are represented by a point on a set of macroscopic continuum fields (see Fig.~\ref{fig:consv}). At the local scale, solid grains, liquid films, and gas bubbles occupy a finite volume and interact across well-defined interfaces. At the system scale, phases are represented by the volume fraction they occupy in a control volume and their interactions are described by averaged process terms. This approach entails that different local phase topologies may result in identical averaged phase fractions, and that local mechanical and thermodynamical phase interactions are reduced to non-unique and often phenomenological constitutive relations and material closures. Mixture models hence strategically prioritize the tractable description of system-scale behaviour over the rigorous and well-resolved representation of local-scale phase interactions.

Building a mixture model requires the formulation of conservation statements and thermodynamic principles, along with constitutive relations for transport processes and coefficient closures prescribing the material response to applied forces. Constitutive relations are functions of independent variables (e.g., phase fractions, velocities, pressures, temperatures, concentrations, etc.) that describe the averaged transport of thermodynamic properties within a phase between control volumes (\textit{fluxes}), and between phases within a control volume (\textit{transfers}) (see Fig.~\ref{fig:consv}). For example, thermal diffusion within a material phase falls into the category of \textit{fluxes} and is a function of the temperature gradient within that phase; the thermal equilibration between two phases is classified as a \textit{transfer} and is a function of the temperature differences between the phases. The constitutive relations further require material closures, which encapsulate how system-scale fluxes and transfers depend on pure-phase material properties (e.g., density, viscosity, diffusivity, etc.), as well as local-scale phase topologies (e.g., size, shape, and arrangement of phase constituents, etc.). Choosing the constitutive relations and material closures constitutes the most consequential steps of model building in a mixture theory framework.

The main difference between different mixture models is the scale at which conservation statements, thermodynamic principles, constitutive relations, and material closures are formulated. Perhaps the most physically consistent approach is to formulate and close the governing equations the local scale, as laid out in the Thermodynamically Constrained Averaging Theory \citep[e.g.,][]{gray2013averaging, gray2014introduction, dye2015multiscale}. Thermodynamic constraints resulting from local-scale energy balances \citep{boruvka1985free} then inform a first principles-based averaging procedure that systematically bridges multiple scales up to the system scale \citep{gray1993mathematical, miller2005thermodynamically}. As a consequence, system-scale variables and processes remain formally related to their local-scale equivalents. This strategy is particularly powerful for problems sensitive to interface energetics \citep[e.g.,][]{gray2010thermodynamically, gray2011algebraic}. 

Thermodynamically Constrained Averaging hence comes close to providing a rigorous and general multi-scale model framework for multi-phase reactive transport, at least in the porous flow limit where it has mostly been applied \citep[e.g.,][]{gray2014introduction, dye2015multiscale}. The resulting equations, however, are rather intricate. In addition to the standard conservation laws, they entail evolution equations for common curves and triple points \citep[e.g.,][]{gray2010thermodynamically}, as well as surface tensions and capillary pressures \citep[e.g.,][]{gray2011tcat}. In igneous systems, detailed, phase-specific observations of micro-scale dynamics are often difficult to obtain and, if available, are typically incomplete, inference-laden, and challenging to interpret. It hence becomes difficult to meaningfully constrain the added complexity required to formally bridge the scales.

We opt to follow the more simplistic approach of classical Non-equilibrium Thermodynamics \citep{degroot84, jou01} and Rational Thermodynamics \citep[e.g.,][]{truesdell84, drew2006theory}. These models formulate system-scale conservation laws for mass, momentum, and energy, as well as an entropy inequality \citep{Coleman1963}, which are derived from spatial averaging \citep[e.g.,][]{lahey1988three,simpson10} or ensemble averaging \citep[e.g.,][]{drew71, drew2006theory,Oliveira2018} of local statements. The entropy inequality is then exploited to constrain admissible constitutive relations for system-scale processes \citep[e.g.,][]{muller1967entropy, muller68, liu72, hassanizadeh1979general}. Once the fundamental conservation laws are written, these models operate exclusively at the system scale and thus forgo a formal, first-principles link to local-scale variables and processes \citep[e.g.,][]{hassanizadeh1980general, hassanizadeh1990mechanics, svendsen95, bennethum2000macroscale}.

The deliberate focus on the system scale constitutes both the key limitation and also the chief strength of these mixture model. Their constitutive relations and material closures are inherently non-unique and inevitably phenomenological in nature. However, they yield expressions formally resembling their more familiar local-scale, single-phase equivalents and maintain a tractable level of mathematical complexity. Hence, this approach allows---indeed, encourages---the formulation of relatively accessible models of a reasonably limited complexity. While less satisfying theoretically, models of this kind have had significant impact, as highlighted by the two-phase melt transport model of \citet{mckenzie84} and the two-fluid model by \citet{drew71} on which the former is based.

In the following, we first summarise pertinent igneous processes and existing models that motivate this study before deriving our model framework in four steps. In the first step, we state the fundamental principles in the form of conservation equations for mass, momentum, energy, and entropy. Second, we invoke thermodynamic principles to construct an expanded entropy inequality that combines all system-scale fluxes and transfers requiring constitutive relations. Third, we exploit the inequality for choosing admissible constitutive relations for these reactive transport terms. Finally, we introduce phenomenological closures for material response coefficients coupling fluxes and transfers to their driving forces and discuss how these may be calibrated to recover relevant limiting cases of two- and three-phase flows.

\section{Multi-phase reactive transports in igneous systems \label{sec:background}}
Igneous processes occupy a wide range of temporal and spatial scales \citep{Crisp1984,Caricchi2014,Papale2018}. As non-linear systems, they can exhibit sudden and dramatic shifts in behavior. One consequential expressions occurs when silicic crustal magma reservoirs, which have remained stagnant in the crust for centuries to millennia, are mobilised towards renewed explosive activity within few years, weeks or even days \citep[e.g.,][]{Burgisser2011,Cooper2014}.

\begin{figure}[htb]
  \centering
  \includegraphics[width=1.0\textwidth]{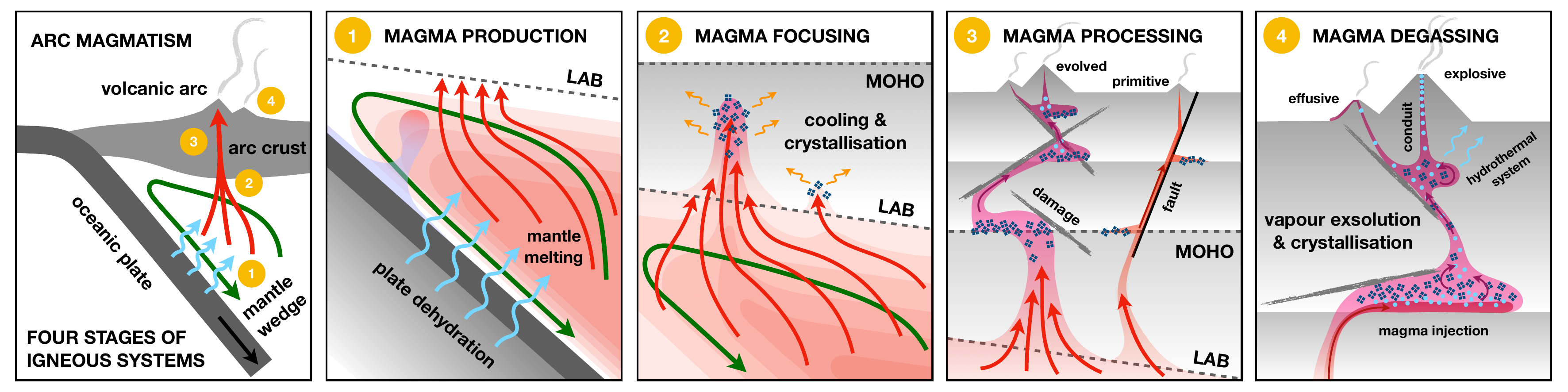}
  \caption{Multi-phase reactive transport processes in a magmatic arc. \textbf{(a)} Stage 1: hydrous melts percolate through and react with mantle wedge to produce primitive arc melts. \textbf{(b)} Stage 2: primitive melts focus into base of lithosphere, where they accumulate, evolve in composition, and thermally erode the lithosphere. \textbf{(c)} Stage 3: Crustal magmas are stored and differentiatied in crystal-rich mush bodies with transient melt lenses, and extracted along fractures. \textbf{(d)} Stage 4: vapour exsolution upon decompression and cooling drives volcanic eruptions and may generate subvolcanic ore deposits.}
  \label{fig:overview}
\end{figure}

A uniquely challenging aspect of igneous systems is that they transition from porous flow in the predominantly solid upper mantle to suspension flow in melt-rich magma bodies feeding volcanic vents. Adding to the challenge are the complex thermodynamics of petrological phase equilibria spanning a wide compositional space including silicates, metal oxides, and volatile species. To motivate our work, we briefly discuss arc magmatism as an example igneous system (see Fig.~\ref{fig:overview}) and review some existing models of the processes involved. We distinguish four stages that are pertinent to other igneous systems as well: (1) melt generation in the asthenosphere, (2) melt focusing into the lithosphere, (3) magma processing in the crust, and (4) shallow magma degassing.

Hydrous fluids sourced from breakdown of hydrated minerals percolate from the subducting plate \citep{cagnioncle07,wilson2014} and react with the asthenospheric mantle wedge to produce primitive arc melts \citep{gaetani98,grove06}. These melts segregate from their mantle residue by reactive porous flow at low melt fractions (Fig.~\ref{fig:overview}, stage 1). McKenzie's model \citep{mckenzie84} and similar formulations \citep{fowler85, ribe85b, Scott1986} based on the earlier work of \citet{drew71}, \citet{sleep74}, and \citet{turcotte78} have been employed to describe the porous flow of hydrous fluids and silicate melts through a compacting rock matrix. \citet{bercovici01a} introduced a phase-symmetrical generalization including surface tensions and a more rigorous treatment of the energetics of the problem \citet{bercovici03}. When coupled to models of thermo-chemical evolution and melt-rock reactions \citep{sramek07,Tirone2009,rudge11,Tirone2017,Oliveira2018}, non-linear reaction-transport feedbacks may emerge. For example, the reactive infiltration instability \citep{chadam86} leads to channelised melt transport \citep{aharonov95, spiegelman01,keller16} with important ramifications for the geochemistry of melts and their solid residue \citep{spiegelman93d,spiegelman03a,hewitt10}.

As mantle melts percolate into the base of the cooler mantle lithosphere, thermal exposure drives crystallization, which deposits latent heat, fertile minerals, and volatiles \citep[e.g.,][]{Keller2017} (Fig.~\ref{fig:overview}, stage 2). Locations of high melt flux may form hot funnels into which melts from a wider source area are focused \citep{England2010,ReesJones2018}. Throughout the lithosphere and crust, magma transport may stall along rheological and density contrasts \citep[e.g.,][]{Ritter2013} leading to melt contents exceeding the disaggregation threshold of the granular matrix \citep[e.g.,][]{arzi1978critical, van1979experimental, renner00}. Depending on the mechanical and thermal structure of the crust, its thickness \citep{Hildreth1988}, and regional tectonic regime \citep{Cembrano2009}, transport towards the shallow crust is thought to be accommodated by diapirs rising through ductile crust \citep{Bateman1984,Cruden1990} or by fractures propagating through brittle rock \citep{Clemens1992,Rubin1993,havlin13}. Porous melt transport models have been extended to include visco-elastic/brittle-plastic rheologies \citep{connolly07,rozhko07,keller13,yarushina15,Oliveira2018}, though their application has remained limited.

Repeated injection or sustained flux of melt into the lower crust followed by crystallization, fractionation, and amalgamation of magma batches has been invoked \citep{Hildreth1988,Annen2006,Cashman2013} to explain differentiation from primitive (low Si, high Mg) to evolved (high Si, low Mg) magma compositions. The latter erupt from major arc volcanic centers or build new continental crust from large plutonic bodies. Fractional crystallization, i.e., the settling out of crystallisation products from melt-rich magma chambers \citep{Bowen1915} has long been the main paradigm of magmatic differentiation and has been investigated by a range of igneous suspension models \citep{Huppert1981,Brandeis1986,martin1988,Rudman1992,
Bergantz1999,dufek10,Molina2012}. Today, consensus is building towards a new paradigm of magma processing in trans-crustal, crystal-rich mush bodies interspersed with transient, crystal-poor magma lenses \citep{Caricchi2015,cashman17} (Fig.~\ref{fig:overview}, stage 3). To test this hypothesis, models are required that bridge the limits of porous and suspension flows including the regime of disaggregated but densely packed crystal mush in between.

In the shallow crust, volatile exsolution from the melt phase marks the final stage of igneous systems: the degassing of sub-volcanic magma bodies that drives volcanic eruptions and sometimes forms magmatic-hydrothermal ore deposits (Fig.~\ref{fig:overview}, stage 4). Volatile exsolution upon magma decompression and cooling \citep{Papale1999} produces droplets of supercritical fluids or brines, and/or bubbles of gaseous vapour \citep{Bachmann2006,Driesner2007a,Driesner2007b,Ruprecht2008}. However, volatile depletion of the melt also facilitates crystallization \citep[e.g.,][]{Metrich2001}. Thus, magma degassing becomes a strongly non-linear reactive transport where volatile exsolution provides buoyancy to drive magma transport \citep{Stevenson1998,Beckett2014} while concurrent crystallisation increases magma viscosity \citep{Caricchi2007,Costa2009,Pistone2012} to resist flow. Both bubbles and crystals may segregate relative to the carrier melt, rendering it a complex three-phase flow problem. Finally, shear stress and gas over-pressure in a stiffening and vesiculating magma may build up to the point of fragmentation and explosive eruption \citep[see e.g.,][and refs therein]{Gonnermann2015}. 

Experiments show that this last stage is prone to flow localization and tipping-point behaviors \citep[e.g.,][]{Oppenheimer2015, Barth2019}. To date, few igneous process models have included a third volatile phase. \citet{Gutierrez2010} include the segregation of crystals and bubbles in their magma chamber model, but focus on the thermo-chemical aspects over the mechanical coupling. \citet{Huber2018} extend the two-phase model of \citep{bercovici03} to include two pore fluids in a compacting matrix. \citet{Afanasyev2018} employ a modified reservoir modelling tool to simulate vapour and brine transport through an undeformable porous rock. \citet{Oliveira2018} provide a general theory for $n$ compressible phases but only discuss applications for two incompressible phases in the porous limit. Hence, existing models do not offer a sufficiently general model framework to investigate and compare the range of processes outlined here. The goal here is therefore to provide a framework for formulating multi-phase reactive transport models specifically for porous, mush, and suspension flows of two, three or more solid, liquid and gas phases in the context of igneous systems.

\section{Fundamental principles \label{sec:fundamental}}

\subsection{Continuum model framework}
We formulate a continuum model for reactive transport in multi-phase aggregates comprising $n$ material phases composed of $m$ thermodynamic components. Material phases are assumed to form locally distinct bodies of mass enclosed by well-defined interfaces and thus can, at least in theory, be mechanically separated. Thermodynamic components are chemical species intermingled at a molecular level that constitute the material phases. At each point in the continuum, we characterise phases by their system-scale velocity, $\vi$, pressure, $\Pi$, temperature, $\Ti$, volume fraction, $\phii$, and component mass fractions, $\cji$, conceptually understood as volume averages of their local-scale equivalents over a finite control volume. The model hence has $n \times (4+m)$ independent variables. Superscripts $i$ denote material phases, and subscripts $j$, thermodynamic components. We assume that phase fractions saturate the aggregate, $\sum_i \phii = 1$, and component concentrations make up the entire phase, $\sum_j \cji = 1$.

Phase materials are taken as compressible, fluid-like materials. They are compressible in the sense that phase densities, $\rhoi$, are permitted to evolve in time ($d \rhoi / d t \neq 0$) according to given equations of state \citep[e.g.,][]{Connolly2009} we will not further discuss here. They are fluid-like in the sense that they deform irreversibly at a strain rate proportional to the stress applied. We distinguish between solids, liquids, and gases based on their relative densities, compressibilities, and resistance to deformation. We approximate solids as high to moderate density, near-incompressible, and minimally to moderately deformable fluids. This approximation of solids as very stiff fluids has ample precedent in geodynamics \citep[e.g.,][]{Turcotte2018}, where the focus is on long-term behaviour and reversible (elastic) deformation is considered negligible. Liquids we characterise as moderate to low density, minimally to moderately compressible, and moderately to highly deformable, and gases as low density, moderately to highly compressible, and very highly deformable fluids. While we formally include chemical components and reaction terms in our derivation, we will keep the discussion of chemical thermodynamics to a minimum. 


\subsection{Generic conservation law}
We begin from a generic continuum-scale conservation law \citep[e.g.,][]{svendsen95, drew2006theory} for an $n$-phase set of mass-specific, vector quantities, $\ai$ [quantity per unit mass],
\begin{linenomath*}
\begin{equation}
	\label{eq:consv-general-vec}
	\ddt{\phii \rhoi \ai} + \Div \left(\qfi \otimes \rhoi \ai \right) + \Div \qai + \Gai + \Qai =  \Uai \ ,
\end{equation}
\end{linenomath*}
where $\partial(\cdot)/\partial t$ is the partial derivative with time, $\Grad = \partial (\cdot) / \partial \xvec$ the partial spatial derivative, and $\otimes$ the outer vector product ($\ai \otimes \ai = \ai \trp{\ai}$). Throughout, tensors are denoted in underlined bold, vectors in bold, and scalars in cursive face. Unless stated otherwise, all variables and parameters are functions of spatial position and time ($\ai = \ai(\xvec,t)$). 

Equation \eqref{eq:consv-general-vec} states that the partial phase property density, $\phii \rhoi \ai$ [phase quantity per aggregate volume], evolves in time due to (terms from left to right) the divergence of mass flux, $\qfi \otimes \rhoi \ai$, carried on the volume flux, $\qfi$, the divergence of other fluxes (i.e., molecular diffusion), $\qai$, transfers between phases, $\sum_i \Gai = 0$, external sources, $\Qai$, and internal production, $\Uai$. The volume flux, $\qfi$, will comprise the advected flux carried on the mean phase velocity, $\phii \vi$, but will also allow for a diffusive component  thought to be important in suspension flows \citep{segre2001effective,mucha2004model}. The production, $\Uai$, is zero for true conservation laws, but must be non-negative to allow writing the entropy equation as a conservation statement. 

The generic conservation law \eqref{eq:consv-general-vec} differs from other continuum-scale conservation equations \citep[e.g.,][]{svendsen95, drew2006theory} in that it separately lists transfers, sources, and production terms, which are often lumped together. We neglect any cross-coupling or fluctuation-related terms (e.g., Reynolds stresses) that may arise in averaging schemes used to derive system-scale conservation laws \citep{lahey1988three,drew2006theory,Oliveira2018}. Unlike some previous theories \citep[e.g.,][]{drew2006theory,bercovici03,sramek07,Oliveira2018} we do not attempt to include interface stresses and surface energies in \eqref{eq:consv-general-vec}. The reason is that the mixture model framework we employ would not appropriately represent directional interface properties and hence surface stress tensors at the continuum scale \citep[e.g.,][]{hassanizadeh1990mechanics, niessner2008model, niessner2011comparison, gray2013averaging}. Nevertheless, some limited but pertinent aspects of phase topology and interface processes are later incorporated into phenomenological material closures.

To write the generic conservation law \eqref{eq:consv-general-vec} in Lagrangian form, we introduce the partial phase material derivatives in the moving reference frame of the phase volume flux,
\begin{linenomath*}
\begin{equation}
	\label{eq:lagr-deriv}
	\DiDt{(\, \cdot \,)} = \phii \ddt{(\, \cdot \,)} + \qfi \cdot \Grad(\, \cdot \,) \ ,
\end{equation}
\end{linenomath*}
and obtain,
\begin{linenomath*}
\begin{equation}
	\label{eq:consv-general-lagr}
	\rhoi \DiDt{\ai} + \ai \left(\DiDt{\rhoi} + \rhoi \left[\Gfi + \Div \qfi \right] \right) + \Div \qai + \Gai + \Qai =  \Uai \ .
\end{equation}
\end{linenomath*}
The partial phase material derivative scales with $\phii$ and hence vanishes if the phase is exhausted. Terms in parentheses represent the phase mass balance apart from mass transfers. The terms in square brackets express the partial volume balance, where we have interpreted the rate of change in phase fractions as the volume transfer rate, $\Gfi = \partial \phii / \partial t$. As required of all transfers, the term sums to zero over all phases.

\begin{table}[htb]
\centering \footnotesize
	\caption{\textbf{Conserved property densities, fluxes, transfers, \& sources}}
	\label{tab:consv}
	\begin{tabular}{lllll}
	 \textbf{Property densities} & $\rhoi \ai$ & \textbf{Fluxes $\qai$} & \textbf{Transfers $\Gai$} & \textbf{Sources $\Qai$} \B\\\hline
	 phase mass & $\rhoi$ & $\qfi \rhoi$: phase mass flux & $\Gmi$: phase-change reaction & -- \T\\
	 component mass & $\rhoi \cji$ & $\qji$: chemical diffusion & $\Gji$: component reaction & -- \\
	 phase momentum & $\rhoi \vi$ & $\qvi$: moment. diffusion (stress) & $\Gvi$: segregation drag & $\Qvi$: gravity body force \\
	phase total energy & $\rhoi e^i$ & $\qei$: total energy flux & $\Gei$: total energy transfer & $\Qei$: total energy source \\
	phase entropy$^{(1)}$ & $\rhoi s^i$ & $\qsi$: thermal diffusion & $\Gsi$: thermal equilibration & $\Qsi$: radiogenic heating \B\\\hline
	\multicolumn{5}{l}{$^{(1)}$ Entropy is not conserved, but produced: $\Usi \geq 0$.}\T\B
	\end{tabular}
\end{table}

\subsection{Multi-phase conservation laws}
We state the conservation laws for phase and component mass, phase momentum, phase total energy, and phase entropy by substituting the conserved property densities, fluxes, transfers, and sources listed in Table~\ref{tab:consv} into \eqref{eq:consv-general-vec},
\begin{linenomath*}
\begin{subequations}
\label{eq:consv-lagr}
\begin{align}
	\label{eq:consv-lagr-mass}
	\DiDt{\rhoi} &+ \rhoi \left(\Gfi + \Div \qfi \right) = - \Gmi \ , \\
	\label{eq:consv-lagr-comp}
	\rhoi \DiDt{\cji} &= - \Div \qji - \Gji + \cji \Gmi \ , \\
	\label{eq:consv-lagr-moment}
	\rhoi \DiDt{\vi} &= - \Div \qvi - \Gvi + \vi \Gmi - \Qvi \ , \\
	\label{eq:consv-lagr-energy}
	\rhoi \DiDt{\ei} &= - \Div \qei - \Gei + \ei \Gmi - \Qei \ , \\
	\label{eq:consv-lagr-entropy}
	\rhoi \DiDt{\si} &= - \Div \qsi - \Gsi + \si \Gmi - \Qsi + \Usi \ .
\end{align}
\end{subequations}
\end{linenomath*}
We do not consider external sources of mass ($\Qmi = \Qji = 0$). Component mass equations \eqref{eq:consv-lagr-comp} must sum to the phase mass conservation \eqref{eq:consv-lagr-mass}, and therefore $\sum_j \qji = 0$, and $\sum_j \Gji = \Gmi$ must hold.

\section{Constitutive relations \label{sec:const-rels}}

\subsection{Limiting assumptions}
The equations \eqref{eq:consv-lagr} express the fundamental principles that underpin our model. However, they merely provide a consistent structure for enforcing conservation laws and entropy production but do not yet specify the reactive transport processes we are interested in. We additionally require constitutive relations for fluxes, $\qai$, and transfers, $\Gai$, which describe the processes that occur in response to their driving forces. We determine sources, $\Qai$, through \textit{a priori} assumptions regarding materials and potential fields externally affecting the system.

In the following constitutive choices, we adhere to the four axiomatic principles laid out in \citet{passman84} (see Appendix \ref{app:axioms} for details). Accordingly, we seek constitutive relations for each phase that are local, frame-invariant, entropy-producing functions of independent variables. The space of admissible functions is expansive. To limit complexity, we restrict our choices by three auxiliary assumptions. First, we seek only \emph{decoupled} relations, where each flux or transfer is a function only of its respective conjugate force that will arise in the entropy inequality. Second, we limit our choices to \emph{linear} relations, with each process a linear function of its conjugate force. And third, we assume that material response coefficients coupling processes to their conjugate forces must be \emph{isotropic} scalars, but they are generally allowed to vary in space and time, and to be non-linear functions of independent  variables. The latter assumption eliminates material anisotropy from our considerations but allows for non-linear closures such as a non-Newtonian rheology. Whereas it is clear that these limited constitutive relations will not rigorously or completely represent the complex nature of igneous processes, we contend that they represents a suitable starting point for devising models of tractable mechanical and thermodynamic complexity.

\subsection{Thermodynamic principles}
In the next step, we assemble an expanded entropy inequality that places all transfers and fluxes explicitly under the thermodynamic constraint of non-negative entropy production \citep{liu72, degroot84, jou01}. Recent works that have followed a similar procedure include the two-phase granular flow model of \cite{monsorno16} or the two-phase porous flow model of \cite{yarushina15}. Here, we modify the approach by introducing a new, phase- and process-wise symmetrical formulation for fluxes within and transfers between $n$ phases. We do not impose thermal equilibrium between phases \citep[e.g.,][]{rudge11}, but retain separate energy conservation and entropy production statements for each phase. 

In analogy to the local-scale definition of total energy, we assume that a phase's specific total energy at the continuum scale is the sum of internal energy, $\ui$, and kinetic energy at the same scale,
\begin{linenomath*}
\begin{equation}
\ei = \ui + \half {\vi}^2 \, .
\end{equation}
\end{linenomath*}
Changes in specific total energy in the reference frame of the phase volume flux are,
\begin{linenomath*}
\begin{equation}
	\label{eq:ineq-total-energy}
	\DiDt{\ei} = \DiDt{\ui} + \vi \cdot \DiDt{\vi} \ .
\end{equation}
\end{linenomath*}
Assuming that pressure-volume work is the only reversible work done on the system, $\ui$ evolves with changes to specific entropy, $\si$, specific volume, $1/\rhoi$, and component concentrations, $\cji$,
\begin{linenomath*}
\begin{equation}
	\label{eq:ineq-internal-energy}
	\DiDt{\ui} = \Ti \DiDt{\si} - \Pi \DiDt{1/\rhoi} + \sum_j \mji \DiDt{\cji} \ .
\end{equation}
\end{linenomath*}
Equation \eqref{eq:ineq-internal-energy} introduces phase temperatures, $\Ti$, pressures, $\Pi$, and chemical potentials, $\mji$, as the thermodynamic conjugates to phase entropy, volume, and component concentrations. In our framework, the thermodynamic pressure is identical to the mechanical pressure introduced as the isotropic or mean stress below. This is justified assuming that irreversible volumetric deformation internal to compressible phase materials remains negligible \citep{Bennethum2004,Oliveira2018,Moulas2018}.

Allowing for other forms of reversible work on the system would result in further contributions to internal energy in \eqref{eq:ineq-internal-energy}. For example, to consider reversible (elastic) deformational work, energy stored in deviatoric and volumetric strain would be added \citep[e.g.,][]{yarushina15} to \eqref{eq:ineq-internal-energy}. Some granular flow models add a further term $\sim D^i \Grad \phii / Dt$ related to reversible energy stored in granular configurations \citep[e.g.,][]{monsorno16}. We neglect these additional complexities here.

Using \eqref{eq:ineq-total-energy} to substitute for $\ui$, and transforming the derivative of specific volume to density, we rewrite \eqref{eq:ineq-internal-energy} as an expression for entropy evolution,
\begin{linenomath*}
\begin{equation}
	\label{eq:ineq-entropy-energy-decomp}
	\Ti \DiDt{\si} = \DiDt{\ei} - \vi \cdot \DiDt{\vi} - \dfrac{\Pi}{(\rhoi)^2} \DiDt{\rhoi} - \sum_j \mji \DiDt{\cji} \ .
\end{equation}
\end{linenomath*}
Equation \eqref{eq:ineq-entropy-energy-decomp} now relates entropy to total energy, momentum, phase mass, and component mass. 

When multiplying entropy production \eqref{eq:consv-lagr-entropy} by absolute temperature, $\Ti \geq 0$, we note that $\Ti \Usi \geq 0$ must hold. Hence, we write the basic entropy inequality as,
\begin{linenomath*}
\begin{equation}
	\label{eq:ineq-basic}
	\Ti \Usi = \rhoi \Ti \DiDt{\si} + \Ti \Gsi - \Ti \si \Gmi + \Ti \Div \qsi + \Ti \Qsi \geq 0 \ .
\end{equation}
\end{linenomath*}
We begin expanding \eqref{eq:ineq-basic} by substituting \eqref{eq:ineq-entropy-energy-decomp} for the entropy evolution term,
\begin{linenomath*}
\begin{align}
	\label{eq:ineq-expand}
	\Ti \Usi &= \rhoi \DiDt{\ei} - \vi \cdot \rhoi \DiDt{\vi} - \dfrac{\Pi}{\rhoi} \DiDt{\rhoi} - \sum_j \mji \rhoi \DiDt{\cji} \\\nonumber 
	&+ \Ti \Gsi - \Ti \si \Gmi + \Ti \Div \qsi + \Ti \Qsi \ \geq \ 0 \ .
\end{align}
\end{linenomath*}
Using conservation of phase total energy \eqref{eq:consv-lagr-energy}, phase momentum \eqref{eq:consv-lagr-moment}, component mass \eqref{eq:consv-lagr-comp}, and phase mass \eqref{eq:consv-lagr-mass}, we substitute for the first four terms on the right hand-side of \eqref{eq:ineq-expand} and write,
\begin{linenomath*}
\begin{align}
	\label{eq:ineq-final}
	\Ti \Usi = &- \Gei + \Ti \Gsi + \sum_j \mji \Gji + \vi \cdot \Gvi + \Pi \Gfi + \Pi \vi \cdot \Grad \phii \\\nonumber
	&- \Div \qei + \Ti \Div \qsi + \sum_j \mji \Div \qji + \vi \cdot \Div \qvi + \Pi \Div \qfi - \Pi \vi \cdot \Grad \phii \\\nonumber
	&- \Qei + \Ti \Qsi + \vi \cdot \Qvi \ \geq \ 0 \ .
\end{align}
\end{linenomath*}
Terms are sorted according to their process categories into transfers, fluxes, and sources, and the identity $u^i = \Ti\si - \Pi / \rhoi + \sum_j \mji \cji$ was used to cancel terms multiplying net mass transfers, $\Gmi$. A term of kinetic energy transferred by phase change, $\frac{1}{2} {\vi}^2 \Gmi$, was dropped as we assume it remains negligible in the non-turbulent flows on which we focus here. Finally, we have introduced two terms, $\pm \Pi \vi \cdot \Grad \phii$, anticipating their utility in formulating frame-invariant constitutive relations. The operation is permitted since the terms cancel out and thus have no net effect on entropy production.

It is apparent from \eqref{eq:ineq-final} that, in our continuum framework, entropy is produced from irreversible phase interactions (transfers), from dissipative phase-internal transport (fluxes), and from interactions of the system with its environment (sources). The second law of thermodynamics strictly requires that entropy production of the entire system be non-negative, $\sum_i \Usi \geq 0$. One way to satisfy this constraint is to require the contribution of each process to be positive in each phase separately \citep{jou01}. We note that this condition does not hold in general \citep[see][]{liu72} and that less constrained treatments exist \citep[e.g.,][]{gray2014introduction}. Nevertheless, we adopt the limitation here as it serves to reduce model complexity and allows to exploit useful patterns and symmetries in the equations. We therefore decompose entropy production into additive contributions from transfers, $(\Usi)^\mathrm{trf} \geq 0$, fluxes, $(\Usi)^\mathrm{flx} \geq 0$, and sources, $(\Usi)^\mathrm{src} \geq 0$, and subject each to the non-negativity constraint separately:
\begin{linenomath*}
\begin{subequations}
\label{eq:ineq-decomp}
\begin{align}
	\label{eq:ineq-decomp-trf}
	\Ti (\Usi)^\mathrm{trf} = &- \Gei + \Ti \Gsi + \sum_j \mji \Gji + \vi \cdot \Gvi + \Pi \Gfi + \Pi \vi \cdot \Grad \phii  \geq 0 \ , \\
	\label{eq:ineq-decomp-flx}
	\Ti (\Usi)^\mathrm{flx} = &- \Div \qei + \Ti \Div \qsi + \sum_j \mji \Div \qji + \vi \cdot \Div \qvi + \Pi \Div \qfi - \Pi \vi \cdot \Grad \phii \geq 0 \ , \\
	\label{eq:ineq-decomp-src}
	\Ti (\Usi)^\mathrm{src} = &- \Qei + \Ti \Qsi + \vi \cdot \Qvi  \geq 0 \ .
\end{align}
\end{subequations}
\end{linenomath*}

We are now ready to leverage the expanded and decomposed entropy inequality \eqref{eq:ineq-decomp} for choosing admissible constitutive relations for fluxes and transfers. 

\subsection{Transfers}
In our model, continuum-scale phase interactions comprise transfers of partial volume, mass, momentum, entropy, and energy between phases. Entropy and component transfers act to thermally and chemically equilibrate phases. Momentum transfers take on the role of the drag resisting phase segregation and volume transfers comprise changes in phase fractions due to what we term phase compaction. 

Segregation and compaction are the mechanical phase interactions that control multi-phase transport. We define segregation as the relative transport of phases in the aggregate with respect to the others, and compaction as the transfer of partial volume between phases. Note that compaction may include volumetric (or bulk) deformation internal to compressible phase materials but is typically dominated by multi-phase interactions. Hence, even mixtures comprising only incompressible phases will generally experience compaction. The term compaction itself is perhaps not optimal, because the process generally involves both the compaction (i.e., partial volume contraction, $\partial \phi /\partial t < 0$) of some phases as well as the simultaneous decompaction (i.e., partial volume expansion, $\partial \phi /\partial t > 0$) of others. Nevertheless, we retain the terminology for the sake of continuity with previous work \citep{mckenzie84}.

Local-scale phase interactions in natural systems are driven by thermodynamic gradients across interfaces. On the system scale, we represent interface gradients by differences between averaged continuum phase fields. For example, if a hotter melt infiltrates a cooler rock matrix, the local temperature gradients across melt-rock interfaces will drive heat transfer and lead to thermal equilibration. On the continuum scale, the same process is expressed by a heat transfer rate driven by the difference in continuum phase temperatures, $T^\mathrm{melt} - T^\mathrm{rock}$.

To avoid taking differences between all possible phase pairs in an $n$-phase system, it is expedient to introduce deviations of phase states from a common reference state. For a generic property, $\ai$, we define a reference state, $\mathbf{a}^* = \sum_i \oai \ai$, which is a weighted mean of phase states with weights, $\sum_i \oai = 1$, left to be determined. Phase deviations from the reference state are denoted $\Dai = \ai - \mathbf{a}^*$. 

We seek phase-wise transfer rates that are proportional to phase deviations, $\Gai \sim \Dai$, and act to bring $\Dai \rightarrow \mathbf{0}$. We note that in the limit of vanishing phase deviations, the reference state can be understood as approaching the state of phase equilibrium, $\mathbf{a}^* \rightarrow \mathbf{a}^\mathrm{eq}$ (i.e., the state where transfers cede). Outside that limit, $\mathbf{a}^*$ represents a dynamically evolving target towards which phase transfers drive the system, while the phase deviations can be understood as a measure of phase disequilibrium.

By substituting $\ai = \mathbf{a}^* + \Dai$ for the conjugate variables, $\vi$, $\Ti$, $\Pi$, $\mji$, in \eqref{eq:ineq-decomp-trf}, and after collecting reference and deviation terms, we obtain,
\begin{linenomath*}
\begin{align}
	\label{eq:trf-ineq-decomp}
	\Ti(\Usi)^\mathrm{trf} &=- \left(\Gei - \Tstar \Gsi - \sum_j \mjstarP \Gji - \vstar \cdot \Gvi - \Pstar \left(\Gfi+ \dfrac{\Gmi}{\rhostar}\right) - \Pstar \vstar \cdot \Grad \phii \right) \\\nonumber
	&+ \DTi \Gsi + \sum_j \DmjiP \Gji + \Dvi \cdot \left(\Gvi + \Pstar \Grad \phii\right) + \DPi \left(\Gfi + \vstar \cdot \Grad \phii + \dfrac{\Gmi}{\rhostar} \right) \ \geq \ 0 \ ,
\end{align}
\end{linenomath*}
where we have expanded the additional pressure-volume term as, 
\begin{linenomath*}
\begin{align}
	\label{eq:trf-PV-decomp}
	\Pi \vi \cdot \Grad \phii = \Pstar \vstar \cdot \Grad \phii + \DPi \vstar \cdot \Grad \phii + \Pstar \Dvi \cdot \Grad \phii + \DPi \Dvi \cdot \Grad \phii \ .
\end{align}
\end{linenomath*}
The nonlinear deviation term in \eqref{eq:trf-PV-decomp}, $\DPi \Dvi \cdot \Grad \phii$, has not previously come up in similar derivations \citep{bercovici03,sramek07,rudge11,Oliveira2018}. We chose to only allow linear constitutive relations here and therefore will not carry the term forward. We have furthermore followed \cite{sramek07} in adding a term $\Pi \Gmi/\rhostar$ (with $\rhostar$ a reference density) to reduce the volume transfer or compaction rate by the volume change of net mass transfer. The term is compensated for by transforming the chemical potentials to $\tilde{\mu}_j^i = \mji - \Pi/\rhostar$, bearing in mind that the net mass transfer is the sum of component mass transfers, $\Gmi = \sum_j \Gji$.

Examining terms on the first line of \eqref{eq:trf-ineq-decomp}, we choose the transfer of total energy such that it does not contribute to entropy production and satisfies $\sum_i \Gei = 0$,
\begin{linenomath*}
\begin{equation}
	\label{eq:trf-rate-energy}
	\Gei = \Tstar \Gsi + \vstar \cdot \Gvi + \sum_j \mjstarP \Gji + \Pstar \left(\Gfi+ \dfrac{\Gmi}{\rhostar}\right) + \Pstar \vstar \cdot \Grad \phii \ .
\end{equation}
\end{linenomath*}

The remaining terms in \eqref{eq:trf-ineq-decomp} form conjugate pairs of transfer rates, $\Gai$, multiplying their forcing deviations, $\Dai$. We choose decoupled, linear, isotropic constitutive relations that satisfy the entropy principle by writing,
\begin{linenomath*}
\begin{equation}
	\label{eq:trf-rate-general}
	\Gai = C_a^i \Dai \ ,
\end{equation}
\end{linenomath*}
introducing scalar transfer coefficients, $C_a^i \geq 0$. This phase- and process-symmetrical form does not yet enforce the zero-sum constraint, $\sum_i \Gai = 0$. For that we choose weights, $\oai$, that cast reference states, $\mathbf{a}^*$, as the coefficient-weighted averages of phase states,
\begin{linenomath*}
\begin{equation}
	\label{eq:trf-weights-general}
	\oai = \normsum{C_a^i}{i} = \dfrac{C_a^i}{\sum_{k} C_a^k} \ ,
\end{equation}
\end{linenomath*}
where we have introduced the short-hand notation, $\normsum{( \ \cdot \ )^i}{i}$, denoting normalization of an indexed quantity by the sum over the specified index, here $i$. The reference state for each conjugate variable thus tends towards the phase state associated with the dominant transfer coefficient. The physical intuition is that phase transfers are driven by deviations from the most rapidly equilibrating phase. In the absence of further constraints we follow \cite{sramek07} and define the reference density by the same weights $\ofi$ as the reference pressure.

Finally, our constitutive relations for transfers of entropy, component mass, momentum, and volume are,
\begin{linenomath*}
\begin{subequations}
\label{eq:trf-rate}
\begin{align}
	\label{eq:trf-rate-entropy}
	\Gsi &= \Csi \, \dfrac{\DTi}{\Tstar} \ , \\
	\label{eq:trf-rate-comp}
	\Gji &= \Cji \, \dfrac{\DmjiP}{R \Tstar} \ ,\\ 
	\label{eq:trf-rate-moment}
	\Gvi &= \Cvi \, \Dvi - \Pstar \Grad \phii \ , \\
	\label{eq:trf-rate-volume}
	\Gfi &= \Cfi \, \DPi - \vstar \cdot \Grad \phii - \dfrac{\Gmi}{\rhostar} \ ,
\end{align}
\end{subequations}
\end{linenomath*}
where we have normalised entropy transfers by $\Tstar$ and mass transfers by $R \Tstar$, with $R$ the universal gas constant. Additional terms $\sim \Grad \phii$ in momentum and volume transfers ensure frame-invariance as discussed in previous work \citep[e.g.,][]{bercovici03}. 

While \eqref{eq:trf-rate} is a set of thermodynamically admissible constitutive relations, more rigorous or general treatments of system-scale phase interactions are possible. We neglect cross-coupling between different transfers and driving forces. There is no formal reason why, for example, heat transfer should not be allowed to depend on chemical potential differences, or phase pressure deviations. We also neglect effects of surface tensions and capillary pressures, as well as material anisotropy, all of which may well be important for some igneous phenomena. Nevertheless, we will demonstrate in section \ref{sec:closures} that, given appropriate material closures for transfer coefficients, $C_a^i$, this new formalism of phase transfers bears out the canonical limits of two-phase porous and suspension flows as special cases.

\subsection{Fluxes}
In our framework, fluxes describe phase-internal transport of partial volume, mass, momentum, entropy, and energy. Entropy and component fluxes act to equilibrate thermal and chemical gradients within each phase. The momentum flux takes the role of the deviatoric stress tensor, and the volume flux that of the advective-diffusive transport of partial phase volume.

To facilitate the choice of constitutive relations, we use the chain rule to transform vector and tensor flux divergences in \eqref{eq:ineq-decomp-flx} according to,
\begin{linenomath*}
$$ a^i \Div \q_a^i = \Div a^i \q_a^i - \q_a^i \cdot \Grad a^i \ , $$
\end{linenomath*}
\begin{linenomath*}
$$ \ai \cdot \Div \qai = \Div \ai \qai - \qai : \Grad \ai \ . $$
\end{linenomath*}
After expanding divergences and grouping of terms, we obtain,
\begin{linenomath*}
\begin{align}
	\label{eq:flx-divergence-transform}
	\Ti (\Usi)^\mathrm{flx} = &- \Div \left(\qei - \Ti \qsi - \sum_j \mji \qji - \vi \qvi - \Pi \qfi + \phii \Pi \vi \right) \\\nonumber 
	&- \qsi \cdot \Grad \Ti - \sum_j \qji \cdot \Grad \mji  - \left(\qvi - \phii \Pi \I\right) : \Grad \vi - \left(\qfi - \phii \vi\right) \cdot \Grad \Pi \geq \ 0 \ ,
\end{align}
\end{linenomath*}
where we have expanded the additional pressure-volume term as,
\begin{linenomath*}
\begin{align} 
	\label{eq:flx-PV-decomp}
	- \Pi \vi \cdot \Grad \phii = - \Div \phii \Pi \vi + \phii \Pi \I : \Grad \vi + \phii \vi \cdot \Grad \Pi \ .
\end{align}
\end{linenomath*}

We choose the flux of total energy such that it does not contribute to entropy production (terms in parentheses cancel out),
\begin{linenomath*}
\begin{align}
	\label{eq:flx-const-rel-energy}
	\qei = \Ti \qsi + \sum_j \mji \qji + \vi \qvi + \Pi \qfi - \phii \Pi \vi\ .
\end{align}
\end{linenomath*}

The remaining terms in \eqref{eq:flx-divergence-transform} form conjugate pairs of fluxes, $\qai$, multiplying their forcing gradients, $\Grad \ai$. We once again choose decoupled, linear, isotropic constitutive relations that satisfy the entropy principle by writing, 
\begin{linenomath*}
\begin{align}
	\label{eq:flx-const-rel-general} 
	\qai = - K_a^i \Grad \ai \ ,
\end{align}
\end{linenomath*}
with scalar flux coefficients, $K_a^i \geq 0$. The resulting constitutive relations for fluxes of entropy, component mass, momentum, and volume are,
\begin{linenomath*}
\begin{subequations}
\label{eq:flx-const-rel}
\begin{align}
	\label{eq:flx-const-rel-entropy}
	\qsi &= - \Ksi \, \dfrac{\Grad \Ti}{\Ti} \ , \\
	\label{eq:flx-const-rel-comp}
	\qji &= - \Kji \, \dfrac{\DGmstari}{R \Ti} \ , \\
	\label{eq:flx-const-rel-moment}
	\qvi &= - \Kvi \, \Di + \phii \Pi \I \ , \\
	\label{eq:flx-const-rel-volume}
	\qfi &= - \Kfi \, \DGPstar + \phii \vi \ ,
\end{align}
\end{subequations}
\end{linenomath*}
where we have normalized entropy fluxes by $\Ti$, and component fluxes by $RT^i$. The added pressure and velocity terms in momentum and volume fluxes again ensure frame invariance. 

Our description of momentum flux or shear stress \eqref{eq:flx-const-rel-moment} does not allow for reversible (elastic) deformation; we instead refer the interested reader to the works of \citep{connolly07}, \cite{keller13}, \cite{yarushina15}, and \cite{Oliveira2018} for different perspectives on visco-elastic/brittle-plastic modes of deformation in multi-phase flows. To write \eqref{eq:flx-const-rel-moment} we have decomposed the velocity gradient into deviatoric symmetrical, volumetric, and anti-symmetrical parts, $\Grad \vi = \Di + \Vi + \Wi$:
\begin{linenomath*}
\begin{subequations}
\label{eq:strainrate-decomp}
\begin{align}
\Di &= \half \left(\Grad \vi + \trp{\Grad \vi}\right) - \third \Div \vi \I \ , \\
\Vi &= \third \Div \vi \I \ , \\
\Wi &= \half \left(\Grad \vi - \trp{\Grad \vi}\right) \ ,
\end{align}
\end{subequations}
\end{linenomath*}
Allowing each component of the forcing gradient to contribute to entropy production separately, we can write the momentum flux as,
\begin{linenomath*}
\begin{align}
\label{eq:flx-moment-general}
	\qvi &= - \Kvi \, \Di - \Lambda_V^i \Vi - \Lambda_W^i \Wi + \phii \mathcal{P}^i \I \ .
\end{align}
\end{linenomath*}

To arrive at \eqref{eq:flx-const-rel-moment} above, we have assumed that $\Lambda_W^i = 0$ and hence dropped the rotational term in \eqref{eq:flx-moment-general}, $\Lambda_W^i \Wi$. This choice implies that there is no resistance to purely rotational motion, and that angular momentum is conserved, since $\qvi = \trp{\qvi}$ now holds. In \eqref{eq:flx-moment-general}, we have made a distinction between mechanical and thermodynamic pressures, where the former is a third of the trace of \eqref{eq:flx-moment-general}, and the latter accounts for irreversible deformation internal to compressible phases, $\phii \mathcal{P}^i = \phii \Pi + \Lambda_V^i \Vi$, as discussed in previous work \citep{Bennethum2004,Moulas2018,Oliveira2018}. To reduce complexity, as well as to emphasise the multi-phase nature of compaction, we choose to set the volumetric coefficient to zero, $\Lambda_V^i = 0$, and hence maintain a single pressure definition, $\mathcal{P}^i = \Pi$, in \eqref{eq:flx-const-rel-moment}. This choice implies that we consider the intrinsic volumetric or bulk viscosity of compressible phases to be negligible. While this assumption is reasonable for gas phases at low pressure \citep[e.g.,][]{cramer2012numerical}, we find it useful to invoke the same assumption for igneous solid and liquid phases as well. The reason is that we assume their system-scale volumetric deformation to be dominated by phase compaction, not phase compressibility (i.e., $\partial \phii / \partial t \gg {\rhoi}^{-1} \partial \rhoi / \partial t$). The two-phase compaction model of \cite{mckenzie84}, and the recent extension to $n$ phases by \cite{Oliveira2018} both use the volumetric term in the stress tensor to formulate phase compaction. Hence, they do not distinguish formally between system-scale volumetric flow resulting from irreversible bulk deformation internal to compressible phases, and local-scale solenoidal flow of phases accommodating changes in phase fractions.

To ensure total mass conservation, the component and volume fluxes are subject to sum constraints, $\sum_j \qji = 0$, and $\sum_i (\qfi - \phii \vi) = 0$, respectively. In analogy to the formalism used to enforce sum constraints on transfers, we expand forcing gradients around a reference gradient,
\begin{linenomath*}
\begin{subequations}
\label{eq:flx-decomp} 
\begin{align}
	\label{eq:flx-decomp-comp} 
	\Grad \mji &= \Gmstari + \DGmstari \ , \\
	\label{eq:flx-decomp-volume} 
	\Grad \Pi &= \GPstar + \DGPstar \ .
\end{align}
\end{subequations}
\end{linenomath*}
The reference gradients are again the coefficient-weighted sums,
\begin{linenomath*}
\begin{subequations}
\label{eq:flx-refgradient}
\begin{align}
	\label{eq:flx-refgradient-comp}
	\Gmstari &= \sum_j \omega_{K_j}^i \Grad \mji \ , \ \ \ \ \ \omega_{K_j}^i = \normsum{\Kji}{j} \ , \\
	\label{eq:flx-refgradient-comp}
	\GPstar &= \sum_i \omega_{K_\phi}^i \Grad \Pi \ , \ \ \ \ \ \omega_{K_\phi}^i = \normsum{\Kfi}{i} \ .
\end{align}
\end{subequations}
\end{linenomath*}
When substituting \eqref{eq:flx-decomp} into the entropy production \eqref{eq:flx-divergence-transform}, terms multiplying reference gradients sum to zero ($\sum_j \qji \cdot \Gmstari = 0$, $\sum_i \qfi \cdot \GPstar = 0$) and are therefore not carried further. In general, $\GPstar \neq \Grad \Pstar$ even if transfer and flux coefficients have identical weights, $\ofi = \omega_{K_\phi}^i$. The physical intuition behind this gradient expansion is that diffusive fluxes subject to a sum constraint effectively diffuse down their deviation from the forcing gradient of the fastest diffusing phase. A somewhat counter-intuitive consequence is that components/phases may experience a finite diffusive flux even if not subject to an internal forcing gradient themselves. Hence, the slowest fluxes may effectively anti-diffuse as they compensate for faster fluxes.

Regarding the diffusive part of the volume flux \eqref{eq:flx-const-rel-volume}, we note that it appears as conjugate to $\Grad \Pi$ in the entropy inequality. We have therefore made the internally consistent choice in \eqref{eq:flx-const-rel-volume} that diffusive fluxes of phase volume are driven down deviations in phase pressure gradients. To our knowledge, a constitutive relation of this form has not previously been proposed. Indeed, it remains open how to best interpret it. One possibility is to reconcile it with suspension flow experiments and theory \citep{mucha2004model,segre2001effective}, which suggest that volume diffusion arises from local-scale fluctuations in particle sedimentation velocity and is driven down gradients in phase fractions,
\begin{linenomath*}
\begin{align}
	\label{eq:flx-volume-simpl}
	\qfi = - \tilde{K}_\phii \DGphistar + \phii \vi \ ,
\end{align}
\end{linenomath*}
where
\begin{linenomath*}
\begin{align}
	\label{eq:flx-comp-simpl-ref}
	\Gphistar = \sum_j \omega_{K_\phi}^i \Grad \phii \ .
\end{align}
\end{linenomath*}
In this case, \eqref{eq:flx-volume-simpl} can be understood to imply that $\DGphistar \approx \DGPstar / p_0$, with $\tilde{K}_\phii \approx \Kfi p_0$, and $p_0$ some appropriate pressure scale. However, this interpretation remains to be validated.

\subsection{Sources}
Source terms are determined from \textit{a priori} assumptions regarding the phase materials and the potential field environment the system is exposed to. We choose the total energy source such that it does not contribute to internal entropy production \eqref{eq:ineq-decomp-src}:
\begin{linenomath*}
\begin{align}
	\label{eq:src-const-rel-energy}
	\Qei = \Ti \Qsi + \vi \cdot \Qvi \ .
\end{align}
\end{linenomath*}
For igneous systems, we take the gravitational acceleration, $\gvec$, as external source of momentum, and the specific radiogenic heating rate of phase materials, $H^i$, as external source of entropy:
\begin{linenomath*}
\begin{subequations}
\label{eq:src-const-rel}
\begin{align}
	\label{eq:src-const-rel-entropy}
	\Qsi &= - \dfrac{\phii \rhoi H^i}{\Ti} \ , \\
	\label{eq:src-const-rel-momentum}
	\Qvi &= - \phii \rhoi \gvec \ . 
\end{align}
\end{subequations}
\end{linenomath*}
We are unaware of modelling scenarios that would call for external sources of phase or component mass and therefore do not include them here ($\Qji = \Qmi = 0$).

\subsection{Entropy production and equilibrium}
Substituting constitutive relations for transfers \eqref{eq:trf-rate}, fluxes \eqref{eq:flx-const-rel}, and sources \eqref{eq:src-const-rel} into the entropy inequality \eqref{eq:ineq-final} yields the total phase-wise rate of entropy production,
\begin{linenomath*}
\begin{align}
	\label{eq:entr-prod-final}
	\Ti \Usi = &~\Csi \dfrac{[\DTi]^2}{\Tstar} + \sum_j \Cji \dfrac{[\Dmji]^2}{R \Tstar} + \Cvi [\Dvi]^2 + \Cfi [\DPi]^2 \\\nonumber
	& \Ksi \dfrac{[\Grad \Ti]^2}{\Ti} + \sum_j \Kji \dfrac{[\DGmstari]^2}{R \Ti} + \Kvi \left[\Di\right]^2 + \Kfi [\DGPstar]^2 \ \geq \ 0 \ .
\end{align}
\end{linenomath*}

Under the convention that equilibrium is the state where cumulative entropy production in the full system reaches its maximum \citep{gibbs1948collected}, and hence the total entropy production rate vanishes, $\sum_i \Usi \rightarrow 0$, we find that our model system reaches equilibrium if and when,
\begin{linenomath*}
\begin{align}
	\label{eq:entr-prod-equilibrium}
	\Ti = \Tstar \ , \ \ \vi = \vstar \ , \ \ \Pi &= \Pstar \ , \ \ \mji = \mjstar \ , \\\nonumber
	\Grad \Ti = \mathbf{0} \ , \ \ \Di = \underline{\mathbf{0}} \ , \ \ \Grad \Pi &= \GPstar \ , \ \ \Grad \mji = \Grad \mstari \ .
\end{align}
\end{linenomath*}
Equilibrium in the sense of \eqref{eq:entr-prod-equilibrium} will not normally be achieved in any natural system we are interested in here. In fact, these systems derive their dynamics from the fact that external sources or boundary fluxes continually supply momentum and heat to the system and thus drive the various transport processes contributing to \eqref{eq:entr-prod-equilibrium}. The key to studying these dynamic systems is to identify which are the dominant fluxes and transfers that govern how a system responds to its drivers mechanically, thermally, and chemically. For example, in igneous systems thermal transport properties may be such that heat transfers relax phase temperature deviations at a much faster rate than heat fluxes relax thermal gradients. In that case, $\DTi \approx 0$ may be a reasonable approximation on the time scale of the heat flux responding to $\Grad \Ti \neq 0$. In igneous multi-phase flows, the assumption of thermal phase equilibrium may justify solving for only one mixture temperature rather than each phase temperature separately \citep[e.g.,][]{bercovici03,sramek07,katz08,rudge11,keller16,Oliveira2018}. Similar arguments can be constructed regarding processes other than thermal equilibration, leading to reduced equations concerned with a limited number of transport processes.

\subsection{Simplified chemical thermodynamics}
The constitutive relations developed above require an equation of state for chemical potentials, $\mji$. Recognizing the substantial complexity of chemical thermodynamic in igneous petrology, we shall not make any attempts at it here. We note instead that it is common practice \citep{ribe85a,aharonov95,spiegelman01,hewitt10,
Liang2010,Hesse2011,rudge11,keller16} to simplify \eqref{eq:trf-rate-comp} by introducing parameterised affinities, $\Delta Z_j^i \approx \DmjiP / R \Tstar$, such that, 
\begin{linenomath*}
\begin{align}
	\label{eq:trf-rate-comp-simple}
	\Gji = \Cji \Delta Z_j^{i*} \ .
\end{align}
\end{linenomath*}
For example, affinities have been prescribed in terms of concentration differences, $Z_j^i = \cji - {\cji}^\mathrm{eq}$, from equilibrium concentrations calculated as a function of pressure, temperature and mixture composition \citep[e.g.,][]{keller16}. Here, we carry reaction rates forward by their symbol only and refrain from further discussing the intricacies of igneous petrology.

Alternatively, the assumption of chemical phase equilibrium ($\DmjiP \approx 0$, $\Rightarrow$ $\phii = {\phii}^\mathrm{eq}, \cji = {\cji}^\mathrm{eq}$) can be enforced directly according to parameterised phase diagrams \citep[e.g.,][]{katz08,weatherley12,Solano2014,Jordan2015,ReesJones2018} or with the help of more elaborate free energy minimisation routines \citep[e.g.,][]{Tirone2009,dufek10,Gutierrez2010,Oliveira2018}. Similar to the example of thermal phase equilibrium above, the chemical phase equilibrium assumption may be used to justify solving for component concentrations in the mixture only, rather than in each phase separately. Reaction rates are then no longer required to write component mass conservation and can be eliminated from the model entirely \citep[e.g.,][]{katz08}, unless it is expedient to reconstruct them analytically or numerically \citep[e.g.,][]{hewitt10,weatherley12} given the rates of change of pressure, temperature, and mixture composition.

Similarly, we assume $\Grad \cji \approx \Grad \mji / R \Ti$ \citep[e.g.,][]{spiegelman01,katz08,rudge11} and thus simplify component fluxes to,
\begin{linenomath*}
\begin{align}
	\label{eq:flx-comp-simpl}
	\qji &= - \Kji \DGcstari \ , 
\end{align}
\end{linenomath*}
where
\begin{linenomath*}
\begin{align}
	\label{eq:flx-comp-simpl-ref}
	\Gcstari = \sum_j \omega_{K_j}^i \Grad \cji \ .
\end{align}
\end{linenomath*}

\section{Final Governing Equations \label{sec:finaleqs}}

\subsection{Mechanical equations}
Given our constitutive choices above, we now assemble the final governing equations. We write ($2 \times n$) mechanical governing equations for the variables $\vi$, $\Pi$, obtained by substituting the momentum flux \eqref{eq:flx-const-rel-moment}, transfer \eqref{eq:trf-rate-moment}, and source \eqref{eq:src-const-rel-momentum} into the phase momentum conservation \eqref{eq:consv-lagr-moment}, and the volume flux \eqref{eq:flx-const-rel-volume}, and transfer \eqref{eq:trf-rate-volume} into the phase mass conservation \eqref{eq:consv-lagr-mass},
\begin{linenomath*}
\begin{subequations}
\label{eq:final-fluidmech}
\begin{align}
	\label{eq:final-fluidmech-segr}
	\vdi &= - \dfrac{\phii}{\Cvi} \left(\rhoi \DiDt{\vi} + \phii \Grad \Pstar + \Grad \Pdi - \Div \Kvi \Di - \vi \Gmi - \phii \rhoi \gvec \right) \ , \\
	\label{eq:final-fluidmech-comp}
	\Pdi &= - \dfrac{\phii}{\Cfi} \left(\dfrac{1}{\rhoi}\DiDt{\rhoi} + \phii \Div \vstar + \Div \vdi - \Div \Kfi \DGPstar + \Gmi \left(\dfrac{1}{\rhoi} - \dfrac{1}{\rhostar}\right) \right) \ .
\end{align}
\end{subequations}
\end{linenomath*}
We have arranged terms to emphasise multi-phase reactive transport processes by introducing \textit{segregation velocities}, $\vdi = \phii \Dvi$, and \textit{compaction pressures}, $\Pdi = \phii \DPi$. These generalise Darcy's segregation velocity \citep{hubbert1957darcy} and Terzaghi's effective pressure \citep{Terzaghi1943,skempton60}, both commonly used in models of porous flow in deformable media applied to partially molten rock \citep{connolly98,katz06,keller13} and water-saturated sediments \citep{Birchwood1994,Morency2007}. The final governing equations express that segregation is driven by velocity evolution (acceleration), gradients in reference and compaction pressures, momentum diffusion (shear stresses), momentum transfer by reaction, and buoyancy, and is modulated by the segregation coefficient, ${\phii}^2/\Cvi$. Compaction is driven by density evolution (compressibility), divergence of reference and segregation velocities, volume diffusion, and volume transfer by reaction, and is modulated by the compaction coefficient, ${\phii}^2/\Cfi$. Our compaction model thus extends previous theories by including other processes than divergent flow \citep[e.g.,][]{mckenzie84,bercovici03,Oliveira2018} and reactive volume change \citep[e.g.,][]{sramek07}.

Summing \eqref{eq:final-fluidmech} over all phases yields the mixture momentum and mass conservation equations,
\begin{linenomath*}
\begin{subequations}
\label{eq:final-fluidmech-total}
\begin{align}
	\label{eq:final-fluidmech-totalmoment}
	\Grad \Pb &= - \sum_i \left(\rhoi \DiDt{\vi} - \Div \Kvi \Di - \vi \Gmi - \phii \rhoi \gvec \right) \ , \\
	\label{eq:final-fluidmech-totalmass}
	 \Div \vb &= - \sum_i \left(\dfrac{1}{\rhoi}\DiDt{\rhoi} + \dfrac{\Gmi}{\rhoi} \right) \ ,
\end{align}
\end{subequations}
\end{linenomath*}
written for the gradient of the mixture pressure, $\Pb = \sum_i \phii \Pi = \Pstar + \sum_i \Pdi$, and the divergence of mixture velocity, $\vb = \sum_i \phii \vi = \vstar + \sum_i \vdi$. 

\subsection{Thermo-chemical equations}
Alongside the mechanical equations, we write a set of thermo-chemical governing equations comprising ($n$) evolution equations for $\phii$, ($m \times n$) for $\cji$, and ($n$) for $\Ti$,
\begin{linenomath*}
\begin{subequations}
\label{eq:final-thermochem}
\begin{align}
	\label{eq:final-thermochem-phaseprop}
	\DstarDt{\phii} &= \Div \tilde{K}_\phii \DGphistar - \phii \Div \vstar - \Div \vdi - \dfrac{1}{\rhoi}\DiDt{\rhoi} - \dfrac{\Gmi}{\rhoi} \ , \\
	\label{eq:final-thermochem-compconc}
	\rhoi \DiDt{\cji} &= \Div \Kji \DGcstari - \Gji + \cji \Gmi \ , \\
	\label{eq:final-thermochem-phasetemp}
	\rhoi c^i_p \DiDt{\Ti} &= \Div \Ksi \Grad \Ti - \Csi \DTi + L^i \Gmi + \alpha^i \Ti \DiDt{\Pi} + \phii \rhoi H^i \\\nonumber 
	&+ \Kvi \left[\Di\right]^2  + \Kfi \left[ \DGPstar \right]^2  + \Cvi [\Dvi]^2 + \Cfi [\DPi]^2 \, ,
\end{align}
\end{subequations}
\end{linenomath*}
where  $D^*( \ \cdot \ )/Dt = \partial ( \ \cdot \ )/\partial t + \vstar \cdot \Grad ( \ \cdot \ ) $ is the material derivative in the reference velocity frame, $\alpha^i$ the phase thermal expansivity, $c_\Pi$ the phase specific heat capacity, and $L^i$ the latent heat of phase change. We have obtained \eqref{eq:final-thermochem} by substituting the modified volume flux \eqref{eq:flx-volume-simpl} into the phase mass conservation \eqref{eq:consv-lagr-mass}, the simplified component flux \eqref{eq:flx-comp-simpl} into the component mass conservation \eqref{eq:consv-lagr-comp} (retaining reaction rates symbolically), and the energy flux \eqref{eq:flx-const-rel-energy}, transfer \eqref{eq:trf-rate-energy}, and source \eqref{eq:src-const-rel-energy} into the phase total energy conservation \eqref{eq:consv-lagr-energy}. For details on writing energy conservation as temperature evolution see Appendix \ref{app:final}. The final governing equations express that phase fractions and component concentrations evolve due to advection, diffusion, and reaction, and that temperature (sensible heat) changes with thermal diffusion, thermal equilibration, latent heat of reactions, adiabatic heating, radiogenic heating, and the dissipation of shear deformation, volume diffusion, segregation, and compaction.

\section{Phenomenological Closures \label{sec:closures}}

\subsection{Local diffusive transport}
The final governing equations \eqref{eq:final-fluidmech}--\eqref{eq:final-thermochem} present an internally consistent and thermodynamically admissible continuum model of multi-phase reactive transport at the system scale. However, without material closures for flux and transfer coefficients, $K_a^i$ and $C_a^i$, these equations remain an empty template. In the constitutive relations above, fluxes and transfers are related to their conjugate forces by non-negative, scalar material response coefficients, which we have allowed to be non-linear functions of independent variables. For the task of choosing closures, we no longer formally rely on fundamental principles but rely on empirical evidence and conceptual thought to guide our choices. 

We propose closures based on the assumption that transfers and fluxes are facilitated by local-scale diffusive transport within and between phase constituents. We will therefore choose closures that depend on the diffusive properties of the pure-phase materials, as well as on a phenomenological representation of how local-scale phase topology effectively permits diffusive relaxation. Note that the advective flux carried on each phase velocity field is part of the volume flux, $\qfi$, and does not require further closure. Advective transfer between phases is captured by terms  in the governing equations multiplying phase change reactions, $\Gmi$. For example, $\vi \Gmi$ is the advective transfer of momentum between phases.  $\Gmi$ itself is the sum of component transfers, $\Gji$, for which we will again propose coefficients based on the assumption of local diffusion.

The pure-phase diffusion parameters, $k_a^i$, for thermal, chemical, and momentum diffusion in igneous materials are relatively well known from experiments and theory: the thermal diffusion parameter or conductivity is $k_s^i = \rhoi c_\Pi \kappa_\Ti$, with $\kappa_\Ti$ the thermal diffusivity; the chemical diffusion parameters are $k_j^i = \rhoi D_j^i$, with $D_j^i$ the component diffusivities; and, the momentum diffusion parameter is $k_v^i = \etai$, the dynamic viscosity. The volume diffusion parameter, $k_\phii$, is less well-defined. We introduce it as $k_\phii = {d^i}^2/\etai$, with $d^i$ the granular size (see paragraph below). It will become more apparent why this choice is appropriate once we apply our model to two-phase porous and suspension flows below. For now, we note that it resembles the compliance (inverse of resistance) of a fluid of viscosity $\etai$ deforming around an obstacle of size $d^i$.

We postulate that the system-scale effects of local-scale phase topology may be reasonably approximated by two phenomenological attributes. The first is a characteristic local length scale, or granular size, $d^i$, which represents the mean grain size, bubble size, or melt film thickness. As the second attribute, we introduce what we term permission functions, $\theta_a^i$, which describe how diffusive transport within and between phases is permitted depending on whether the local phase topologies are contiguous/connected or disaggregated/disconnected.

\subsection{Flux and transfer coefficients}
Based on the above assumptions, we choose the following closures for flux coefficients,
\begin{linenomath*}
\begin{subequations}
\label{eq:flx-coeff}
\begin{align}
	\label{eq:flx-coeff-entr}
	\Ksi &= \phii k_s^i \theta_s^i = \phii \rhoi c_\Pi \kappa_\Ti \theta_s^i \ , \\
	\label{eq:flx-coeff-comp}
	\Kji &= \phii k_j^i \theta_j^i = \phii \rhoi D_j^i \theta_j^i \ , \\
	\label{eq:flx-coeff-moment}
	\Kvi &= \phii k_v^i \theta_v^i = \phii \etai \theta_v^i \ , \\
	\label{eq:flx-coeff-volume}
	K_\phii &= \phii k_\phii \theta_\phii = \dfrac{\phii {d^i}^2 \theta_\phii}{\etai} \ ,
\end{align}
\end{subequations}
\end{linenomath*}
and transfer coefficients,
\begin{linenomath*}
\begin{subequations}
\label{eq:trf-coeff}
\begin{align}
	\label{eq:trf-coeff-entr}
	\Csi &= \dfrac{\phii (1-\phii) k_s^i \theta_s^i}{{d^i}^2} = \dfrac{\phii (1-\phii) \rhoi c_\Pi \kappa_\Ti \theta_s^i}{{d^i}^2} \ , \\
	\label{eq:trf-coeff-comp}
	\Cji &=  \dfrac{\phii (1-\phii) k_j^i \theta_j^i}{{d^i}^2}  = \dfrac{\phii (1-\phii) \rhoi D_j^i \theta_j^i}{{d^i}^2}\ , \\
	\label{eq:trf-coeff-moment}
	\Cvi &=  \dfrac{\phii (1-\phii) k_v^i \theta_v^i}{{d^i}^2} = \dfrac{\phii (1-\phii) \etai \theta_v^i}{{d^i}^2}   \ , \\
	\label{eq:trf-coeff-volume}
	\Cvi &= \dfrac{\phii (1-\phii) k_\phii \theta_\phii}{{d^i}^2} =  \dfrac{\phii (1-\phii) \theta_\phi^i}{\etai} \ .
\end{align}
\end{subequations}
\end{linenomath*}
We scale flux coefficients with $\phii$, and transfer coefficients with $\phii (1-\phii)$ to ensure that fluxes and transfers cease when the phase is exhausted ($\phii \rightarrow 0$), and that transfers also cease in the pure-phase limit ($\phii \rightarrow 1$). Transfer coefficients are directly related to flux coefficients by a factor $(1-\phii)/{d^i}^2$.

To limit the number of independent model choices, we exploit model symmetry and propose volume transfer coefficients \eqref{eq:trf-coeff-volume} based on the volume diffusion parameter, $k_\phii$. By making this choice, we do not imply a correspondence between compaction and volume diffusion other than that they share the same dimensionality and both conceptually relate to system-scale effects of one phase locally deforming around other constituents. As noted in section \ref{sec:closures}, our approach to compaction contrasts with the model of \cite{mckenzie84} and its extension to $n$-phase media by \cite{Oliveira2018}, which both identify their compaction coefficients as volumetric or bulk viscosities. We choose our volume transfer coefficient independent of phase bulk viscosities to emphasise the multi-phase aspect of compaction, which may include but is not typically dominated by volumetric deformation internal to compressible phases.

\subsection{Permission functions}
The permission functions are a set of phase- and process-specific functions, $\theta_a^i \geq 0$, which prescribe the effective system-scale behavior of a multi-phase system and isolate the effective medium aspects of our closures \citep[e.g.,][]{Mavko2009}. Their formulation and calibration represents a highly consequential step of application-oriented model building within our framework.

If permissions are equal to unity, the coefficients depend linearly on phase proportions and thus represent diffusion along straight, connected pathways within a phase. Such simple volumetric weighting of effective transport is not usually sufficient to describe the effective media behaviour of interest here. Instead, permissions are introduced to express the effective increase or decrease of transport rates depending on whether a phase is internally well-connected, or whether its disconnected constituents are connected to adjacent phases of higher or lower diffusivity. One way to formalise this concept is to write permissions as averages of pure-phase diffusive properties weighted by a measure of connectivity within and between phases. 

Different types of averages can be employed to prescribe effective media coefficients. To use momentum diffusion as an example, if total stress accommodated by an aggregate is the sum of partial phase stresses while all phases experience the same strain rates, an arithmetic average is appropriate; it represents an upper limit to effective aggregate viscosity \citep[Voigt bound,][]{Hill1963}. In contrast, weighted harmonic averaging is appropriate if the total strain rate accommodated is equal to the sum of partial strain rates while each phase supports the same stress. This provides a lower limit to effective rheology \citep[Reuss bound,][]{Hill1963}. Stricter bounds have been derived \citep{Hashin1963}, but they still leave considerable room for different effective behavior in between, particularly when pure-phase end-members have strongly contrasting viscosities. 

For a given set of weights, geometric averaging produces effective properties in between upper and lower bounds in logarithmic space, a convenient property when dealing with up to ten or more orders of magnitude phase viscosity contrasts not uncommon in igneous aggregates. We therefore write permission functions as weighted geometric averages, 
\begin{linenomath*}
\begin{equation}
	\label{eq:flx-permission}
	\theta_a^i = \prod_k {M_a^{ik}}^{X_\phi^{ik}} \ ,
\end{equation}
\end{linenomath*}
of pure-phase diffusive parameter contrasts,
\begin{linenomath*}
\begin{equation}
	\label{eq:flx-permission}
  	M_a^{ik} = \dfrac{k_a^k}{k_a^i} \ ,
\end{equation}
\end{linenomath*}
with permission weights, $X_\phi^{ik}$, which we will now describe in more detail.

We propose phase-wise sets of $n$ permission weights: intra-phase weights, $X_\phi^{ii}$, describe the relative availability of connected pathways along which diffusive transport is permitted within a phase; inter-phase weights, $X_\phi^{ik}$ for $i \neq k$, express relative connectivity to other phases by which diffusive transport between adjacent phases is permitted. Permission weights must sum to unity within each phase, $\sum_k X_\phi^{ik} = 1$. 

As a phenomenological metric of local phase connectivity, permission weights should depend on phase fractions, sizes and shapes, sizes, compositions and surface energies of phase constituents, the rate and state of deformation, and more. To limit complexity, we prescribe them as functions of phase fractions only,
\begin{linenomath*}
\begin{align}
	\label{eq:closures-connect}
	X_\phi^{ik} &= \left(\sum_k A^{ik} S_\phi^{ik}\right) \phii + \left(1-\sum_k A^{ik} S_\phi^{ik} \right) S_\phi^{ik} \ ,
\end{align}
\end{linenomath*}
where $S_\phi^{ik}$ are smooth step functions in phase space,
\begin{linenomath*}
\begin{align}
	\label{eq:closures-step}
	S_\phi^{ik} &=  \normsum{ \left(\dfrac{\phik}{B^{ik}} \right) ^ {1/C^{ik}} }{k} \ .
\end{align}
\end{linenomath*}
The square brackets again denote normalisation to the sum over index $k$. The three sets of fitting parameters are: slopes $0 \leq A^{ik} \leq 1$ describing a gradual change in phase connectivity of phase $i$ between pure-phase limits and a smooth step function centred about critical phase fractions $0 \leq B^{ik} \leq 1$ ($\sum_k B^{ik} = 1$), with weights  $C^i \geq 0$ setting the relative step width. The step function describes the substantial decline in internal phase connectivity that marks a disaggregation or percolation threshold.

\begin{figure}[htb]
  \centering
  \includegraphics[width=0.9\textwidth]{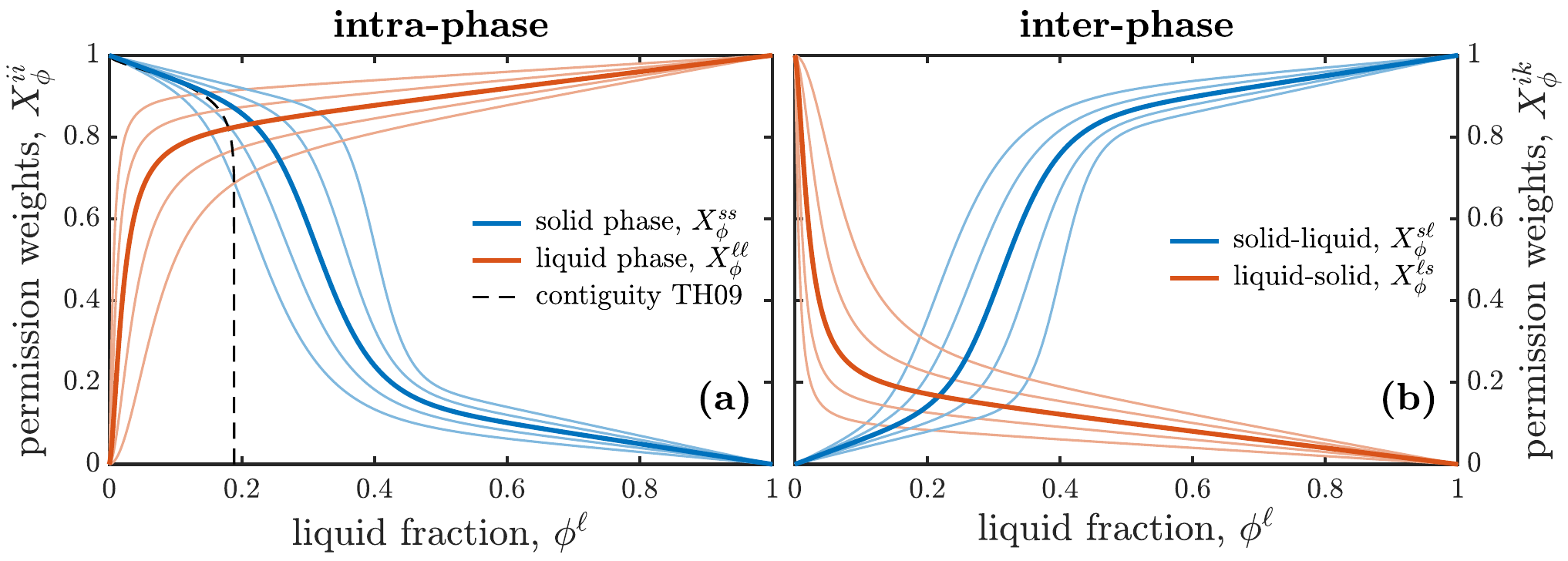}
  \caption{Sample family of permission weights for an igneous two-phase aggregate. Intra-phase weights for the solid (blue) and liquid (red) phases in \textbf{(a)} are unity at respective pure-phase limits, first decrease linearly before dropping steeply along smooth step functions centred around a solid disaggregation threshold ($\phi^\ell = 0.30$ for heavy reference curve), and a liquid percolation threshold ($\phi^ell = 0.02$ for heavy reference). Inter-phase weights in \textbf{(b)} are complementary to intra-phase ones in two-phase mixtures. Fine lines show a range of possible variability. Suitably scaled granular contiguity (black line) for basalt in olivine \citep[TH09:][]{Takei2009} given for comparison}.
  \label{fig:2connect}
\end{figure}

Figure~\ref{fig:2connect} shows a sample family of permission weights broadly applicable to an igneous solid-liquid aggregate. A high solid intra-phase weight, $X_\phi^{ss}$, is indicative of a contiguous granular matrix, while a high liquid intra-phase weight, $X_\phi^{\ell \ell}$, suggests that melt forms a connected liquid network (see Fig.~\ref{fig:2connect}(a)). The highlighted curves in Fig. \ref{fig:2connect} illustrate a scenario of a low dihedral angle ($\sim 30^\circ$) between melt and rock phases, as for basaltic melt in olivine-rich mantle rock. The liquid permission weight drops off steeply towards a liquid percolation threshold at $\phil=0.02$ to express that the pore liquid is allowed to remain connected along grain boundaries even at low liquid fraction \citep[e.g.,][]{VanBargen1986,Rudge2018a}. The melt wets solid grains relatively well, as indicated by the solid weight dropping off about a solid disaggretion threshold at $\phil=0.3$. 

Due to the phase-wise unity sum constraint on permission weights, inter-phase weights in Fig.~\ref{fig:2connect}(b) exactly complement intra-phase ones in Fig.~\ref{fig:2connect}(a). High solid-liquid permission weights, $X_\phi^{s \ell}$, at high melt fractions express that disaggregating solid grains become well connected to the carrier melt. Conversely, the liquid-solid weights, $X_\phi^{\ell s}$, increase sharply towards very low melt fractions as remaining melt films disconnect and become enclosed by solid grains. The fitting parameter values for the highlighted curves in Fig.~\ref{fig:2connect} are listed in Appendix Table \ref{tab:fitting2}; a script to reproduce this figure is available online \citep{scripts-repo}.

Our permission weights are, at least conceptually, related to more quantitative geometric attributes of phase topology, such as granular contiguity, relative contact area, volumetric connectivity, and others. Such attributes have been empirically quantified by microtomographic analysis \citep{Zhu2011,Miller2014,Colombier2018} or theoretically derived  for idealised equilibrium melt-grain textures \citep[e.g.,][]{Takei1998,Takei2009,Zhu2003,Ghanbarzadeh2014,Rudge2018a}. Based on this body of work we can define phase connectivities, $x_\phi^{ik}$, as the fractions of total surface area of constituents of phase $i$ connecting within the same phase, $x_\phi^{ii}$, and to adjacent phases $k$, $x_\phi^{ik}$. Within reason, we can relate our permission weights to such geometrically well-defined metrics. For example, we show the permission weight-equivalent form of the semi-empirical granular contiguity model of \citet{Takei2009} in Fig.~\ref{fig:2connect}(a) (black line: $1 + \mathrm{log}_{10}( {\varphi^s}^2) / 16$, with $\varphi^s = 1-2 {\phil}^{0.5}$ the granular contiguity). In comparison, our solid permission weights not only capture the loss of contiguity as the solid disaggregates, but additionally describe a smooth transition from intra-phase connectivity within the solid, to inter-phase connectivity between solid and liquid phases.

\begin{figure}[ppp]
  \centering
  \includegraphics[width=0.9\textwidth]{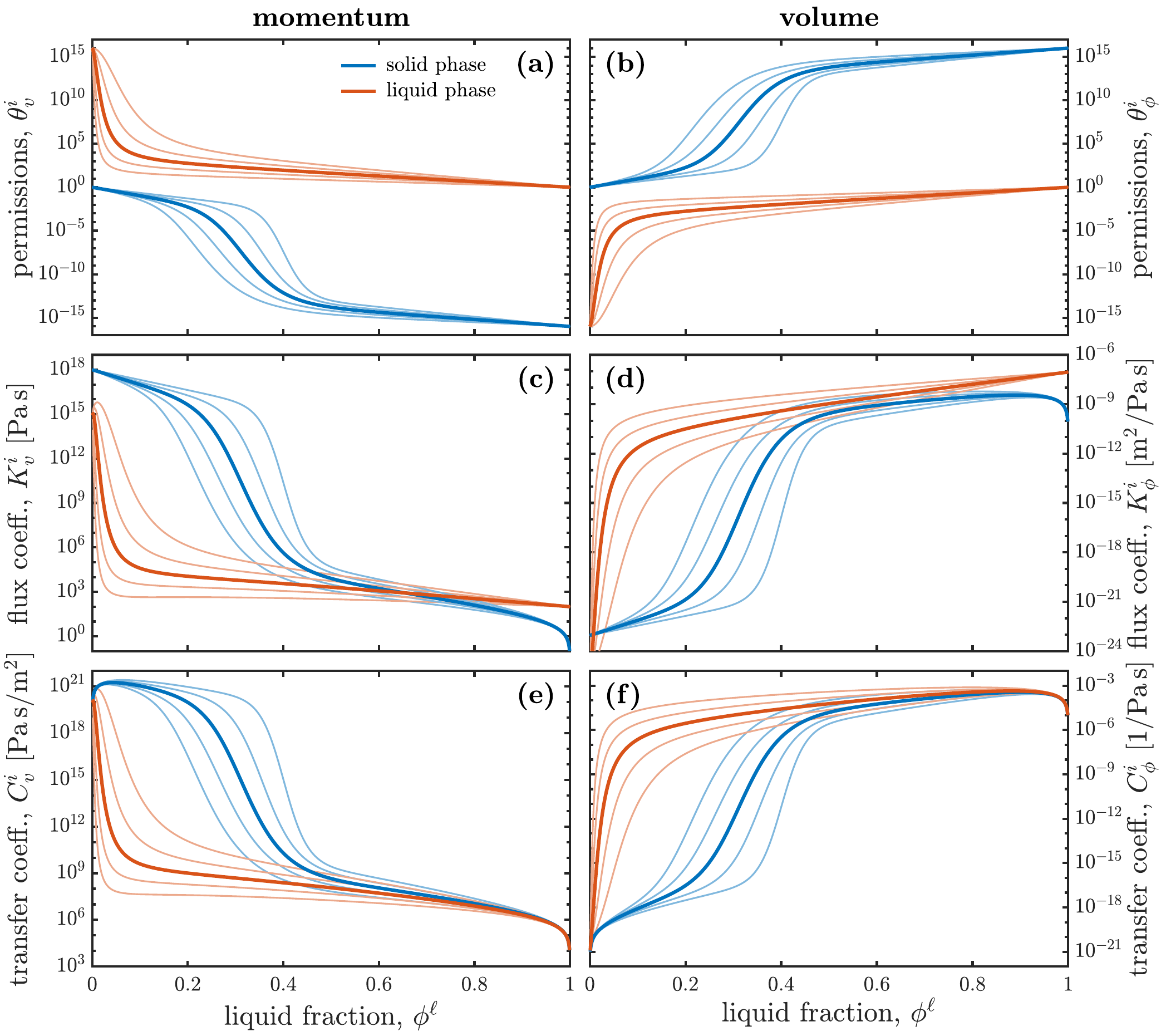}
  \caption{Sample family of permission functions, and flux and transfer coefficients for momentum and volume transport, based on sample family of permission weights in Fig. \ref{fig:2connect}. Functions are calculated for igneous two-phase system with 3 mm granular size, and solid and liquid viscosities of 10$^{18}$ Pas, and 10$^2$ Pas, respectively. The shape of momentum \textbf{(a)} and volume permissions \textbf{(b)} for the solid (blue) and liquid (red) phases reflect the shape of permission weights, while their amplitude is controlled by phase viscosity contrasts. Resulting coefficients for momentum \textbf{(c)} and volume \textbf{(d)} fluxes, and momentum \textbf{(e)} and volume \textbf{(f)} transfers show effective medium behaviour as prescribed by permissions.}
  \label{fig:2coeff}
\end{figure}

Figure \ref{fig:2coeff} shows the momentum and volume permissions (Fig. \ref{fig:2coeff}(a) \& (b)) along with the resulting flux and transfer coefficients (Fig. \ref{fig:2coeff}(c)--(f)) given the permission weights in Fig.~\ref{fig:2connect}; a script to reproduce the figure is available online \citep{scripts-repo}. The two phases are given a (constant) granular size of $d_0$ = 3 mm, and (constant) pure-phase viscosities of $\etas_0$ = 10$^{18}$ Pas, and $\etal_0$ = 10$^{2}$ Pas, respectively. Owing to the strong phase viscosity contrast, permissions vary over sixteen orders of magnitude. Due to the shape of permission weights, much of that variation occurs over a limited region of two-phase space.

The momentum permissions (Fig. \ref{fig:2coeff}(a)) describe the relative variation of effective phase viscosities. After an initially log-linear decrease, the solid momentum permission steeply drops across the solid disaggregation limit at $\phil = 0.3$. The liquid momentum permission first steeply plunges from a high value at the percolation limit, which reflects the effectively stiff rheology of isolated melt pockets, before gradually approaching unity in the pure-melt limit. The permissions for phase volume diffusion (Fig. \ref{fig:2coeff}(b)) show inverted trends since volume diffusivity contrasts are dominated by the inverse ratio of phase viscosities, $M_\phi^{ik} \approx {M_v^{ik}}^{-1}$.

Inspecting the resulting momentum and volume flux coefficients in Fig. \ref{fig:2coeff}(c), we note that our models predict momentum diffusion (i.e., shear stress) to be dominated by the solid phase up to a relatively high liquid content of $\sim$ 60\% despite disaggregation at 30\%. The volume diffusion coefficients in Fig. \ref{fig:2coeff}(d) are most significant at high melt fractions, with both phase coefficients dropping to negligible values at low melt fractions. The momentum and volume transfer coefficients (i.e., modulators of segregation drag and compaction compliance, respectively) in Fig. \ref{fig:2coeff}(e) \& (f) show strong similarity in shape to their corresponding flux coefficients, but drop off sharply at either end of the phase space due to their scaling with $\phi (1-\phii)$. 

The permissions demonstrated for mechanical coefficients here can be applied similarly to thermal and chemical coefficients.  The resulting closures provide a set of phase- and process-wise symmetrical transport coefficients to close the governing equations.

\section{Special limits for igneous systems \label{sec:discussion}}
With a closed set of governing equations in place, we can proceed to discuss the application of our model framework to specific two- and three-phase limits pertinent to igneous systems. In the following, we apply our model to the limit of two incompressible phases, solid rock or crystals, ($i = s$), and liquid melt ($i = \ell$). We first obtain a generalized set of two-phase equations valid across all phase proportions. We then calibrate a set of permission functions that recover existing empirical and theoretical models in the limits of two-phase porous and suspension flows. By inspecting the reference velocity, $\vstar$, and pressure, $\Pstar$, arising from the calibrated transfer coefficients, we show how to reduce the general two-phase equations to the porous and suspension limits, and how to treat the mush transition in between. Finally, we add a compressible volatile phase ($i = v$, a supercritical fluid, magmatic brine, or gaseous vapour), and briefly discuss a possible coefficient calibration for igneous three-phase flows and its ramifications for magma degassing regimes.

For simplicity, we will limit the following discussion by assuming thermal phase equilibrium ($\Ti = \Tstar$) and an isochemical aggregate ($\cji = c_0$). We continue to include reactions by their symbol only and hence assume a known melting rate, $\Gamma_\rho^{s \ell}$ (mass transfer from solid to liquid phase), and degassing rate, $\Gamma_\rho^{\ell v}$ (mass transfer from liquid to vapour phase).

\subsection{Two incompressible phases}
The governing equations for low-Reynolds number, incompressible,  two-phase reactive transport in a solid-liquid aggregate are a special case of the general equations \eqref{eq:final-fluidmech}--\eqref{eq:final-thermochem}:
\begin{linenomath*}
\begin{subequations}
\label{eq:twophase-mechanical}
\begin{align}
	\label{eq:twophase-segr-solid}
	\vds &= - \dfrac{\phis}{C_v^s} \left(\phis \Grad \Pstar + \Grad \Pds - \Div K_v^s \Ds - \phis \rhos \gvec \right) \ , \\
	\label{eq:twophase-segr-liquid}
	\vdl &= - \dfrac{\phil}{C_v^\ell} \left(\phil \Grad \Pstar + \Grad \Pdl - \Div K_v^\ell \Dl - \phil \rhol \gvec \right) \ , \\
	\label{eq:twophase-comp-solid}
	\Pds &= - \dfrac{\phis}{C_\phis} \left(\phis \Div \vstar + \Div \vds - \Div K_\phis \Delta (\Grad P)^{s*} + \Gamma_\rho^{s \ell}\left(\dfrac{1}{\rhos}-\dfrac{1}{\rhostar}\right) \right) \ , \\
	\label{eq:twophase-comp-liquid}
	\Pdl &= - \dfrac{\phil}{C_\phil} \left(\phil \Div \vstar + \Div \vdl - \Div K_\phil  \Delta (\Grad P)^{\ell *} - \Gamma_\rho^{s \ell}\left(\dfrac{1}{\rhol}-\dfrac{1}{\rhostar}\right) \right) \ , \\
	\label{eq:twophase-solidevo}
	\DstarDt{\phis} &= - \phis \Div \vstar - \Div \vds + \Div \tilde{K}_\phis \Delta (\Grad \phi)^{s*} - \dfrac{\Gamma_\rho^{s \ell}}{\rhos} \ , \\
	\label{eq:twophase-liquidevo}
	\DstarDt{\phil} &= - \phil \Div \vstar - \Div \vdl + \Div \tilde{K}_\phil \Delta (\Grad \phi)^{\ell*} + \dfrac{\Gamma_\rho^{s \ell}}{\rhol} \ , \\
	\label{eq:twophase-tempevo}
	c_p \rhob \DbDt{T} &= \Div \bar{K}_s \Grad T + \Delta L^{s\ell} \Gamma_\rho^{s \ell} + \alpha_0 T \gvec \cdot \overline{\rho \vel} + \bar{\Psi}_d + \rhob H_0 \ .
\end{align}
\end{subequations}
\end{linenomath*}
Note that in saturated two-phase systems ($\phil = 1 - \phis$), the solid and liquid phase evolution equations are exactly equivalent, and in practice only one of them is required. In the temperature equation \eqref{eq:twophase-tempevo}, we have assumed constant and equal heat capacity, $c_P$, thermal expansivity, $\alpha_0$, and internal heating rate, $H_0$, for all phases. $\bar{K}_s = \sum_i \Ksi$ is the aggregate thermal conductivity, $\rhob = \sum_i \phii \rhoi$ the aggregate density, $\Delta L^{s\ell}=L^s-L^\ell$ the latent heat of melting, and $\bar{\Psi}_d$ the total dissipation of shear, segregation and compaction flow. The aggregate material derivative in the temperature equation is,
\begin{linenomath*}
\begin{equation}
	\label{eq:lagr-deriv}
	\DbDt{(\, \cdot \,)} = \ddt{(\, \cdot \,)} + \vstar \cdot \Grad(\, \cdot \,) + \sum_i \dfrac{\rhoi}{\rhob} \vdi \cdot \Grad(\, \cdot \,) \ .
\end{equation}
\end{linenomath*}
We have further assumed that pressure changes in the adiabatic term in \eqref{eq:twophase-tempevo} are dominated by flow against gravity, $D_\phii \Pi / Dt \approx \phii \rhoi \vi \cdot \gvec$ and have written it in terms of the aggregate mass flux, $\overline{\rho \vel} = \rhob \vstar + \sum_i \rhoi \vdi$.

The two momentum equations above, \eqref{eq:twophase-segr-solid} \& \eqref{eq:twophase-segr-liquid}, are equivalent to those given by \citet{bercovici03} when neglecting surface tensions. Their two-phase drag coefficient, $c$, relates to our momentum transfer coefficients as,
\begin{linenomath*}
\begin{equation}
	\label{eq:twophase-dragcoeff}
	c = \omega_{C_v}^\ell C_v^s = \omega_{C_v}^s C_v^\ell =  \dfrac{C_v^s C_v^\ell}{C_v^s + C_v^\ell} \ .
\end{equation}
\end{linenomath*}
Our model further suggests a corresponding two-phase compaction compliance coefficient,
\begin{linenomath*}
\begin{equation}
	\label{eq:twophase-volumecoeff}
	c_\phi = \omega_{C_\phi}^\ell C_\phis = \omega_{C_\phi}^s C_\phil = \dfrac{C_\phi^s C_\phil}{C_\phis + C_\phil} \ .
\end{equation}
\end{linenomath*}
The two mass equations, \eqref{eq:twophase-comp-solid} \& \eqref{eq:twophase-comp-liquid}, differ formally from \citet{bercovici03} in that they are written as equations for phase compaction pressures. Our formulation also includes volume diffusion terms that theirs did not.

\subsection{Closures bridging porous and suspension regimes}
To apply the phase-symmetrical two-phase limit to porous and suspension flows in igneous systems, we calibrate the permissions for the mechanical flux and transfer coefficients. We do so for the example of a basaltic melt and olivine-rich rock, a relatively well-characterised system representing mantle magmatism, and continue to use the same granular size and pure-phase viscosities as in the examples above ($d_0$ = 3 mm, $\etas_0 = 10^{18}$ Pas, $\etal_0 = 10^2$ Pas). A similar procedure could be followed for other melt or rock types.

We achieve the calibration through an automatic fitting routine. Given a set of pure-phase material parameters, the routine randomly samples fitting parameters $A^{ik}$, $B^{ik}$, and $C^{ik}$ from suitable normal distribution. The resulting realisations of two-phase permission weights in turn define coefficients, which we compare to a set of reference curves by their least-squares misfit norm. The mean and standard deviations of the sampling distributions are adjusted as the misfit norm is reduced to narrow in on the best-fit region of the parameter space. New random samples are taken until the misfit norm drops below a specified tolerance or a set number of samples are taken; a script to reproduce the calibration routine is available online \citep{scripts-repo}.

\begin{figure}[ppp]
  \centering
  \includegraphics[width=0.9\textwidth]{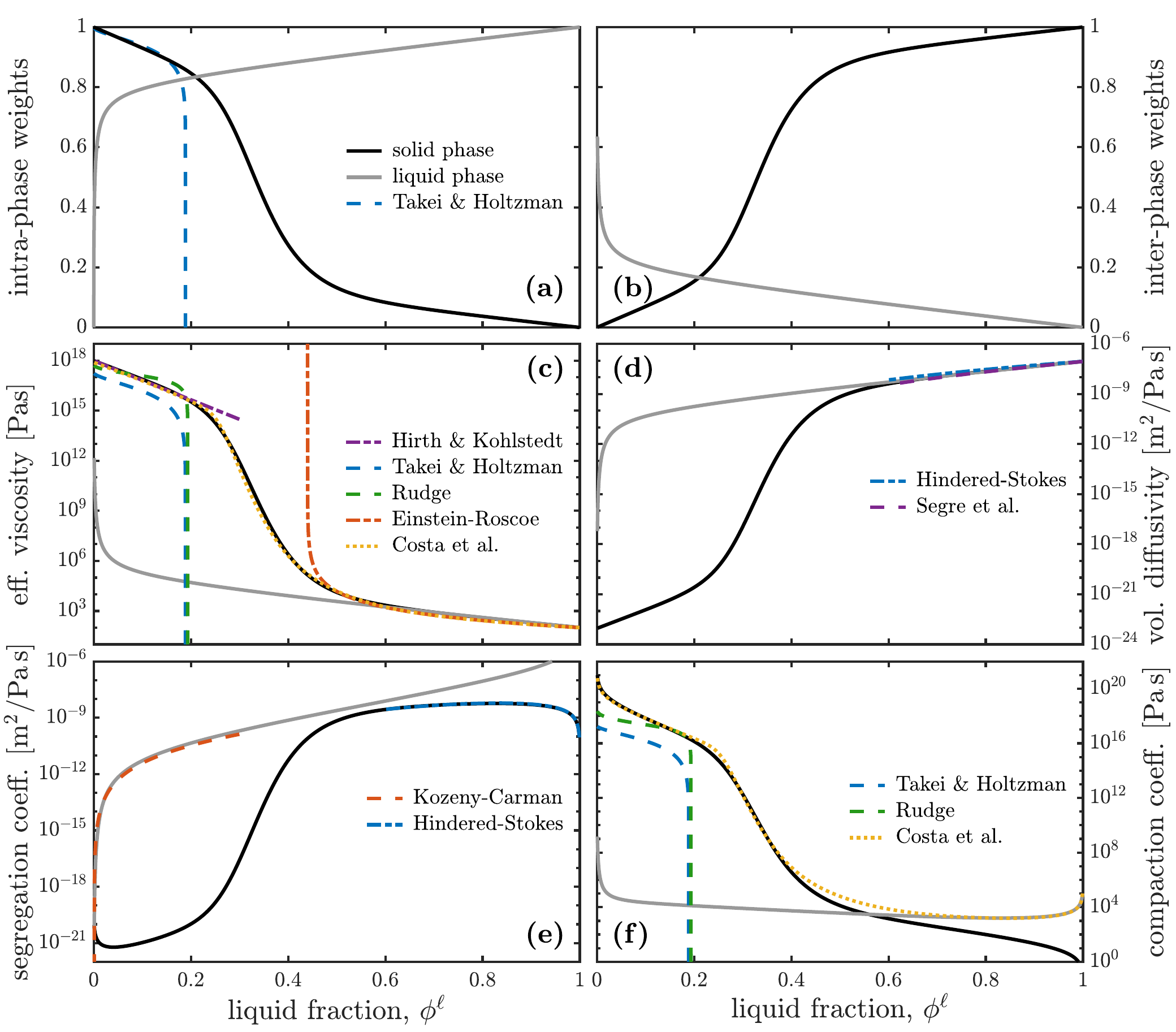}
  \caption{Mechanical coefficients fitted to reference closures for basalt and olivine rock. Permission weights in \textbf{(a)} \& \textbf{(b)} show solid disaggregation at $\phil \approx 27 \%$, and liquid disconnection at $\phil \approx 0.1 \%$. Effective shear viscosities ($\Kvi/\phii$) in \textbf{(c)} fitted to reference parameterisation after \cite{Costa2009} (yellow); effective shear viscosities from \cite{hirth03}, \cite{Takei2009}, and \cite{Rudge2018b}, and \citep{Roscoe1952} shown for comparison. Effective volume diffusivities ($\Kfi/\phii$) in \textbf{(d)} fitted to models related to hindered-Stokes speed, and experiments and theory of \cite{segre2001effective}. Segregation coefficients (${\phii}^2/\Cvi$) in \textbf{(e)} fitted to Kozeny-Carman-type Darcy percolation and hindered-Stokes settling coefficients. Effective compaction viscosities (${\phii}^2/\Cfi$) in \textbf{(f)} compared to reference curve from \cite{Costa2009} divided by $\phil (1-\phil)$, and compaction viscosity models by \cite{Takei2009} and \cite{Rudge2018b}.}
  \label{fig:fitcoeff}
\end{figure}

Figure \ref{fig:fitcoeff} shows the obtained best-fit coefficients in forms equivalent to the reference curves used for calibration, as well as a number of further closures from the literature for comparison. The automatic calibration routine was stopped after $2 \times 10^5$ samples taken. The best-fit permission weights in Fig.~\ref{fig:fitcoeff}(a)--(b) indicate disaggregation of the solid at a melt fraction of $\sim$30\%, and a relatively sudden disconnection of the liquid towards a percolation limit at $\sim$0.1\% melt content. The reference models we calibrate against include an effective shear viscosity curve across phase space (Fig.~\ref{fig:fitcoeff}(c)) \citep{Roscoe1952,hirth03,Costa2009,Takei2009,Rudge2018b}, a Darcy segregation coefficient for percolating liquid (Fig.~\ref{fig:fitcoeff}(e)) \citep[e.g.,][]{hubbert1957darcy, mckenzie84} and effective compaction viscosity for compacting solid (Fig.~\ref{fig:fitcoeff}(f)) \citep{Costa2009,Takei2009,Rudge2018b} in the porous limit, and a volume diffusivity (Fig.~\ref{fig:fitcoeff}(d)) \citep{segre2001effective} and hindered-Stokes segregation coefficient for settling particles in the suspension limit (Fig.~\ref{fig:fitcoeff}(e)). 

For the purpose of calibrating our flux and transfer coefficients to reference curves in common two-phase flow notation, we relate our momentum flux coefficients to effective phase and mixture viscosities as, $\Kvi/\phii = \etai \theta_v^i$, and, $\sum_i \Kvi$, respectively. In analogy, we relate our volume flux coefficients to effective phase and mixture volume diffusivities as, $\Kfi/\phii = k_\phii \theta_v^i$, and $\sum_i \Kfi$. We further relate our momentum transfer coefficients to phase segregation coefficients, ${{\phii}^2}/{\Cvi}$, which formally correspond to the canonical Darcy percolation and hindered-Stokes settling coefficients. Lastly, we relate our volume transfer coefficients to phase compaction coefficients, ${{\phii}^2}/{\Cfi}$, corresponding to the effective compaction (or bulk) viscosity in \cite{mckenzie84} and similar works. 

\citet{Costa2009} proposes a semi-empirical parameterisation for the effective mixture viscosity between the limits of partially molten rock and crystal-bearing magma. We use it to define a reference curve (yellow dotted, Fig.~\ref{fig:fitcoeff}(c)) that fits with a range of end-member models \citep{Roscoe1952,hirth03,Takei2009,Rudge2018b} applicable to basalt and olivine. Our calibration minimises the misfit between $\sum_i \Kvi$ and the reference curve across the entire two-phase space, as well as between $K_v^s/\phis$, and $K_v^\ell/\phil$ and the reference curve on the intervals of $\phil \in [0,0.3]$, and $\phil \in [0.6,1]$, respectively.

At low melt fraction, experiments suggest an exponential decrease in effective viscosity, $\etas \exp(-\lambda \phi^\ell)$ \citep[e.g.,][]{hirth03}. Our reference curve traces a slope of $\lambda = 27$ \citep{mei02} (purple dash-dotted, Fig.~\ref{fig:fitcoeff}(c)). Effective viscosity laws have been derived analytically for grain-boundary diffusion (Coble) creep \citep{Takei2009}, and volume diffusion (Nabarro-Herring) creep \citep{Rudge2018b} of idealised basalt-in-olivine textures. Our reference curve first follows a similar slope to that of \citet{Rudge2018b} (green dashed, Fig.~\ref{fig:fitcoeff}(c)), but continues on to a smooth step across the disaggregation threshold rather than dropping to zero. The model of \citet{Takei2009} (blue dashed, Fig.~\ref{fig:fitcoeff}(c)) shows a similar initial slope and disaggregation drop-off, but has a weakening offset at very low melt fractions, a consequence of melt along grain boundaries providing fast diffusion pathways  in their model. \citet{Schmeling2012} proposes an alternative viscosity model based on effective media theory, which produces curves of similar shape (not shown here).

For crystal-bearing magmas, the increase in effective mixture viscosity with solid fraction has been described by the Einstein-Roscoe relation, $\eta^\ell (1-\phi^s/\phi^s_c)^{-\phi^s_c B}$ \citep{Einstein1906,Einstein1911,Roscoe1952,Krieger1959}. Our reference curve is calibrated to fit an Einstein-Roscoe law with $\phi^s_c = 0.58$, and $B = 4.0$ (red dash-dotted, Fig.~\ref{fig:fitcoeff}(c)). Whereas this simple model of particle stiffening becomes ill-defined at the critical solid fraction, $\phi^s_c$, our reference curve continues along a smooth step function towards the solid end of phase space. 

As mentioned above, the diffusive part of the phase volume flux, $\qfi$, can be interpreted in the context of sedimentation experiments and theory \citep[e.g.,][]{nicolai1995particle,segre2001effective,mucha2004model}, which suggest that local fluctuations in settling velocity lead to diffusion of the suspended particle fraction. In general, a diffusivity is the product of a mean velocity fluctuation times a mean free path length (or correlation length), $ \kappa_\phii = \delta v^i h^i$. To a reasonable approximation, sedimentation studies indicate that $\delta v^i \sim | \Delta \vel^{s \ell} |$, and $h^i \sim 10 \times d^i$ \citep{nicolai1995particle,segre2001effective,mucha2004model}. In dilute suspensions, $| \Delta \vel^{s \ell} |$ between suspended particles and the carrier liquid can be approximated by the hindered-Stokes settling speed, 
\begin{linenomath*}
\begin{align}
	\label{eq:hindered-Stokes}
	| \Delta \vel^{s \ell} | = \dfrac{2 {d_0}^2}{9 \etal} \Delta \rho^{s \ell} g (1-\phis)^{m} \ ,
\end{align}
\end{linenomath*}
with $\Delta \rho^{s \ell} = \rhos - \rhol$ the solid-liquid density contrast, and a hindering exponent $4 \leq m \leq 6$. Taking \eqref{eq:hindered-Stokes} to the dilute limit ($\phis \rightarrow 0$), we recover a volume diffusion parameter in the form we introduced above,
\begin{linenomath*}
\begin{align}
	\label{eq:volume-diff}
	k_\phis = \dfrac{\kappa_\phis}{p^s_0} = \dfrac{{d_0}^2}{\etal}  \ ,
\end{align}
\end{linenomath*}
where we have divided $\kappa_\phis = 10 | \Delta \vel^{s \ell} | d_0$ by the granular pressure scale, $p^s_0 = |\Delta \rho^{s \ell}| g d_0$ and set the remaining geometric prefactor to unity.

We calibrate our volume flux coefficients by fitting the effective mixture volume diffusivity, $\sum_i \Kfi$, to two reference curves. The first is $d_0^2 (1-\phis)^5 / \etal$ (blue dash-dotted, Fig.~\ref{fig:fitcoeff}(d)), obtained by multiplying $k_\phis$ in \eqref{eq:volume-diff} by the hindering function in \eqref{eq:hindered-Stokes} at $m=5$. The second is the theoretically derived and experimentally confirmed volume diffusivity relation of \citet{segre2001effective} based on Brownian suspension theory \citep[e.g.,][]{AlNaafa1993} (purple dashed, Fig.~\ref{fig:fitcoeff}(d)). The latter produces a broadly similar but slightly steeper curve. For our calibration, we evaluate the misfit between $\sum_i \Kfi$ and both reference curves on the interval $\phil \in [0.6,1]$. We are not aware of volume diffusivity models to compare against in the porous limit. Our best-fit  coefficients predict negligibly small values in that limit, particularly for the solid phase. 

We calibrate our liquid momentum transfer coefficient, $C_v^\ell$, on the interval $\phil \in [0,0.3]$ by minimising the misfit between the  liquid segregation coefficient, ${\phil}^2/C^\ell_v$, and a reference Darcy percolation coefficient,
\begin{linenomath*}
\begin{equation}
\dfrac{{\phil}^2}{C_v^\ell} = \dfrac{k_\mathrm{KC}}{\etal} = \dfrac{{d_0}^2}{50} \, \dfrac{1}{\etal} \, \left(\phil-0.001\right)^{2.75} (1-\phil)^{-2} \ ,
\end{equation}
\end{linenomath*} 
were $k_\mathrm{KC}$ is a Kozeny-Carman permeability following a powerlaw of melt fraction within the range of theory and experiments \citep[e.g.,][]{VanBargen1986,Zhu2011,Miller2014,Rudge2018b}. The solid momentum transfer coefficient we calibrate on the interval $\phil \in [0.6,1]$ by minimising the misfit between our solid segregation coefficient ${\phis}^2/C^s_v$ and a hindered-Stokes settling coefficient,
\begin{linenomath*}
\begin{equation}
\dfrac{{\phis}^2}{C_\phi^s} = \dfrac{{d_0}^2}{\eta^\ell} \phis (1-\phis)^{5} \ . 
\end{equation}
\end{linenomath*} 
Our calibrated closures not only reconcile these previously disconnected end-member models but notably achieve this fit by employing the same permissions as used to fit effective phase viscosities above.

The defining feature of effective compaction viscosity models for partially molten rock is an inverse dependence on melt fraction, $\zetas \sim \etas / \phil$ \citep{mckenzie84,bercovici03}, necessary to approach infinite resistance to compaction in the pure-solid limit. \cite{simpson10b} recover this behaviour by homogenisation across a range of idealised pore geometries, and so do \cite{Schmeling2012} in their parameterisation based on effective medium theory. In contrast, \citet{Takei2009} do not obtain such a dependence in their analytical Coble creep model, while \citet{Rudge2018b} finds a less pronounced dependence of $\sim 1/\log \phil$ for Nabarro-Herring creep.

In fact, we do not need to further calibrate our volume transfer coefficients against any reference models here. Rather, Fig. \ref{fig:fitcoeff}(f) shows our compaction coefficients, ${\phii}^2/\Cfi$, directly resulting from the calibrated permissions fitted to other reference curves above. By virtue of the factor $\phii(1-\phii)$ included in all transfer coefficients (see \eqref{eq:trf-coeff}), our compaction coefficients recover the inverse dependence on phase fraction in both the porous and suspension limits.  Other than that, the compaction coefficients in Fig. \ref{fig:fitcoeff}(f) follow the shape of the effective shear viscosity curves in Fig. \ref{fig:fitcoeff}(c). The correspondence is apparent when comparing our calibrated compaction coefficients to the reference shear viscosity in Fig.~\ref{fig:fitcoeff}(c) divided by $\phil (1-\phil)$ (yellow dotted, Fig.~\ref{fig:fitcoeff}(f)). 

Compaction is not typically considered in suspension models \citep[e.g.,][]{mucha2004model}, where it is implicitly assumed that phase pressure deviations become negligible. Nevertheless, our framework defines compaction coefficients across all of phase-space, including the liquid-dominated limit. However, the magnitude of the liquid compaction coefficient remains small compared to its solid equivalent in the porous regime. Equation \eqref{eq:final-fluidmech-comp} suggests that pressure deviations scale with the compaction coefficient. Hence, the low coefficient values bear out the assumption of negligible compaction pressures in the suspension limit.

\begin{figure}[ppp]
  \centering
  \includegraphics[width=0.5\textwidth]{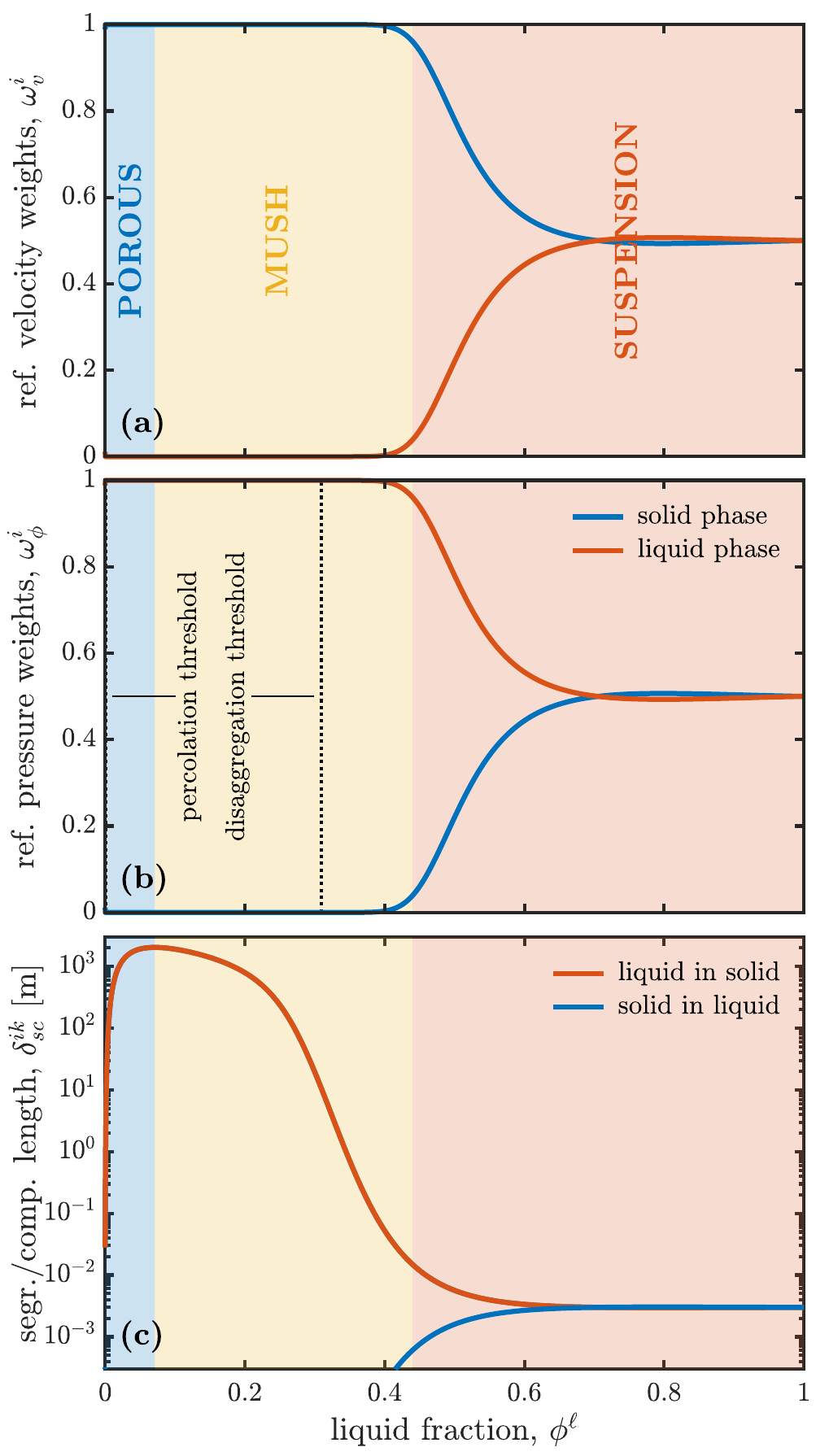}
  \caption{Porous to suspension flow transition for calibrated two-phase system of basalt and olivine. Reference weights for velocity \textbf{(a)} and pressure \textbf{(b)} for the solid (blue) and liquid (red) according to calibration shown in Fig. \ref{fig:fitcoeff}. Segregation-compaction lengths for liquid-in-solid (red) and solid-in-liquid (blue) shown in \textbf{(c)}. Dotted lines indicate liquid percolation ($\phi=0.001$) and solid disaggregation ($\phil = 0.3$) thresholds. Porous (blue), Mush (yellow), and Suspension (red) regimes indicated as defined by transitions in segregation-compaction length scales.}
  \label{fig:2transition}
\end{figure}

\subsection{Two-phase porous flow limit}
The calibrated set of material closures now bridges the regimes of two-phase porous and suspension flows of basalt and olivine. We can now use these coefficients to find reduced equations that apply in either of the limits and for the mush regime in between. We do so by inspecting the coefficient-based reference velocity and pressure, $\vstar$ and $\Pstar$. According to \eqref{eq:trf-weights-general}, $\vstar$ and $\Pstar$ are the weighted sums of phase velocities and pressures with weights $\ovi$ and $\ofi$ based on momentum and volume transfer coefficients. Figure \ref{fig:2transition} shows the weights calculated for the transfer coefficients calibrated as in Fig. \ref{fig:fitcoeff}; the script to reproduce this figure is available online \citep{scripts-repo}.

In the porous flow limit, the solid velocity weight, $\omega_{C_v}^s$, approaches unity, whereas $\omega_{C_v}^\ell$ vanishes. As a result, $\vstar \rightarrow \vs$ for $\phil \rightarrow 0$, and the solid velocity emerges as reference point for segregation velocities. We find the opposite holds for pressure weights, $\omega_{C_\phi}^{s,\ell}$, and therefore $\Pstar \rightarrow \Pl$, meaning the liquid pressure becomes reference point for compaction pressures. Both are a consequence of the momentum and volume permissions depending on phase viscosity contrasts and their inverse, respectively. Based on conceptual reasoning, \citet{bercovici03} anticipated similar weights for what they termed interfacial velocity and pressure. Their weights in our notation, $\omega_{C_v}^i = \normsum{\phii \etai}{i}$, are the same as ours if momentum permissions are unity. Their interface pressure weights were given simply as $\omega_{C_\phi}^i = 1-\omega_{C_v}^i$.

Examining segregation velocities and compaction pressures for both phases we find that $\vds$ and $\Pdl$ vanish in the porous limit. Our choice of transfer coefficients therefore implies that the solid velocity, $\vs$, the liquid pressure, $\Pl$, the liquid segregation velocity, $\vdl$, and the solid compaction pressure, $\Pds$, are suitable variables to pose reduced porous flow equations in. The latter two are equivalent to the canonical Darcy segregation velocity, $\vdl = \phil (\vl - \vs)$ \citep{Darcy1856,hubbert1957darcy}, and Terzaghi's effective pressure, $\Pds = \phis (\Ps - \Pl)$ \citep{Terzaghi1943,skempton60}, which have previously been employed for formulating two-phase porous flow equations \citep[e.g.,][]{connolly07,Morency2007,keller13}. Our framework now formalises this choice as a consequence of the adopted coefficient calibration.

We write the reduced governing equations in the porous flow limit as:
\begin{linenomath*}
\begin{subequations}
\label{eq:twophase-porous}
\begin{align}
	\label{eq:twophase-porous-moment}
	\Grad \Pstar &= - \Grad \Pds + \Div \bar{K}_v \Dstar + \rhob \gvec  \ , \\
	\label{eq:twophase-porous-mass}
	\Div \vstar &= - \Div \vdl - \Gamma_\rho^{s \ell} \left(\dfrac{1}{\rhos} - \dfrac{1}{\rhol} \right) \ ,\\
	\label{eq:twophase-porous-segr}
	\vdl &= - \dfrac{{\phil}^2}{C_v^\ell} \left(\Grad \Pstar - \rhol \gvec \right) \ , \\
	\label{eq:twophase-porous-comp}
	\Pds &= - \dfrac{(1-\phil)^2}{C_\phis} \left(\Div \vstar + \dfrac{\Gamma_\rho^{s \ell}}{1-\phil} \left(\dfrac{1}{\rhos}-\dfrac{1}{\rhol}\right) \right) \ , \\
		\label{eq:twophase-porous-phaseevo}
	\DstarDt{\phil} &= (1-\phil) \Div \vstar + \dfrac{\Gamma_\rho^{s \ell}}{\rhos} \ , \\
	\label{eq:twophase-porous-tempevo}
	c_p \rhob \DbDt{T} &= \Div \bar{K}_s \Grad T + \Delta L^{s\ell} \Gamma_\rho^{s \ell} + \alpha_0 T \gvec \cdot \left(\rhob \vstar + \rhol \vdl \right) + \rhob H_0 \\\nonumber 
	&+ \bar{K}_v \left[\Dstar\right]^2 + \dfrac{C_v^\ell \left[\vdl\right]^2}{{\phil}^2} + \dfrac{C_\phis \left[\Pds\right]^2}{(1-\phil)^2} \ .
\end{align}
\end{subequations}
\end{linenomath*}
In \eqref{eq:twophase-porous-moment}, $\bar{K}_v = \sum_i \Kvi \approx K_v^s$ is the effective aggregate viscosity. The relatively small magnitude of $K_v^\ell$, $K_\phi^s$, and $K_\phi^\ell$ in this limit (see Fig. \ref{fig:2coeff}) justifies dropping shear stress terms for the liquid and volume diffusion terms for both phases.

The reduced equations for porous flow in a deformable matrix \eqref{eq:twophase-porous} express Stokes flow of the mixture \eqref{eq:twophase-porous-moment}--\eqref{eq:twophase-porous-mass}, coupled to the segregation of the pore liquid \eqref{eq:twophase-porous-segr} and the compaction of the solid matrix \eqref{eq:twophase-porous-comp}. The aggregate momentum and mass equations, \eqref{eq:twophase-porous-moment}--\eqref{eq:twophase-porous-mass} recover the incompressible Stokes equations if $\vdl$, $\Pds$ and $\Gamma_\rho^{s \ell}$ vanish. The third and fourth equations recover Darcy's law for segregation \citep{Darcy1856} and McKenzie's law for compaction \citep{mckenzie84}, but are now written in terms of momentum and volume transfers and their coefficients. For phase evolution \eqref{eq:twophase-porous-phaseevo} we retain the solid equation, in which the divergence of $\vds$ drops out, leaving only reference velocity terms. In the temperature evolution \eqref{eq:twophase-porous-tempevo}, we have spelled out the non-negligible contributions to viscous dissipation, which are of solid shear, liquid segregation, and solid compaction flow. 

The compaction equation \eqref{eq:twophase-porous-comp} marks a slight shift from previous models \citep{sramek07,katz08,rudge11,weatherley12,keller16,Turner2017}. In \eqref{eq:twophase-porous-comp}, the solid compaction pressure relates both to mechanical ($\Div \vstar$) and reactive ($\Gamma_\rho^{s \ell} (1/\rhos-1/\rhol)$) rates of partial volume change, whereas most previous models had it as a function of the former only. We note that the reactive contribution would drop out under the assumption that $\rhos \approx \rhol$, which has often been invoked in previous work \citep[e.g., see extended Boussinesq approximation in][]{katz08}. 

\subsection{Two-phase suspension flow limit}
In the suspension flow limit, both the velocity and pressure weights in Fig.~\ref{fig:2transition}(a)--(b) gradually approach values near one half, at least for the calibration adopted here. For some alternative calibrations within the range shown in Figs \ref{fig:2connect} \& \ref{fig:2coeff}, the liquid velocity weight becomes dominant. Therefore, we interpret that, to a reasonable approximation, the liquid can be adopted as the velocity reference frame, $\vstar \rightarrow \vl$, and hence $\vdl \approx 0$. However, the approximately equal velocity weights (see Fig.~\ref{fig:2transition}(a)) imply that phase velocity differences driven by gravity or other forces on the system relax at similar rates in both phases, thus keeping both phase velocities close to the reference state. We may therefore expect that solid segregation remains small, $|\vds| \leq |\vstar|$. This holds at least for low-Re flows in systems much larger than the granular scale, as was recognized before in the context of magma chamber convection \citep{martin1988,Brandeis1986,Rudman1992,Bergantz1999}.

The same reasoning regarding reference pressure weights suggests that compaction pressures remain negligible in the suspension limit. Moreover, with solid grains fully submerged in the carrier liquid, buoyancy-driven phase pressure deviations will scale as $\sim \Drho_0 g_0 d_0 \ll 1$ Pa for typical igneous parameters. Therefore, we recover that $\Pstar = \Ps = \Pl$, and $\Pds = \Pdl = 0$ are good approximations, and that compaction pressure effects become negligible in the suspension limit. Of course, in general phase fractions can still be expected to evolve, $\partial \phii / \partial t \neq 0$, but volume transfers will not produce significant phase pressure deviations in suspensions.

We write the reduced equations for two-phase suspension flow in the variables of $\vstar$, $\Pstar$, $\vds$, $\phis$, and $T$:
\begin{linenomath*}
\begin{subequations}
\label{eq:twophase-suspend}
\begin{align}
	\label{eq:twophase-suspend-moment}
	\Grad \Pstar &= \Div \bar{K_v} \Dstar + \rhob \gvec \ , \\
	\label{eq:twophase-suspend-mass}
	\Div \vstar &= - \Div \vds - \Gamma_\rho^{s \ell} \left(\dfrac{1}{\rhos} - \dfrac{1}{\rhol} \right) \ , \\
	\label{eq:twophase-suspend-segr}
	\vds &= - \dfrac{{\phis}^2}{C_v^s} \left(\Grad \Pstar - \rhos \gvec \right) \ , \\
	\label{eq:twophase-suspend-phaseevo}
	\DstarDt{\phis} &= (1-\phis) \Div \vstar + \Div \tilde{K}_\phis \Grad \phi^s + \dfrac{\Gamma_\rho^{s \ell}}{\rhol} \ , \\
	\label{eq:twophase-suspend-tempevo}
	c_p \rhob \DbDt{T} &= \Div \bar{K}_s \Grad T + \Delta L^{s\ell} \Gamma_\rho^{s \ell} + \alpha_0 T \gvec \cdot \left(\rhob \vstar + \rhos \vds\right) + \rhob H_0 \\\nonumber 
	&+ \bar{K}_v \left[\Dstar\right]^2 + \dfrac{C_v^s \left[\vds\right]^2}{{\phis}^2} \ .
\end{align}
\end{subequations}
\end{linenomath*}
We have assumed $\tilde{K}_\phi^\ell \approx 0$ consistent with $\vdl \approx 0$. Hence, volume diffusion in \eqref{eq:twophase-suspend-phaseevo} depends on the solid volume diffusivity, $\tilde{K}_\phis = K_\phis \Drho_0 g_0 d_0$, and gradients in solid fraction only. 

The reduced equations in the suspension limit broadly agree with previous theoretical \citep[e.g.,][]{drew71,monsorno2016two,mucha2004model} and applied models for igneous \citep[e.g.,][]{dufek10} and engineering problems \citep[e.g.,][]{Ho2005}. 

\subsection{Two-phase mush regime}
Igneous systems are thought to occupy the entire spectrum from porous flow at low degrees of partial melting to suspension flows in volcanic conduits and lava lakes. There is increasing evidence that crystal-mush bodies at intermediate melt fractions may make up significant portions of crustal magma processing systems \citep[e.g.,][]{cashman17}. By means of a brief analysis of physical scales in the mechanical governing equations we can assess the boundaries of the reduced porous and suspension flow regimes, and discuss how to treat the mush transition in between.

Scaling analysis of equations \eqref{eq:final-fluidmech} and \eqref{eq:final-fluidmech-total} reveals two pertinent sets of scales: the characteristic speed and pressure of buoyancy-driven flow of the aggregate, $u^*_\mathrm{conv}$, and $p^*_\mathrm{buoy}$, and the phase-wise segregation speeds and compaction pressures, $u^i_\mathrm{segr}$, and $p^i_\mathrm{comp}$ (derivation in Appendix \ref{app:scaling}). Two useful dimensionless numbers are found by taking ratios of the phase segregation by the aggregate convection speed, and of the phase compaction to the aggregate buoyancy pressure:
\begin{linenomath*}
\begin{subequations}
\label{eq:twophase-scale-Rgeneral}
\begin{align}
	\label{eq:twophase-scale-Rsegrgeneral}
	\mathrm{R}^{i}_\mathrm{segr} &= \dfrac{u^i_\mathrm{segr}}{u^*_\mathrm{conv}} = \dfrac{{\phii_0}^2 \bar{K}_{v,0}}{C_{v,0}^i \, \ell_0^2} \ , \\
	\label{eq:twophase-scale-Rcompgeneral}
	\mathrm{R}^i_\mathrm{comp} &= \dfrac{p^i_\mathrm{comp}}{p^*_\mathrm{buoy}} = \dfrac{{\phii_0}^2}{\bar{K}_{v,0} C_{\phi,0}^i} \ ,
\end{align}
\end{subequations}
\end{linenomath*}
where $\ell_0$ is a length scale characteristic of the system (e.g., diapir radius, layer depth, domain height, etc.), $\bar{K}_{v,0}$ is the characteristic effective viscosity of the aggregate, and $C^i_{v,0}$ and $C^i_{\phii,0}$ the scales of the mechanical transfer coefficients, taken as their functional value at the characteristic phase fractions, $\phii_0$.  The segregation and compaction numbers in \eqref{eq:twophase-scale-Rgeneral} express the relative importance of phase segregation and compaction with respect to the collective flow dynamics of the aggregate. If they take values much less than one, mechanical multi-phase effects are expected to remain negligible, effectively reducing the problem to a one-phase flow.

Multiplying the segregation number of one phase with the compaction number of another, we write the segregation-compaction number,
\begin{linenomath*}
\begin{subequations}
\begin{align}
	\label{eq:twophase-scale-Rscgeneral}
	\mathrm{R}^{ik}_\mathrm{sc} &= \mathrm{R}^i_\mathrm{segr} \, \mathrm{R}^k_\mathrm{comp} = \dfrac{{\delta^{ik}_\mathrm{sc}}^2}{\ell_0^2} \ , \\ 
	\label{eq:twophase-scale-deltageneral}
	\delta^{ik}_\mathrm{sc} &= \sqrt{\dfrac{{\phii_0}^2 {\phik_0}^2}{C_{v,0}^i C_{\phi,0}^k}} \ .
\end{align}
\end{subequations}
\end{linenomath*}
In \eqref{eq:twophase-scale-Rscgeneral}, we have interpreted the segregation-compaction number as the squared ratio of an emergent length scale, $\delta^{ik}_\mathrm{sc}$, over the system scale. The former is the inherent physical scale associated with the segregation of one phase accommodated by the compaction of another, a generalisation of the scale known as the compaction length in two-phase porous flow models following \cite{mckenzie84}. For example, for buoyancy-driven liquid segregation through a compacting matrix, $\delta_\mathrm{sc}^{\ell s}$ expresses the characteristic length over which segregation velocity and compaction pressure decay away from heterogeneities such as a permeability barrier. In $n$-phase systems, several such length scales emerge, each characterising the mechanical interactions between two phases.

The segregation-compaction lengths can be used as a diagnostic for assessing the pertinent dynamic regimes in multi-phase flows. If the inherent length scale is much larger than the system scale, segregation rates far exceed, and thus become decoupled from, compaction rates. For example, that limit applies for Darcy flow through a rigid matrix. If the inherent scale is much smaller than the system scale, both segregation and compaction can be considered negligible compared to system-scale dynamics. This applies, for example, to convection in dilute suspensions at low Reynolds number.

In Fig. \ref{fig:2transition}(c), we show $\delta^{\ell s}_\mathrm{sc}$ (liquid segregating through compacting solid, red line) as well as its counterpart, $\delta^{s \ell}_\mathrm{sc}$ (solid segregating through compacting liquid, blue line), truncated an order of magnitude below the granular scale. At low melt fractions, the liquid-in-solid length rapidly increases from the granular scale to order 10 m just above the percolation threshold at $\phil = 0.001$. We interpret this steep initial increase as the onset of the porous flow regime (blue shading, \ref{fig:2transition}). The inherent length scale continues its steep increase until it saturates above order 1 km at $\phil \approx 0.07$. Up to this point, the addition of melt increases the inherent length and lets the speed of segregation grow relative to compaction. Qualitatively, therefore, melt extraction becomes more efficient and melt drainage systems grow more extensive as more melt is added in this regime.

Towards intermediate melt fractions, $\delta_\mathrm{sc}^{\ell s}$ begins to contract again as the disaggregating solid phase is weakening. Addition of melt now leads to rates of solid compaction picking up relative to melt segregation. The contraction of the inherent length implies that localised melt-rich lenses may arise from positive perturbations in melt content. If such melt lenses are buoyant they may drive overturn instabilities within larger mush bodies \citep[e.g.,][]{cashman17,Seropian2018}. We interpret the turning point from growing to contracting segregation-compaction length as the transition from the porous to the mush regime (yellow shading in \ref{fig:2transition}). With further addition of melt, $\delta^{\ell s}_\mathrm{sc}$ drops gradually back to the granular scale and indeed falls to the value of its counterpart, the solid-through-liquid segregation-compaction length, $\delta^{s \ell}_\mathrm{sc}$. In our example, we choose the point where $\delta^{\ell/s}_\mathrm{sc} = 5 d_0$ at $\phil \approx 0.44$ to mark the transition from the mush to the suspension flow regime (red shading in Fig.~\ref{fig:2transition}).

Interestingly, the reference pressure and velocity weights in Fig.~\ref{fig:2transition}(a)--(b) suggest that the reduced porous flow equations \eqref{eq:twophase-porous} remain a reasonable approximation across the mush transition. Our assessment is in line with current thinking that igneous systems evolve from porous melt segregation in solid-dominated drainage systems of up to many kilometres, to potentially extensive, relatively mobile mush bodies at intermediate melt fractions, with interspersed and likely transient liquid-dominated magma bodies. However, the dramatic variations in the inherent length scale---from several kilometres down to the granular scale---over a relatively modest increase of $\sim$20\% in melt fraction implies that the assumed separation of local and system scales may not hold well across this transition, and that a more rigorous multi-scale approach may become necessary \citep[e.g.,][]{gray1989averaging,gray1993mathematical}.

\subsection{Three-phase flows}
Upper crustal magma processing, volcanic activity, and the genesis of mineral resources are, to a large degree, consequences of the exsolution of a third, volatile-rich, liquid to gas phase during decompression, cooling, and crystallisation. The solubility of volatiles, most notably H$_2$O, CO$_2$, and SO$_2$, in igneous melts is a strong function of pressure but also depends on temperature and composition \citep{Papale1999,Papale2006}. As magma rises through the upper crust or cools in shallow magma bodies, volatiles exsolve from the melt to form magmatic volatile phases: supercritical fluids, brines, hydrothermal liquids, and gaseous vapours \citep[e.g.,][]{Driesner2007a,Driesner2007b,Sverjensky2014}. For simplicity, we collectively refer to volatile vapours (superscript $v$) and characterise them as very low-viscosity, low-density, compressible fluids.

The main challenge of building a three-phase model is to prescribe appropriate permission weights to represent local-scale phase topologies of vapour, liquid and solid phases. Ideally, the model would be calibrated to experiments or field observations that shed light on the equilibrium textures between mineral grains, melt films, and vapour bubbles. Here, we demonstrate the procedure by a simple, qualitative example. We use the same solid and liquid phase parameters as for the highlighted curves in Figs \ref{fig:2connect} \& \ref{fig:2coeff} above and add a third phase. We assume that the vapour preferentially wets the liquid rather than the solid (i.e., the dihedral angle between vapour and solid is larger than between liquid and solid). As a consequence, the vapour should have a higher percolation threshold than the liquid in a solid-dominated matrix, and the solid should have a higher disaggregation threshold in a solid-vapour than in a solid-liquid mixture. If sufficient liquid is present, the vapour phase should form disconnected bubbles up to high vapour fractions where a connected vapour phases should emerge. These assumptions form the phenomenological basis for our choice of permission weights.

\begin{figure}[htb]
  \centering
  \includegraphics[width=1.0\textwidth]{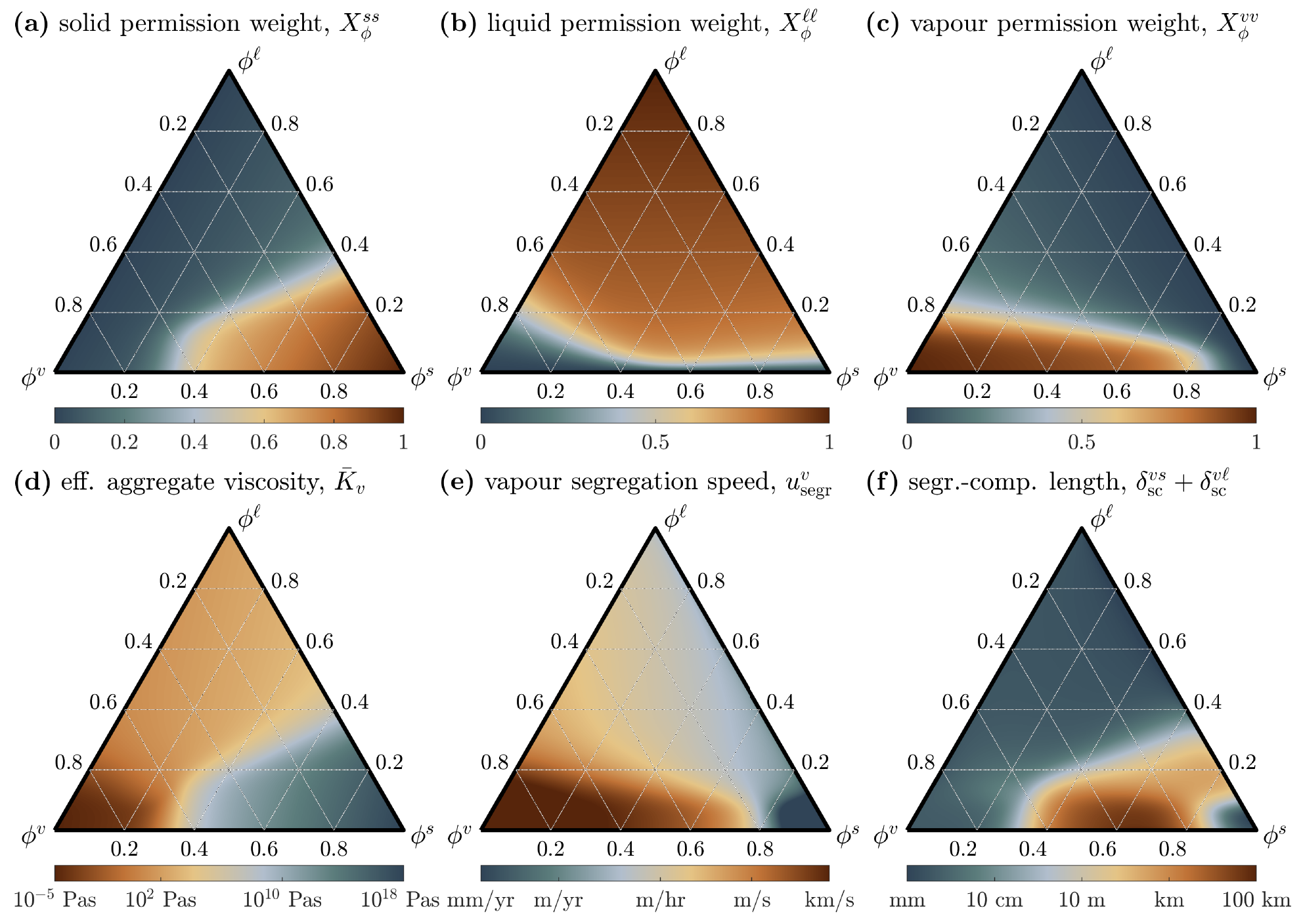}
  \caption{Example igneous three-phase system of solid rock/crystals, liquid melt/droplets, and gaseous vapour/bubbles. Intra-phase permission weights for the solid \textbf{(a)}, liquid \textbf{(b)}, and vapour \textbf{(c)} show regions of connected  (brown) and disconnected (teal) phase topology. The effective aggregate viscosity \textbf{(d)} shows three regimes of convective mobility: solid-dominated immobile (teal), liquid-dominated mobile (amber), and vapour-dominated turbulent (brown). Vapour segregation speed \textbf{(e)} varies from slow migration at mm/yr to turbulent flow at km/s. The length scale for segregating vapour \textbf{(f)} peaks at low liquid and moderate solid fraction, where vapour segregates as a connected phase in a solid-dominated porous matrix.}
  \label{fig:3phase}
\end{figure}

Figure \ref{fig:3phase}(a)--(c) shows the intra-phase permission weights of our example plotted over the three-phase space. To reflect that the vapour is less effectively wetting grain boundaries than the liquid, the solid disaggregation threshold in Fig.~\ref{fig:3phase}(a) is shifted further along the solid-vapour than along the solid-liquid axis. The liquid weight in Fig.~\ref{fig:3phase}(b) prescribes the disconnection of liquid films between solid grains at very low liquid fractions, while melt films between bubbles are set to collapse around $\phi^v \approx 0.8$. The vapour permission weight in Fig.~\ref{fig:3phase}(c) shows a percolation threshold for a solid-dominated matrix at $\sim$10\% vesicularity. The transition from disconnected bubbles to connected vapour films is fixed at $\phi^v \approx 0.8$ along the liquid-vapour axis, congruous with the choice of liquid disconnection around that vapour fraction. The fitting parameters for this example are listed in the Appendix Table \ref{tab:fitting3}, and the full set of permission weights are shown in Appendix Fig. \ref{fig:3connect}.

Based on these permission weights, we calculate mechanical flux and transfer coefficients for the three-phase example. We choose granular size, and solid and liquid viscosities as above, and assume a vapour viscosity of 10$^{-5}$ Pas. The effective aggregate viscosity in Fig. \ref{fig:3phase}(d) shows a region of high strength where the solid remains contiguous (teal); a second region of lower strength (amber), and a small third regime of lowest strength (brown) mark where the melt and vapour assume the role of carrier phase. All effective phase viscosities, phase segregation, and phase compaction coefficients for the three-phase example are given in Appendix Fig.~\ref{fig:3coeff}.

The characteristic buoyancy-driven vapour segregation speed, $u^v_\mathrm{segr}$, shown in Fig. \ref{fig:3phase}(e) is calculated for $\rhos_0$ = 3000 kg/m$^3$, $\rhol_0$ = 2500 kg/m$^3$, $\rhov_0$ = 200 kg/m$^3$. It shows a region of low mobility ($<$ m/yr) near the solid limit where vapour is locked in isolated vesicles and porous flow is limited. Much of the phase space at moderate to high melt fractions is characterised by an intermediate vapour segregation speed (m/yr--m/hr), where vapour bubbles segregate by displacing melt-dominated magma around them. At moderate to low melt fractions ($< 0.2$) and vapour fractions above the percolation threshold, our choice of permission weights prescribes that the vapour phase forms a connected topology in a solid-dominated matrix. As a consequence, the vapour segregation speed grows significantly to order m/s. Such a transition from bubbly to porous vapour segregation as a function of both the vesicularity and crystallinity is indicated by analogue experiments \citep[e.g.,][]{Colombier2018}. If further corroborated, such a regime transition could prove to be important in the degassing of subvolcanic magma reservoirs. 

The inherent length scale of vapour segregation through the compacting magma (Fig. \ref{fig:3phase}(f)) varies from the granular scale to order 100 km. The maximum values are found where connected vapour films form in a contiguous, solid-dominated matrix. In such systems, efficient degassing may occur over long distances without significant matrix deformation. Outside that region, our example prescribes vapour segregation by bubble migration through a crystal-rich mush to crystal-poor magma. The low segregation-compaction length in the liquid-dominated region of phase space suggests that segregation by bubble migration is slow compared to magma creep. In Appendix \ref{app:threephase}, we show reduced equations for these two limits of three-phase porous flow of connected melt and vapour phases through a contiguous rocky matrix and of three-phase suspension flow of disaggregated crystals and bubbles in a carrier melt.  All reference velocity and pressure weights along with pertinent segregation-compaction lengths are given in Appendix Fig.~\ref{fig:3transition}. 

\begin{figure}[htb]
  \centering
  \includegraphics[width=0.5\textwidth]{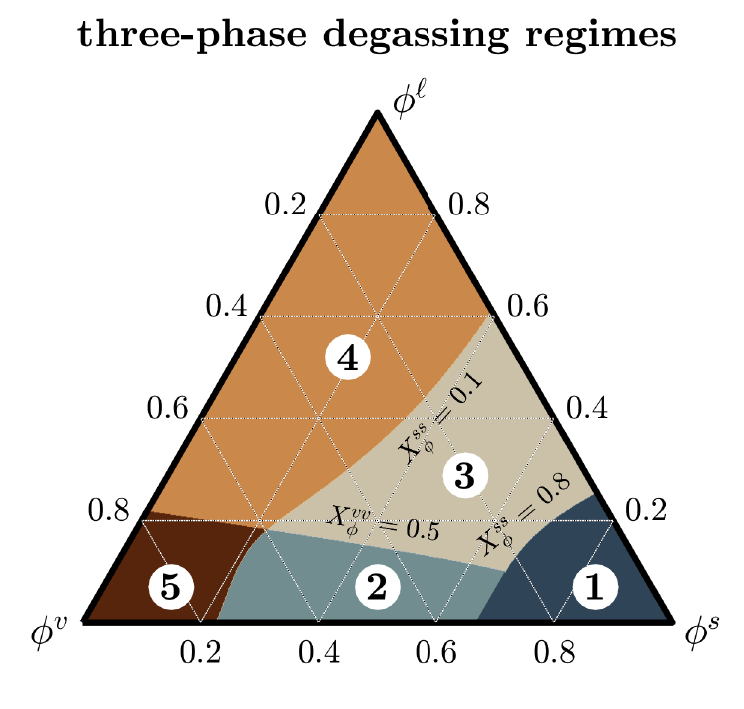}
  \caption{Tentative regime diagram for igneous three-phase dynamics of magma degassing: (1) porous flow; (2) granular flow; (3) bubbly mush flow; (4) bubbly suspension flow; (5) ash flow. Regime boundaries are drawn based on solid and vapour permission weights.}
  \label{fig:3regimes}
\end{figure}

Figure~\ref{fig:3regimes} shows a tentative regime diagram marking the regions of three-phase space where our example calibration predicts different regimes of magma degassing. A first regime (1), sees the onset of degassing by porous flow in a solid-dominated matrix. This regime is marked by a strong increase in segregation-compaction length, meaning that an addition of vapour increases both the segregation speed as well as the length scale of pressure-communicating drainage systems. In the second regime (2), the porous segregation speed through a disaggregating, melt-poor granular medium increases continuously. Whereas degassing is effective in this regime, the matrix becomes more deformable as the vapour fraction increases, until it may fragment, fluidise, and/or become entrained in the fast-flowing vapour. The third regime (3) is one of bubbly flow in a moderately crystalline to crystal-rich mush. In this regime, degassing rates become sluggish and the compaction length drops down to the granular scale as the mush becomes more mobile. The fourth regime (4) is characterised by bubbly suspension flow in crystal-poor magma, where the segregation-compaction length remains near the granular scale, and convective mobility may well exceed segregation mobility of vapour bubbles. The final regime (5) is the gas-supported suspension or ash cloud regime, where turbulence will likely become important and the present model therefore no longer applies.

\section{Summary and conclusions \label{sec:conclusions}}
We have derived a continuum-scale mixture theory framework for multi-phase reactive transport motivated by the need to better understand igneous processes across the regimes of porous, mushy, and suspension flows. Our derivation draws on concepts from Rational Thermodynamics and procedures from Non-equilibrium Thermodynamics. The approach entails that we forgo a first principles-based connection between local-scale phase interactions and system-scale reactive transport. Instead, we base our model on continuum-scale conservation equations for interpenetrating phase continuum fields and assume that energies and entropies are meaningfully defined for the mixture of phases at the same scale.

We express system-scale mechanical and thermodynamic transport processes in a set of linear, decoupled, isotropic constitutive relations. The constitutive relations define fluxes and transfers as proportional to their conjugate gradients and phase deviations. The proportionality is set by material response coefficients, for which we propose a set of internally consistent but non-unique phenomenological closures, which require problem-specific calibration, ideally against field observations or laboratory experiments. The proposed closures are based on the assumption that fluxes and transfers are facilitated by the diffusion of heat, chemical species, momentum, and partial volume within and between phases at the local scale.

Applying our model to a specific natural or engineering context requires specification of pure-phase diffusive transport properties as well as a set of permission weights. The latter give a phenomenological metric of how the local-scale phase topology is connected within each phase and with adjacent phases. Describing effects of phase connectivity is particularly important in the context of vapour-bearing three-phase flows where it allows to distinguish between bubbly and thin-film flow regimes at the same relative vapour fraction. Ratios of pure-phase properties assembled according to the permission weights define permission functions that encapsulate the system-scale effects of how local-scale material properties and phase topologies permit reactive transport.

Models of igneous processes are inevitably underdetermined by observations and experiments \citep{Oreskes1994}. Available data constraints typically comprise indirect observations of subsurface processes, incomplete geological and historical evidence, and laboratory experiments of limited scope and applicability. As a consequence, the model framework developed here is not intended for constructing representations of reality but for formulating testable hypotheses that can be falsified in the field or laboratory. 

Observational and experimental constraints in fact take a dual role with respect to building and validating models in our framework. On the one hand, laboratory or field data are necessary for calibrating problem-specific model closures. The data best suited for this purpose is measured at the granular scale (e.g., the dihedral angle between basalt and olivine). Such constraints inform the formulation of auxiliary hypotheses regarding what local-scale phase interactions are relevant for system-scale behavior and should therefore be included in the calibrated continuum model \citep{hempel1948studies, hempel1965aspects}.

On the other hand, the calibrated system-scale model itself represents a hypothesis regarding how the assumed local-scale phase interactions are reflected in system-scale observables. Falsifying this central model hypothesis requires data collected at an analogue system scale in the laboratory (e.g., average velocity of particles settling out of suspension), or ideally at the natural system scale in the field. Field observations may include, for example, data obtained by geochemical sampling, geological mapping, geophysical imaging, remote sensing, or volcano monitoring. One subtle yet important intricacy is that the consistency between model and observables can only be assessed contingent upon the auxiliary hypotheses made in calibrating the model, implying that the calibration cannot be validated independently from the model itself \citep{hempel1948studies, hempel1965aspects,Oreskes1994}.

Finally, we have used our framework to construct reduced equations for special igneous two- and three-phase limits and discussed ramifications for modelling and understanding the various stages of igneous processes from source to surface. With the appropriately calibrated phenomenological closures, the general governing equations reduce to the canonical limits of two-phase porous and suspension flow. We find that it is possible to fit our mechanical flux and transfer coefficients to previous parameterizations including a Kozeny-Carman permeability, a hindered Stokes settling coefficient, and various empirical and theoretical relations for effective shear and compaction viscosities and volume diffusivity, by calibrating a single set of permission weights. Whereas we have focused on viscous materials here, our model framework lends itself to extension into the visco-elastic and brittle-plastic deformational domains. The framework therefore serves as a generalised starting point for the formulation of process-based, hypothesis-driven models of igneous systems from source to surface.

\subsection*{Author Contributions}
TK conceived the study and took the lead in both model formulation and writing. JS provided theoretical background on mixture models and their connection to local-scale dynamics, and contributed to model formulation and writing.

\subsection*{Acknowledgments}
The authors owe much gratitude to David Bercovici for his time and invaluable insights that were instrumental for initiating this work. We further thank Be\~nat Oliveira for his thorough and constructive review, and Gary Mavko, John Rudge, David Rees-Jones, and Gregor Golabek for their insightful remarks that have helped improve the study. TK acknowledges support from the Swiss National Science Foundation Postdoc.Mobility Fellowship 177816, and JS from the US Army Research Office grants W911NF-18-1-0092, and W911NF-12-R-0012-04 (ECASE award). The authors declare no conflicts of interest related to this work. 

\bibliographystyle{gji}
\bibliography{references} 

\cleardoublepage
\appendix

\renewcommand{\theequation}{\Alph{section}\arabic{equation}}
\renewcommand{\thefigure}{\Alph{section}\arabic{figure}}
\renewcommand{\thetable}{\Alph{section}\arabic{table}}

\setcounter{equation}{0}
\setcounter{figure}{0}
\setcounter{table}{0}

\section*{Appendix A: Axiomatic principles for constitutive choices \label{app:axioms}}
\setcounter{section}{1}

Rational Thermodynamics \citep{truesdell84} invokes axiomatic principles that express \textit{a priori} requirements functions must satisfy to be admitted as constitutive relation for thermodynamic processes. First, the \emph{entropy principle} requires that constitutive relations must not violate the non-negativity of entropy production. Second, the principle of \emph{local action} states that all thermodynamic processes respond to thermodynamic forces acting within the local control volume or on its surfaces, requiring constitutive relations to be functions of local states and spatial gradients of independent variables. And third, the principle of \emph{frame invariance} specifies that constitutive relations must be Galilean frame-invariant, that is, independent of the inertial reference frame.

The fourth principle of \emph{separation of phases} understands material phases in the mixture to be discrete bodies separated by well-defined interfaces at the local scale. It therefore stipulates that constitutive relations for processes internal to a phase must be phase-wise functions chosen as formally separate from the mixture. \cite{passman84} states that therefore constitutive relations for fluxes must be functions of independent variables and material properties pertaining to the respective phase only, whereas transfers can be function of all independent variables. We argue that this interpretation cannot hold in general for two reasons. First, to satisfy conservation principles, some fluxes must obey a zero sum constraint and thus be a function of thermodynamic forces and transport properties in all phases. Second, fluxes passing through disconnected constituents of a disaggregated phase must be functions of transport properties in the surrounding carrier phase, which sets the local boundary conditions along shared interfaces. We therefore interpret the separation of phases to the effect that constitutive relations for all processes are chosen as phase-wise separate functions of independent variables and material properties of all phases in the mixture.

\section*{Appendix B: Assembling the temperature equation \label{app:final}}
\setcounter{section}{2}

It is expedient to express energy conservation in terms of the evolution of phase temperature, $\Ti$. We begin assembling the final energy equation from the basic conservation of phase energy \eqref{eq:consv-lagr-energy}, into which we substitute \eqref{eq:ineq-total-energy},
\begin{linenomath*}
\begin{align}
	\label{eq:energy-internal}
	\rhoi \DiDt{\ui} = - \vi \cdot \rhoi \DiDt{\vi} - \Div \qei - \Gei + \ei \Gmi - \Qei \ .
\end{align}
\end{linenomath*}
Whereas in \eqref{eq:ineq-internal-energy}, internal energy is given as a function of specific entropy, volume, and chemical potentials we now write it in terms of temperature, pressure and component concentrations \citep[e.g.,][]{rudge11},
\begin{linenomath*}
\begin{align}
	\label{eq:energy-internal-variables}
	\DiDt{\ui} = c^i_p \DiDt{\Ti} - \dfrac{\alpha^i \Ti}{\rhoi} \DiDt{\Pi} + \dfrac{\Pi}{{\rhoi}^2} \DiDt{\rhoi} + \sum_j \mji \DiDt{\cji} \ ,
\end{align}
\end{linenomath*}
with $\alpha^i$ the thermal expansivity, and $c_\Pi$ the heat capacity of the phase. We substitute \eqref{eq:energy-internal-variables} into \eqref{eq:energy-internal} to write the evolution of sensible heat coupled to that of momentum, phase mass and component mass,
\begin{linenomath*}
\begin{align}
	\label{eq:energy-expanded}
	\rhoi c^i_p \DiDt{\Ti} = &- \vi \cdot \rhoi \DiDt{\vi} - \dfrac{\Pi}{\rhoi} \DiDt{\rhoi} - \sum_j \mji \rhoi \DiDt{\cji} \\\nonumber
	&- \Div \qei - \Gei + \ei \Gmi - \Qei + \alpha^i \Ti \DiDt{\Pi} \ .
\end{align}
\end{linenomath*}
Next, we substitute conservation of momentum \eqref{eq:consv-lagr-moment}, component mass \eqref{eq:consv-lagr-comp}, and phase mass \eqref{eq:consv-lagr-mass} to obtain,
\begin{linenomath*}
\begin{align}
	\label{eq:energy-expanded-mass}
	\rhoi c^i_p \DiDt{\Ti} = &- \Div \qei + \vi \cdot \Div \qvi + \sum_j \mji \Div \qji + \Pi \Div \qfi \\\nonumber
	&- \Gei + \vi \cdot \Gvi + \sum_j \mji \Gji + \Pi \Gfi + \Ti \si \Gmi \\\nonumber
	&- \Qei + \vi \cdot \Qvi + \alpha^i \Ti \DiDt{\Pi} \ ,
\end{align} 
\end{linenomath*}
where we have once more used $u^i = \Ti\si - \Pi / \rhoi + \sum_j \mji \cji$ and dropped a term $\frac{1}{2} {\vi}^2 \Gmi$. We can now substitute the constitutive relations for fluxes, \eqref{eq:flx-const-rel-energy} and \eqref{eq:flx-const-rel}, transfers, \eqref{eq:trf-rate-energy} and \eqref{eq:trf-rate}, and sources \eqref{eq:src-const-rel}, 
\begin{linenomath*}
\begin{align}
	\label{eq:energy-expanded-trfall}
	\rhoi c^i_p \DiDt{\Ti} = &\Div \Ksi \Grad \Ti + \Kvi \left[ \Di \right]^2 + \sum_j \Kji \, \dfrac{\left[ \DGmstari \right]^2}{R \Ti} + \Kfi \left[ \DGPstar \right]^2 \\\nonumber
	&- \Csi \DTi + \Cvi {\Dvi}^2 + \sum_j \Cji \dfrac{{\DmjiP}^2}{R \Tstar} + \Cfi {\DPi}^2 + \left(\Ti \si + \dfrac{\Pi}{\rhostar}\right) \Gmi \\\nonumber
	&+ \alpha^i \Ti \DiDt{\Pi} + \phii \rhoi H^i \ .
\end{align}
\end{linenomath*}
To arrive here we have canceled out two terms $\pm \Pi \vi \cdot \Grad \phii$ and dropped a nonlinear disequilibrium term, $\DPi \Dvi \cdot \Grad \phii$. We further simplify \eqref{eq:energy-expanded-trfall} by assuming dissipation terms of component fluxes and transfers remain negligible, and by introducing the latent heat of phase change, $L^i = \Ti \si + \Pi/\rhostar$, and thus obtain the final form,
\begin{linenomath*}
\begin{align}
	\label{eq:final-energy-simple}
	\rhoi c^i_p \DiDt{\Ti} = &\Div \Ksi \Grad \Ti + \Kvi \left[ \Di \right]^2 + \Kfi \left[ \DGPstar \right]^2 \\\nonumber 
	&- \Csi \DTi + \Cvi {\Dvi}^2 + \Cfi {\DPi}^2 + L^i \Gmi \\\nonumber
	&+ \alpha^i \Ti \DiDt{\Pi} + \phii \rhoi H^i \ .
\end{align}
\end{linenomath*}
The final tally of processes contributing to the evolution of sensible heat are (from left to right), thermal diffusion, dissipation of fluxes and transfers of momentum and volume, latent heat of phase-change reactions, and adiabatic and radiogenic heating.

\section*{Appendix C: Scales of mechanical governing equations \label{app:scaling}}
\setcounter{section}{3}
\setcounter{equation}{0}

We write a set of mechanical governing equations for $n$ incompressible, unreactive phases assuming that mechanical transfer coefficients have relative weights, $\ovi$ and $\ofi$, such that one phase ($i = 1$) dominates the reference velocity, whereas another phase ($i = n$) dominates the reference pressure. This condition holds if the former is the most competent ($K^1_v \gg K_v^{i\neq 1}$) and the latter the most compliant phase ($K^n_v \ll K_v^{i\neq n}$). For that case, it is expedient to write one equation each for $\vstar$ and $\Pstar$, and $n-1$ equations each for $\vdi$ ($i \in [2,n]$) and $\Pdi$ ($i \in [1,n-1]$):
\begin{linenomath*}
\begin{subequations}
\label{eq:analysis-dim-governing}
\begin{align}
	\label{eq:analysis-dim-mixmoment}
	\Grad p^* &= - \sum_{i=1}^{n-1} \Grad \Pdi + \Div \bar{K}_v \Dstar + \Delta \bar{\rho} \gvec \ , \\
	\label{eq:analysis-dim-mixmass}
	\Div \vstar &= - \sum_{i=2}^{n} \Div \vdi \ , \\
	\label{eq:analysis-dim-phasemoment}
	\vdi &= - \dfrac{{\phii}^2}{\Cvi} \left(\Grad p^* + \dfrac{1}{\phii} \Grad \Pdi - \Delta \rho^{i*} \gvec \right) \ , \ \ \mathrm{for} \ \ i \in [2,n] \ , \\
	\label{eq:analysis-dim-phasemass}
	\Pdi &= - \dfrac{{\phii}^2}{\Cfi} \left(\Div \vstar + \dfrac{1}{\phii} \Div \vdi \right) \ , \ \ \mathrm{for} \ \ i \in [1,n-1] \ .
\end{align}
\end{subequations}
\end{linenomath*}
We have opted to reduce the reference pressure gradient to its dynamic part by subtracting a static reference pressure gradient, $\Grad p^* = \Grad \Pstar - \rho^* \gvec$. Accordingly, we have introduced the reference density difference, $\Delta \bar{\rho} = \bar{\rho} - \rho^* = \sum_{i=2}^{n} \phii \Delta \rho^{i*}$. We have dropped volume diffusion terms to further simplify the following scaling analysis. 

To non-dimensionalise the problem, we introduce physical scales ($[\ai] = a_0$ denotes that $a_0$ is the characteristic physical scale of $\ai$),
\begin{linenomath*}
\begin{align}
	\label{eq:analysis-scales}
	[\vstar] = u^*_0 \ ; \ \ \ [\vdi] = u^i_0 \ ; \ \ \ [p^*] &= p^*_0 \ ; \ \ \ [\Pdi] = p^i_0 \ ; \ \ \ [\Grad,\mathbf{\underline{D}}] = 1/\ell_0 \ ; \\
	[\phii] = \phii_0 \ ; \ \ \ [\Delta \bar{\rho}, \Delta \rho^{i*}] &= \Delta \rho_0 \ ; \ \ \ [\gvec] = g_0 \ ; \\
	[\bar{K}_v] = \bar{K}_{v,0} = \bar{K}_v(\phii_0) \ ; \ \ \ [\Cvi] &= C^i_{v,0} = \Cvi(\phii_0) \ ; \ \ \ [\Cfi] = C_{\phi,0}^i = \Cfi(\phii_0) \ .
\end{align}
\end{linenomath*}
We scale reference and segregation velocities and reference and compaction pressures with separate scales to reflect that they represent distinct physical processes. The scales of coefficients are taken as their functional value at the characteristic phase fractions.

Substituting variables and parameters in \eqref{eq:analysis-dim-governing} by their scales \eqref{eq:analysis-scales} multiplied by dimensionless variables ($\ai = a_0 {\ai}'$), we obtain,
\begin{linenomath*}
\begin{subequations}
\label{eq:analysis-nondim-governing}
\begin{align}
	\label{eq:analysis-nondim-mixmoment}
	[1] \left(\Grad {p^*} \right) &= - \sum_{i=1}^{n-1} \left[ \dfrac{p^i_0}{p^*_0} \right] \left( \Grad {\Pdi} \right) + \left[ \dfrac{u^*_0 \bar{K}_{v,0}}{p^*_0 \ell_0} \right] \left(\Div \bar{K}_v \Dstar\right) + \left[\dfrac{\Delta \rho_0 g_0 \ell_0}{p^*_0}\right] \left(\Delta \bar{\rho} \hat{\mathbf{z}} \right) \ , \\
	\label{eq:analysis-nondim-mixmass}
	[1] \left( \Div \vstar \right) &= - \sum_{i=2}^{n} \left[ \dfrac{u^i_0}{u^*_0} \right] \left( \Div \vdi \right) \ , \\
	\label{eq:analysis-nondim-phasemoment}
	[1] (\vdi) &= - \left[ \dfrac{{\phii_0}^2 p^*_0}{u^i_0 C^i_{v,0} \ell_0}\right] \dfrac{{\phii}^2}{\Cvi} \left(\Grad p^* + \left[\dfrac{p^i_0}{\phii_0 p^*_0} \right] \dfrac{1}{\phii} \Grad \Pdi - \left[\dfrac{\Delta \rho_0 g_0 \ell_0}{p^*_0} \right] \Delta \rho^{i*} \hat{\mathbf{z}} \right) \ , \ \ \mathrm{for} \ \ i \in [2,n] \ , \\
	\label{eq:analysis-nondim-phasemass}
	[1] \left(\Pdi\right) &= - \left[ \dfrac{{\phii_0}^2 u^*_0}{p^i_0 C^i_{\phi,0} \ell_0} \right]\dfrac{{\phii}^2}{\Cfi} \left(\Div \vstar + \left[\dfrac{u^i_0}{\phii_0 u^*_0} \right] \dfrac{1}{\phii} \Div \vdi \right) \ , \ \ \mathrm{for} \ \ i \in [1,n-1] \ .
\end{align}
\end{subequations}
\end{linenomath*}
For ease of reading we have divided each equation through by their left hand side scale, have grouped dimensional scales in square brackets, and dropped all primes. The scale of $[1]$ remaining on the left hand side indicates that the equations are now dimensionless.

Inspecting the resulting (dimensionless) groups of scales, we can identify various speed and pressure scales emerging from the problem. The scales of the buoyancy terms in \eqref{eq:analysis-nondim-mixmoment} and \eqref{eq:analysis-nondim-phasemoment} can be reduced to unity by choosing a buoyancy pressure scale for $p^*$,
\begin{linenomath*}
\begin{align}
	\label{eq:analysis-pres-scale}	
	p^*_\mathrm{buoy} \equiv p^*_0 = \Delta \rho_0 g_0 \ell_0 \ .
\end{align}
\end{linenomath*}
Using $p^*$, the scale of the viscous stress term in \eqref{eq:analysis-nondim-phasemoment} can be canceled by choosing the buoyancy-driven convective speed as scale for $\vstar$,
\begin{linenomath*}
\begin{align}
	\label{eq:analysis-vel-scale}
	u^*_\mathrm{conv} \equiv u^*_0 = \dfrac{\Delta \rho_0 g_0 \ell^2_0}{\bar{K}_{v,0}} \ .
\end{align}
\end{linenomath*}
These first two scales are the same as for a one-phase Stokes flow driven by a buoyancy force $\sim \Delta \rho_0 g_0$ and resisted by an effective viscosity $\sim \bar{K}_{v,0}$.

We can further eliminate the leading scales on the right hand sides of \eqref{eq:analysis-nondim-phasemoment} and \eqref{eq:analysis-nondim-phasemass} by choosing scales for phase-wise segregation speeds and compaction pressures, respectively, of the form:
\begin{linenomath*}
\begin{subequations}
\label{eq:analysis-segr-comp-scales}
\begin{align}
	u^i_\mathrm{segr} &\equiv u^i_0 = \dfrac{{\phii_0}^2 \Delta \rho_0 g_0}{C^i_{v,0}} \ , \\
	p^i_\mathrm{comp} &\equiv p^i_0 = \dfrac{{\phii_0}^2 \Delta \rho_0 g_0 \ell_0}{C^i_{\phi,0} \bar{K}_{v,0}} \ .
\end{align}
\end{subequations}
\end{linenomath*}

Substituting $u^i_\mathrm{segr}$ and $p^i_\mathrm{comp}$ into the scales multiplying the compaction pressure gradient and segregation velocity divergence in \eqref{eq:analysis-nondim-mixmoment} and \eqref{eq:analysis-nondim-mixmass} we obtain two dimensionless numbers,
\begin{linenomath*}
\begin{subequations}
\label{eq:analysis-segr-comp-scales}
\begin{align}
	\mathrm{R}^i_\mathrm{segr} &\equiv \dfrac{u^i_\mathrm{segr}}{u^*_\mathrm{conv}} = \dfrac{{\phii_0}^2 \bar{K}_{v,0}}{C^i_{v,0} \ell^2_0} \ , \\
	\mathrm{R}^i_\mathrm{comp} &\equiv \dfrac{p^i_\mathrm{comp}}{p^*_\mathrm{buoy}} = \dfrac{{\phii_0}^2}{C^i_{\phi,0} \bar{K}_{v,0}} \ .
\end{align}
\end{subequations}
\end{linenomath*}

Multiplying the segregation and compaction numbers between phase pairs we define a segregation-compaction number, $R^{ik}_\mathrm{sc}$, which can be interpreted as the square ratio of an emergent length scale, $\delta^{ik}_\mathrm{sc}$, to the system scale, $\ell_0$:
 \begin{linenomath*}
\begin{subequations}
\label{eq:analysis-segr-comp-scales}
\begin{align}
	\mathrm{R}^{ik}_\mathrm{sc} &\equiv \mathrm{R}^i_\mathrm{segr} \, \mathrm{R}^k_\mathrm{comp} = \dfrac{{\delta^{ik}_\mathrm{sc}}^2}{\ell^2_0}\ , \\
	\delta^{ik}_\mathrm{sc} &= \sqrt{\dfrac{{\phii}^2 {\phik}^2}{C^i_{v,0} C_{\phi,0}^k}} \ .
\end{align}
\end{subequations}
\end{linenomath*}
This emergent segregation-compaction length scale generalises on the compaction length arising in two-phase porous flow models \citep[e.g.,][]{mckenzie84}. 

With these scales and dimensionless numbers, we rewrite the non-dimensional governing equations as,
\begin{linenomath*}
\begin{subequations}
\label{eq:analysis-nondim-governing}
\begin{align}
	\label{eq:analysis-nondim-mixmoment}
	\Grad {p^*} &= - \sum_{i=1}^{n-1} \mathrm{R}^i_\mathrm{comp} \Grad {\Pdi} + \Div \bar{K}_v \Dstar + \Delta \bar{\rho} \hat{\mathbf{z}} \ , \\
	\label{eq:analysis-nondim-mixmass}
	\Div \vstar &= - \sum_{i=2}^{n} \mathrm{R}^i_\mathrm{segr} \Div \vdi \ , \\
	\label{eq:analysis-nondim-phasemoment}
	\vdi &= - \dfrac{{\phii}^2}{\Cvi} \left(\Grad p^* + \dfrac{\mathrm{R}^i_\mathrm{comp}}{\phii_0 \phii} \Grad \Pdi - \Delta \rho^{i*} \hat{\mathbf{z}} \right) \ , \ \ \mathrm{for} \ \ i \in [2,n] \ , \\
	\label{eq:analysis-nondim-phasemass}
	\Pdi &= - \dfrac{{\phii}^2}{\Cfi} \left(\Div \vstar + \dfrac{\mathrm{R}^i_\mathrm{segr}}{\phii_0 \phii} \Div \vdi \right) \ , \ \ \mathrm{for} \ \ i \in [1,n-1] \ .
\end{align}
\end{subequations}
\end{linenomath*}

\section*{Appendix D: Three-phase porous and suspension limits \label{app:threephase}}
\setcounter{section}{4}
\setcounter{equation}{0}


In analogy to the two-phase porous and suspension flow limits, we write reduced equations for the special limits of three-phase porous flow in solid-dominated, and suspension flow in liquid-dominated igneous aggregates. Although equations of state for volatile phases can be complex \citep[e.g.,][]{Sverjensky2014}, we choose a linearised pressure-dependent vapour density for simplicity here: 
\begin{linenomath*}
\label{eq:vapour-eos}
\begin{equation}
\rho^v = \rho^v_0 (1 + \beta^v P^v) \ ,
\end{equation} 
\end{linenomath*}
with $\beta^v$ [1/Pa] the isothermal vapour compressibility.

\subsection{Three-phase porous flow limit}
We define a limit of three-phase porous flow characterised by connected liquid and gas phases in a contiguous solid matrix. In this limit, the solid velocity assumes the role of reference velocity, and the liquid and vapour pressures sustaining only negligible pressure differences between them together take the role as reference pressure. Hence, the mechanical governing equations can be reduced to,
\begin{linenomath*}
\begin{subequations}
\label{eq:porous3}
\begin{align}
	\label{eq:porous3-mixmoment}
	\Grad \Pstar &= - \Grad \Pds + \Div  \bar{K}_v \Dstar + \bar{\rho} \gvec \ , \\
	\label{eq:porous3-mixmass}
	\Div \vstar &= - \Div \vdl - \Div \vdv \\\nonumber 
	&- \Gamma_\rho^{s \ell} \left( \dfrac{1}{\rhos} - \dfrac{1}{\rhol} \right) - \Gamma_\rho^{\ell v} \left( \dfrac{1}{\rhol} - \dfrac{1}{\rhov} \right) - \beta^v \bar{\rho} (\phiv \vstar + \vdv) \cdot \gvec \ \\
	\label{eq:porous3-liquidsegr}
	\vdl &= -\dfrac{{\phil}^2}{C_v^\ell} \left(\Grad \Pstar - \rhol \gvec \right) \ , \\
	\label{eq:porous3-gassegr}
	\vdv &= -\dfrac{{\phiv}^2}{C_v^v} \left(\Grad \Pstar - \rhov \gvec \right) \ , \\
	\label{eq:porous3-solidcompact}
	\Pds &= -\dfrac{{\phis}^2}{C_\phis} \left( \Div \vstar + \dfrac{\Gamma_\rho^{s \ell}}{\phis} \left(\dfrac{1}{\rhos}-\dfrac{1}{\rhol}\right) \right) \ , \\
	\label{eq:porous3-meltevo}
	\DstarDt{\phil}  &= - \phil \Div \vstar - \Div \vdl + \dfrac{\Gamma_\rho^{s \ell} - \Gamma_\rho^{\ell v}}{\rhol} \ , \\
	\label{eq:porous3-gasevo}
	\DstarDt{\phiv}  &= - \phiv \Div \vstar - \Div \vdv + \dfrac{\Gamma_\rho^{\ell v}}{\rhov} - \beta^v \bar{\rho} (\phiv \vstar + \vdv) \cdot \gvec \ .
\end{align} 
\end{subequations}
\end{linenomath*}
For the compressibility terms, we have assumed that pressure changes are dominated by flow against gravity.

\subsection{Three-phase suspension flow limit}
We define a limit of three-phase suspension flow characterised by disconnected solid particles and vapour bubbles suspended in a connected carrier melt. Consistent with our choice in the two-phase suspension limit above, we choose the liquid velocity as reference and consider compaction pressures negligible. Hence, the mechanical governing equations can be reduced to,
\begin{linenomath*}
\begin{subequations}
\label{eq:suspend3}
\begin{align}
	\label{eq:suspend3-mixmoment}
	\Grad \Pstar &=  \Div \bar{K}_v \Dstar - \bar{\rho} \gvec \ , \\
	\label{eq:suspend3-mixmass}
	\Div \vstar &= - \Div \vds - \Div \vdv \\\nonumber
	&-\Gamma_\rho^{s \ell} \left( \dfrac{1}{\rhos} - \dfrac{1}{\rhol} \right) - \Gamma_\rho^{\ell v} \left( \dfrac{1}{\rhol} - \dfrac{1}{\rhov} \right) - \beta^v \bar{\rho} (\phiv \vstar + \vdv) \cdot \gvec \ , \\
	\label{eq:suspend3-solidsegr}
	\vds &= -\dfrac{{\phis}^2}{C_v^s} \left(\Grad \Pstar - \rhos \gvec \right) \ , \\
	\label{eq:suspend3-solidsegr}
	\vdv &= -\dfrac{{\phiv}^2}{C_v^v} \left(\Grad \Pstar - \rhov \gvec \right) \ , \\
	\label{eq:suspend3-solidevo}
	\DstarDt{\phis}  &= \Div \tilde{K}_\phis \Grad \phis - \phis \Div \vstar - \Div \vds - \dfrac{\Gamma_\rho^{s \ell}}{\rhos} \ , \\
	\label{eq:suspend3-solidevo}
	\DstarDt{\phiv}  &= \Div \tilde{K}_\phiv \Grad \phiv - \phiv \Div \vstar - \Div \vdv + \dfrac{\Gamma_\rho^{\ell v}}{\rhov} - \beta^v \bar{\rho} (\phiv \vstar + \vdv) \cdot \gvec \ .
\end{align} 
\end{subequations}
\end{linenomath*}

\setcounter{section}{1}
\setcounter{table}{0}
\begin{table}[ppp]
\centering
	\caption{\textbf{Material \& fitting parameters for two-phase examples}}
	\label{tab:fitting2}
	\begin{tabular}{llll}
	\textbf{Parameter} & \textbf{Symbol} & \textbf{Units} & \textbf{Value} \B \\ \hline
	 \multicolumn{4}{l}{\textbf{Phase materials}} \T \\
	 Solid viscosity & $\etas_0$ & Pas & 10$^{18}$ \\
	 Liquid viscosity & $\etal_0$ & Pas & 10$^2$ \\
	 Granular length scale & $d_0$ & m & 0.003 \B \\ \hline
	  \multicolumn{4}{l}{\textbf{Example permission weights}} \T \\
	 Solid slopes & $A^{s k}$ & - & [0.60, 0.25] \\
	 Liquid slopes & $A^{\ell k}$ & - & [0.20, 0.20] \\
	 Solid thresholds & $B^{s k}$ & - & [0.3, 0.7] \\
	 Liquid thresholds & $B^{\ell k}$ & - & [0.98, 0.02] \\
	 Solid weights & $C^{s k}$ & - & [0.2, 0.2] \\
	 Liquid weights & $C^{\ell k}$ & - & [0.6, 0.6] \B \\ \hline
	 \multicolumn{4}{l}{\textbf{Permission weights calibrated for basalt \& olivine}} \T \\
	 Solid slopes & $A^{s k}$ & - & [0.6945, 0.1832] \\
	 Liquid slopes & $A^{\ell k}$ & - & [0.5360, 0.1834] \\
	 Solid thresholds & $B^{s k}$ & - & [0.6906, 0.3094] \\
	 Liquid thresholds & $B^{\ell k}$ & - & [0.9993, 0.0007] \\
	 Solid weights & $C^{s k}$ & - & [0.6889, 0.1750] \\
	 Liquid weights & $C^{\ell k}$ & - & [0.8154, 1.5642] \B \\ \hline
	 \multicolumn{4}{l}{\textbf{Reference viscosity curve after Costa et al. (2009)}} \T \\
	 Critical solid fraction & $\phi_*$ & - & 0.62 \\
	 Initial slope & $\delta$ & - & 24 \\
	 Step function sharpness & $\gamma$ & - & 3.25\\
	 Phase offset & $\xi$ & - & 4e-5 \\
	 Einstein coefficient & $B$ & - & 4.0 \B \\ \hline
	\end{tabular}
\end{table}

\begin{table}[ppp]
\centering
	\caption{\textbf{Material \& fitting parameters for three-phase example}}
	\label{tab:fitting3}
	\begin{tabular}{llll}
	 \textbf{Parameter} & \textbf{Symbol} & \textbf{Units} & \textbf{Value} \B \\ \hline
	 \multicolumn{4}{l}{\textbf{Phase materials}} \T \\
	 Solid viscosity & $\etas_0$ & Pas & 10$^{18}$ \\
	 Liquid viscosity & $\etal_0$ & Pas & 10$^2$ \\
	 Vapour viscosity & $\etav_0$ & Pas & 10$^{-5}$ \\
	 Granular length scale & $d_0$ & m & 0.003 \B \\ \hline	 
	 \multicolumn{4}{l}{\textbf{Example permission weights}} \T \\
	 Solid slopes & $A^{s k}$ & - & [0.60, 0.25, 0.30] \\
	 Liquid slopes & $A^{\ell k}$ & - & [0.20, 0.20, 0.20] \\
	 Vapour slopes & $A^{v k}$ & - & [0.20, 0.20, 0.20] \\
	 Solid thresholds & $B^{s k}$ & - & [0.30, 0.15, 0.55] \\
	 Liquid thresholds & $B^{\ell k}$ & - & [0.48, 0.02, 0.50] \\
	 Vapour thresholds & $B^{v k}$ & - & [0.80, 0.08, 0.12] \\
	 Solid weights & $C^{s k}$ & - & [0.20, 0.20, 0.20] \\
	 Liquid weights & $C^{\ell k}$ & - & [0.60, 0.60, 0.12] \\
	 Vapour weights & $C^{v k}$ & - & [0.20, 0.25, 0.50] \B \\ \hline 
	\end{tabular}
\end{table}

\newpage

\setcounter{figure}{0}
\begin{figure}[ppp]
  \centering
  \includegraphics[width=0.9\textwidth]{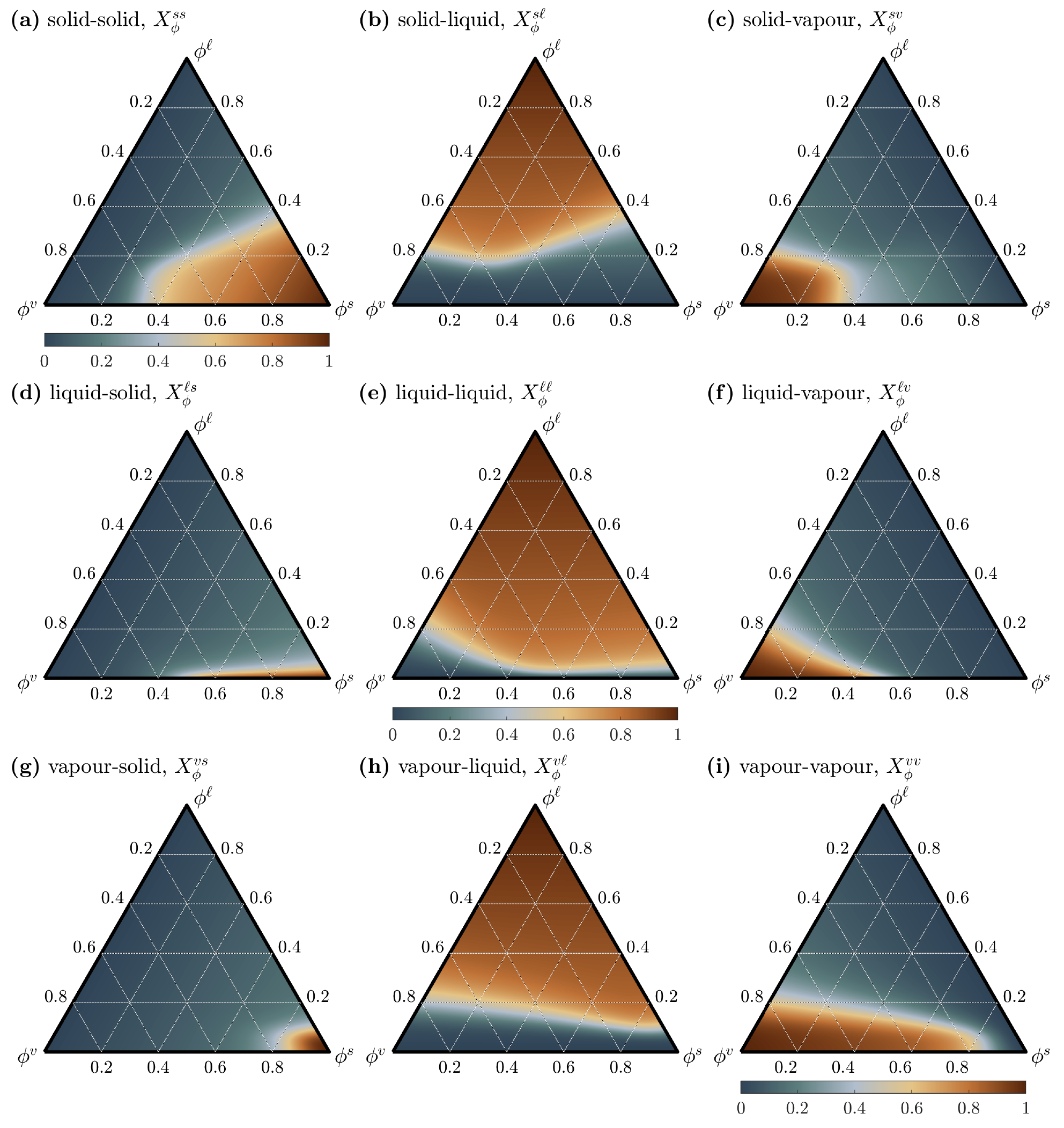}
  \caption{Permission weights for the example igneous three-phase calibration referenced in the main text. Brown color denotes high connectivity, teal denotes disconnected phases.}
  \label{fig:3connect}
\end{figure}

\begin{figure}[htb]
  \centering
  \includegraphics[width=0.9\textwidth]{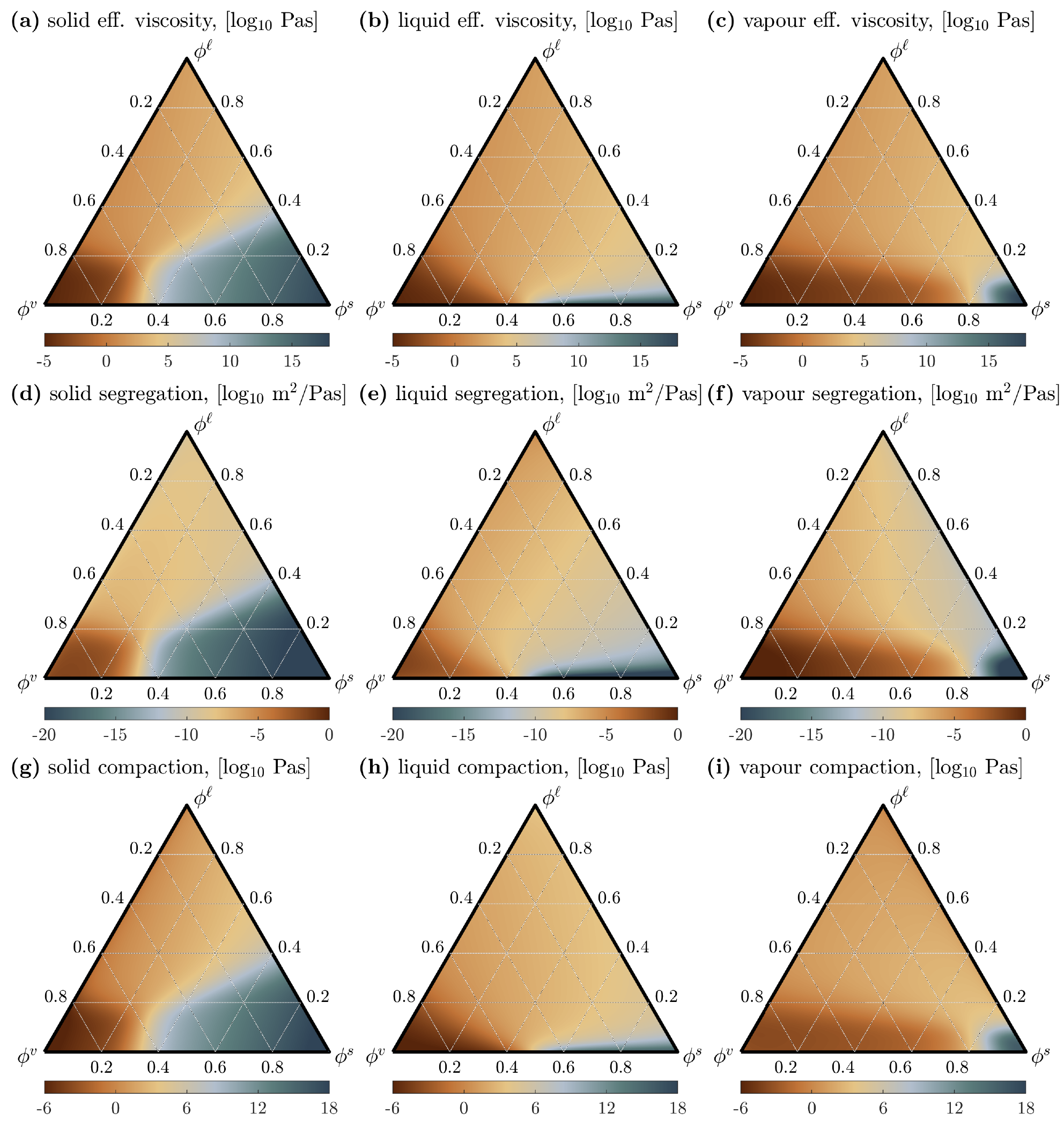}
  \caption{Effective viscosities, segregation, and compaction coefficients for the example igneous three-phase calibration referenced in the main text. Brown color denotes high phase mobility, teal denotes high resistance to flow or deformation.}
  \label{fig:3coeff}
\end{figure}

\begin{figure}[htb]
  \centering
  \includegraphics[width=0.9\textwidth]{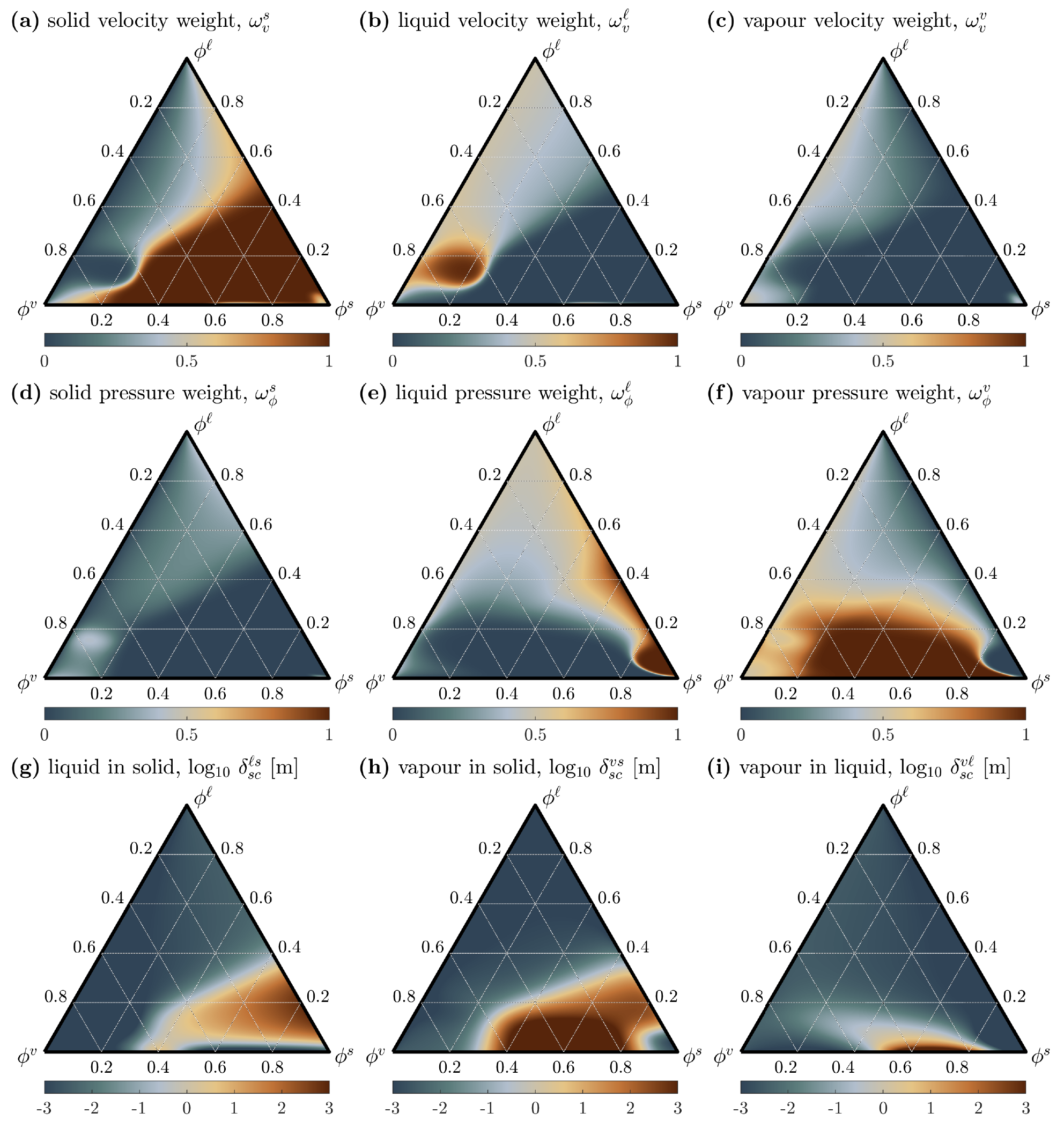}
  \caption{Coefficient-based weights for reference velocity and pressure, and the liquid-in-solid, vapour-in-solid, and vapour-in-liquid segregation-compaction lengths for the example igneous three-phase calibration referenced in the main text.}
  \label{fig:3transition}
\end{figure}

\end{document}